\documentclass[twocolumn,showkeys,showpacs,preprintnumbers,prd,superscriptaddress,nofootinbib]{revtex4-1}
\usepackage{graphicx}
\usepackage{epsf}
\usepackage{bm}
\usepackage{amsmath}
\usepackage{amsfonts}
\usepackage{amssymb}
\usepackage{epstopdf}
\usepackage{natbib}
\usepackage{hyperref}
\usepackage{color}
\usepackage{verbatim}
\usepackage{multirow}
\usepackage{bm}
\usepackage{hyperref}

\begin{document}

\title{Arbitrating the $S_8$ discrepancy with growth rate measurements from Redshift-Space Distortions}

\author{Rafael C. Nunes}
\email{rafadcnunes@gmail.com}
\affiliation{Divis\~{a}o de Astrof\'{i}sica, Instituto Nacional de Pesquisas Espaciais, Avenida dos Astronautas 1758, S\~{a}o Jos\'{e} dos Campos, 12227-010, S\~{a}o Paulo, Brazil}

\author{Sunny Vagnozzi}
\email{sunny.vagnozzi@ast.cam.ac.uk}
\affiliation{Kavli Institute for Cosmology (KICC) and Institute of Astronomy,\\University of Cambridge, Madingley Road, Cambridge CB3 0HA, United Kingdom}

\begin{abstract}
\noindent Within the $\Lambda$CDM model, measurements from recent Cosmic Microwave Background (CMB) and weak lensing (WL) surveys have uncovered a $\sim 3\sigma$ disagreement in the inferred value of the parameter $S_8 \equiv \sigma_8\sqrt{\Omega_m/0.3}$, quantifying the amplitude of late-time matter fluctuations. Before questioning whether the $S_8$ discrepancy calls for new physics, it is important to assess the view of measurements other than CMB and WL ones on the discrepancy. Here, we examine the role of measurements of the growth rate $f(z)$ in arbitrating the $S_8$ discrepancy, considering measurements of $f\sigma_8(z)$ from Redshift-Space Distortions (RSD). Our baseline analysis combines RSD measurements with geometrical measurements from Baryon Acoustic Oscillations (BAO) and Type Ia Supernovae (SNeIa), given the key role of the latter in constraining $\Omega_m$. From this combination and within the $\Lambda$CDM model we find $S_8 = 0.762^{+0.030}_{-0.025}$, and quantify the agreement between RSD+BAO+SNeIa and \textit{Planck} to be at the $2.2\sigma$ level: the mild disagreement is therefore compatible with a statistical fluctuation. We discuss combinations of RSD measurements with other datasets, including the $E_G$ statistic. This combination increases the discrepancy with \textit{Planck}, but we deem it significantly less robust. Our earlier results are stable against an extension where we allow the dark energy equation of state $w$ to vary. We conclude that, from the point of view of combined growth rate and geometrical measurements, there are hints, but no strong evidence yet, for the \textit{Planck} $\Lambda$CDM cosmology over-predicting the amplitude of matter fluctuations at redshifts $z \lesssim 1$. From this perspective, it might therefore still be premature to claim the need for new physics from the $S_8$ discrepancy.
\end{abstract}



\maketitle

\section{Introduction}
\label{sec:introduction}

The concordance $\Lambda$CDM model provides a wonderful fit to a wide variety of observations~\cite{Riess:1998cb,Perlmutter:1998np,Aghanim:2018eyx,Aiola:2020azj,Alam:2020sor}. However, the increase in precision and sensitivity of recent surveys, brought about by remarkable experimental developments, has uncovered intriguing discrepancies among parameters inferred from independent measurements.

One of these discrepancies is the well-known $H_0$ tension, referring to discrepancies between various late- and early-time independent measurements of the Hubble constant $H_0$~\cite{Aghanim:2018eyx,Riess:2019cxk,Wong:2019kwg,Freedman:2019jwv,Aiola:2020azj}. Whether the $H_0$ tension calls for new physics, and what this new physics might be, are the subject of an ongoing and rapidly evolving research direction (see e.g. Refs.~\cite{Bernal:2016gxb,Mortsell:2018mfj,Poulin:2018cxd,Kreisch:2019yzn,Vagnozzi:2019ezj,Visinelli:2019qqu,Sakstein:2019fmf,Hill:2020osr,Ballesteros:2020sik,Braglia:2020iik,Efstathiou:2020wxn,Das:2020xke,Choudhury:2020tka,Brinckmann:2020bcn,Efstathiou:2021ocp,De_Felice_2021} for discussions, and Refs.~\cite{Verde:2019ivm,DiValentino:2020zio,DiValentino:2021izs} for recent reviews). A milder yet not less enduring discrepancy is also present between Cosmic Microwave Background (CMB) and low-redshift probes of the amplitude of matter fluctuations, affecting $\sigma_8$ (the present day linear theory amplitude of matter fluctuations averaged in spheres of radius $8\,h^{-1}{\rm Mpc}$) and the matter density parameter $\Omega_m$: this discrepancy is best captured by the parameter $S_8 \equiv \sigma_8\sqrt{\Omega_m/0.3}$, which reflects the main degeneracy direction of weak lensing measurements.

Within the context of the $\Lambda$CDM model, CMB anisotropy measurements from \textit{Planck} and ACT+WMAP indicate $S_8$ values of $0.834 \pm 0.016$~\cite{Aghanim:2018eyx} and $0.840 \pm 0.030$~\cite{Aiola:2020azj} respectively. On the other hand, the value of $S_8$ inferred by a host of weak lensing and galaxy clustering measurements is typically lower than the CMB-inferred values, ranging between $0.703$ and $0.782$: examples of surveys reporting lower values of $S_8$ include CFHTLenS~\cite{Joudaki:2016mvz}, KiDS-450~\cite{Joudaki:2016kym}, KiDS-450+2dFLenS~\cite{Joudaki:2017zdt}, KiDS+VIKING-450 (KV450)~\cite{Hildebrandt:2018yau}, DES-Y1~\cite{Troxel:2017xyo}, KV450+BOSS~\cite{Troster:2019ean}, KV450+DES-Y1~\cite{Joudaki:2019pmv,Asgari:2019fkq}, a re-analysis of the BOSS galaxy power spectrum~\cite{Ivanov:2019pdj}, KiDS-1000~\cite{Asgari:2020wuj}, and KiDS-1000+BOSS+2dFLenS~\cite{Heymans:2020ghw}. \textit{Planck} Sunyaev-Zeldovich cluster counts also infer a rather low value of $S_8=0.774 \pm 0.034$~\cite{Ade:2015fva}. To balance the discussion, it is also worth remarking that KiDS-450+GAMA~\cite{vanUitert:2017ieu} and HSC SSP~\cite{Hamana:2019etx} indicate higher values of $S_8$, of $0.800^{+0.029}_{-0.027}$ and $0.804^{+0.032}_{-0.029}$ respectively. These measurements are summarized in Fig.~\ref{fig:S8_summary} alongside our key results.

While the status of the $S_8$ discrepancy is perhaps somewhat less clear than that of the $H_0$ tension, it is beyond question that there overall is some disagreement between high- and low-redshift probes of the amplitude of matter fluctuations (see for instance Ref.~\cite{DiValentino:2020vvd} for a concise review of the problem). It is thus worthwhile to investigate whether new physics might solve or at least alleviate the $S_8$ discrepancy, a possibility which has been investigated in several works. Models which have been contemplated in this sense include for example active and sterile neutrinos~\cite{Battye:2013xqa,MacCrann:2014wfa,Vagnozzi:2017ovm,McCarthy:2017csu,Feng:2017nss}, ultra-light axions~\cite{Hlozek:2014lca}, decaying dark matter (DM)~\cite{Enqvist:2015ara,DiValentino:2017oaw,Chudaykin:2017ptd,Pandey:2019plg,Xiao:2019ccl,Abellan:2020pmw,Chen:2020iwm,Abellan:2021bpx}, extended or exotic DM and/or dark energy (DE) models and interactions~\cite{Kunz:2015oqa,Pourtsidou:2016ico,Kumar:2016zpg,Gariazzo:2017pzb,Benetti:2017juy,Buen-Abad:2017gxg,Poulin:2018zxs,Kumar:2018yhh,Lambiase:2018ows,Dutta:2018vmq,Kumar:2019gfl,Kumar:2019wfs,Archidiacono:2019wdp,DiValentino:2019ffd,DiValentino:2019jae,Vagnozzi:2019kvw,Chamings:2019kcl,Jimenez:2020ysu,Heimersheim:2020aoc,Choi:2020pyy} including unified dark sector models~\cite{Camera:2017tws}, modified gravity models~\cite{De_Felice_2017,Dossett:2015nda,Nesseris:2017vor,Kazantzidis:2018rnb,Kazantzidis:2019dvk,Skara:2019usd,Zumalacarregui:2020cjh,Barros:2020bgg,Marra:2021fvf,De_Felice_2020}, and more generally extended parameter spaces~\cite{DiValentino:2018gcu,DiValentino:2019dzu}, among the others. It is also worth noting that most of the models invoked to address the $S_8$ discrepancy do so at the expense of worsening the $H_0$ tension, and vice-versa, ~\cite{Vagnozzi:2018jhn,Poulin:2018zxs,Kumar:2019wfs,Hill:2020osr,Alestas:2021xes}, highlighting the importance of a conjoined analysis of the two tensions~\cite{DiValentino:2020kha,DiValentino:2020vvd}.

The possibility that the $S_8$ discrepancy might be at least partially due to systematics cannot be completely excluded, as discussed for instance in Ref.~\cite{Efstathiou:2017rgv} in the context of the KiDS-450 measurements. In this sense, it is important to look at the $S_8$ discrepancy through different eyes, \textit{i.e.} through datasets other than CMB and weak lensing measurements, which might be able to arbitrate the discrepancy or at least point us towards the ingredients needed to resolve it. To draw a parallel with the $H_0$ tension, the inverse distance ladder take on the tension has been instrumental towards narrowing down plausible solutions~\cite{Bernal:2016gxb,Lemos:2018smw,Aylor:2018drw,Schoneberg:2019wmt,Knox:2019rjx}. Broadly speaking, the question we are then interested in is: ``\textit{Is there strong evidence from data other than weak lensing measurements for the \textit{Planck} $\Lambda$CDM cosmology over-predicting the amplitude of matter fluctuations at $z \lesssim 1$}?'' In other words, we want to compare the CMB and weak lensing inferences of $S_8$ against other techniques which can also measure the amplitude of the spectrum of matter fluctuations. Anticipating the answer to the previous question, we will find that there are indeed hints from combined growth and geometrical measurements, but no strong evidence.

We shall address this question making use of measurements of the growth rate of matter density perturbations $f(z)$, as inferred from the peculiar velocities arising from Redshift Space Distortions (RSD) measurements~\cite{Kaiser:1987qv}, which typically constrain the combination $f\sigma_8(z)$. We will combine RSD measurements with two additional classes of probes: \textit{a)} geometrical probes of distances and expansion rates such as Baryon Acoustic Oscillation (BAO), uncalibrated Supernovae Type Ia (SNeIa), and cosmic chronometer (CC) measurements; \textit{b)} the $E_G$ statistic~\cite{Zhang:2007nk}, which measures a combination of gravitational lensing, galaxy clustering, and redshift-space distortions, probing a combination of the two metric potentials, and which is insensitive to galaxy bias and $\sigma_8$ in the linear regime. We will assess the status of the $S_8$ discrepancy in light of the RSD+BAO+SNeIa(+CC) and RSD+$E_G$ dataset combinations, both within the concordance $\Lambda$CDM model and within the 1-parameter $w$CDM extension where the DE equation of state (EoS) $w$ is allowed to vary, to check whether the discrepancy can be alleviated within this extension. We note that related analyses have been conducted in e.g.~\cite{Nesseris:2017vor,Efstathiou:2017rgv,Kazantzidis:2018rnb,Quelle:2019vam,Skara:2019usd,Li:2019nux,Benisty:2020kdt,Garcia-Quintero:2020bac}.

The rest of this paper is then structured as follows. In Section~\ref{sec:data} we present the datasets and statistical methodology used in our analysis. Our results are discussed in Section~\ref{sec:results}, with Section~\ref{subsec:wcdm} reporting the results within the $w$CDM model. We draw concluding remarks in Section~\ref{sec:conclusions}. We invite the busy reader to skip to Fig.~\ref{fig:S8LCDM}, Tab.~\ref{tab:LCDM}, and especially Fig.~\ref{fig:S8_summary}, where they will find the main results of this paper conveniently summarized.

\section{Datasets and Methodology}
\label{sec:data}

In the following, we first discuss the datasets we make use of. We then discuss our analysis methods, in particular our choice of cosmological parameters and tension metric used to assess the concordance or discordance between the adopted datasets and the \textit{Planck} CMB measurements, within the context of the cosmological models being considered.

\subsection{$f\sigma_8$ measurements}
\label{subsec:fs8}

As discussed in the Introduction, the key dataset we will use to try and arbitrate the $S_8$ discrepancy, independently of CMB and weak lensing measurements, are Redshift Space Distortions (RSD) measurements. Recall that RSD are a velocity-induced mapping from real- to redshift-space due to line-of-sight peculiar motions of objects, which introduce anisotropies in their clustering patterns~\cite{Kaiser:1987qv}. This effect depends on the growth of structure, making RSD probes sensitive to the combination $f\sigma_8$, with $f$ the logarithmic derivative of the linear growth rate $D(a)$ with respect to the scale factor $a$:
\begin{eqnarray}
f(a) \equiv \frac{d\ln D(a)}{d\ln a}\,.
\label{eq:f}
\end{eqnarray}
On sub-horizon scales and in the linear regime, the evolution equation for $f(a)$ is given by:
\begin{eqnarray}
\frac{df(a)}{d\ln a} + f^2 + \left ( 2 + \frac{1}{2} \frac{d \ln H(a)^2}{d \ln a} \right ) f- \frac{3}{2} \Omega_m(a) = 0\,,
\end{eqnarray}
where $\Omega_m(a) \equiv \Omega_{m,0}a^{-3}H_0^2/H(a)^2$, with $\Omega_{m,0} \equiv \Omega_m$ the matter density parameter today, and $H(a)$ is the Hubble rate as a function of scale factor. Within the $\Lambda$CDM model, and assuming gravity is described by General Relativity (GR), $f(a)$ scales to good approximation as $f(a) \propto \Omega_m(a)^{0.55}$~\cite{Lahav:1991wc}.

Let us now consider the matter over-density field $\delta_m$. On sub-horizon scales, and assuming that DE does not cluster, the growth equation which governs the evolution of $\delta_m$ is given by:
\begin{eqnarray}
\begin{aligned}
{\delta}_m''(a) + &\left(\frac{3}{a}+\frac{H'(a)}{H(a)}\right){\delta}_m'(a)-\frac{3}{2} \frac{\Omega_m(a)}{a^2} \delta_m(a) = 0,
\end{aligned}
\label{eq:deltam}
\end{eqnarray}
with the prime denoting a derivative with respect to the scale factor $a$. It is worth noting that Eq.~(\ref{eq:deltam}) admits a closed-form solution in terms of the Gaussian hypergeometric function $_2F_1$:
\begin{eqnarray}
\delta_m(a)= a\,_2F_1 \left [ \frac{1}{3}\,,1\,;\frac{11}{6}\,;a^3 ·\left ( 1-\frac{1}{\Omega_m} \right ) \right ]\,.
\label{eq:deltamsolution}
\end{eqnarray}
Redshift surveys can constrain the quantity $f(a)\sigma_8(a) \equiv f\sigma_8(a)$ [or equivalently $f\sigma_8(z)$], which is given by:
\begin{eqnarray}
f\sigma_8(a)=a\frac{\delta_m'(a)}{\delta_m(a_0)}\sigma_{8,0}\,,
\label{eq:fs8}
\end{eqnarray}
with $f$ given by Eq.~(\ref{eq:f}), and $\sigma_8(a)$ given by:
\begin{eqnarray}
\sigma_8(a)=\frac{\delta_m(a)}{\delta_m(1)}\sqrt{\int_0^{\infty}dk\,\frac{k^2P(k)W^2_R(k)}{2\pi^2}}\,.
\label{eq:sigma8a}
\end{eqnarray}
where $W_R(k) = 3[\sin(kR)/kR - \cos(kR)]/(kR)^2$ is the Fourier transform of the top-hat window function, with $R$ the appropriate scale over which the RMS normalization of matter fluctuations is being computed.

Several measurements of $f\sigma_8(a)$ from a variety of different surveys, each making different assumptions (in particular assumptions on the reference value of $\Omega_m$) and subject to different systematics, exist in the literature. Before using any one of them, it is imperative to assess their internal consistency. Such an analysis was recently performed in the context of a Bayesian model comparison framework in Ref.~\cite{Sagredo:2018ahx}, which was able to identify potential outliers as well as subsets of data affected by systematics or new physics. It is worth noting that, within a $\Lambda$CDM+GR framework, RSD measurements of $f\sigma_8$ essentially measure the combination $\sigma_8\Omega_m^{0.55}$, which up to a known constant is closely related to $S_8$.

In this work, we shall make use of the RSD measurements of $f\sigma_8(z)$ provided in Tab.~I of Ref.~\cite{Sagredo:2018ahx}, consisting of 22 measurements of $f\sigma_8(z)$ in the redshift range $0.02<z<1.944$ obtained from the following surveys: 2dFGRS~\cite{Song:2008qt}, 2MASS~\cite{Davis:2010sw}, SDSS-II LRGs~\cite{Samushia:2011cs}, First Amendment SNeIa+IRAS~\cite{Turnbull:2011ty,Hudson:2012gt}, WiggleZ~\cite{Blake:2012pj}, GAMA~\cite{Blake:2013nif}, BOSS DR11 LOWZ~\cite{Sanchez:2013tga}, BOSS DR12 CMASS~\cite{Chuang:2013wga}, SDSS DR7 MGS~\cite{Howlett:2014opa} and SDSS DR7~\cite{Feix:2015dla}, FastSound~\cite{Okumura:2015lvp}, Supercal SNeIa+6dFGS~\cite{Huterer:2016uyq}, VIPERS PDR-2~\cite{Pezzotta:2016gbo}, and eBOSS DR14 quasars~\cite{Zhao:2018gvb}. We refer to these measurements as \textit{RSD}, and further note that these are commonly referred to as the ``Gold 2018'' sample in the literature.

We note that in principle many more measurements of $f\sigma_8$ other than the adopted ones are available (see e.g. Tab.~II of Ref.~\cite{Nesseris:2017vor}). However, as noted in Refs.~\cite{Nesseris:2017vor,Sagredo:2018ahx}, within this enlarged set, not all the measurements are independent, and hence should not be used at the same time without a proper modelling of the cross-covariance. The extensive analyses of Refs.~\cite{Nesseris:2017vor,Sagredo:2018ahx} have allowed for the overlap between these measurements to be minimized, while in turn ensuring that their independence is maximized. Our analysis properly accounts for the covariance between measurements at different redshifts originating from the same analysis (as e.g. in the case of the WiggleZ and eBOSS measurements). Finally, we note that our analysis properly accounts for the so-called ``growth correction'', first discussed in Ref.~\cite{Nesseris:2017vor}, which corrects for the different assumptions of each survey concerning the fiducial values of $\Omega_m$ and $\sigma_8$.

\subsection{$E_G$ measurements}
\label{subsec:eg}

The $E_G$ statistic was first proposed in Ref.~\cite{Zhang:2007nk} as a means of testing deviations from GR, while avoiding potential degeneracies with galaxy bias and $\sigma_8$. $E_G$ measures a combination of gravitational lensing, galaxy clustering, and redshift-space distortions, probing a combination of the two metric potentials, and is insensitive to galaxy bias and $\sigma_8$ in the linear regime. These measurements will be of interest to us given their dependence on the growth factor $f$. $E_G$ is defined as the expectation value of the estimator $\hat{E_G}$, originally defined as~\cite{Zhang:2007nk}:
\begin{eqnarray}
\hat{E_G} = \frac{aC_{\kappa g}(\ell\,,\Delta\ell)}{3H_0^2\sum_{\alpha}j_{\alpha}(\ell\,,\Delta\ell)P_{vg}^{\alpha}}\,,
\label{eq:estimatoreg}
\end{eqnarray}
where for a given multipole $\ell$ and bin of size $\Delta\ell$, and wavenumbers labelled by $k_{\alpha}$, $C_{\kappa g}$ is the lensing convergence-galaxy overdensity cross-correlation, $P_{vg}$ is the galaxy velocity-overdensity cross-spectrum, and $j_{\alpha}$ is an appropriate weighting function which transforms $P_{vg}$ to an angular power spectrum. The expectation value of Eq.~(\ref{eq:estimatoreg}), and hence $E_G$, is given by~\cite{Zhang:2007nk}:
\begin{eqnarray}
E_G = \left [ \frac{a\nabla^2( \Psi + \Phi)}{3H_0^2f\delta_m} \right ]\,,
\label{eq:egexpectation}
\end{eqnarray}
where $\Psi$ and $\Phi$ are the two Newtonian potentials, which appear in the perturbed FLRW metric in conformal Newtonian gauge, and are equal to each other in GR and in the absence of anisotropic stress. For other works examining important theoretical or observational aspects of $E_G$ as a means of testing fundamental physics, we refer the reader for instance to Refs.~\cite{Reyes:2010tr,Amendola:2012ky,Pullen:2014fva,Blake:2015vea,Leonard:2015cba,Pullen:2015vtb,Alam:2016qcl,delaTorre:2016rxm,Amon:2017lia,Singh:2018flu,Blake:2020mzy,Zhang:2020vru}.

Assuming that on the largest scales gravity is correctly described by GR, Eq.~(\ref{eq:egexpectation}) reduces to~\cite{Zhang:2007nk,Amendola:2012ky,Leonard:2015cba}:
\begin{eqnarray}
E_G(z) = \frac{\Omega_m}{f(z)}\,,
\label{eq:eg}
\end{eqnarray}
which is clearly independent of $\sigma_8$ and linear bias. Moreover, note that $E_G$ is expected to be scale-independent not only within GR, but more generally within any theory of gravity captured by a scale-dependent effective Newtonian constant, with a scale-independent relationship between $\Phi$ and $\Psi$. Note, however, that $E_G$ is strictly speaking scale-independent only at linear level. On smaller scales, non-linearities associated to galaxy clustering, galaxy biasing, and weak lensing, make $E_G$ slightly scale-dependent (see e.g. Ref.~\cite{Leonard:2015cba} for further discussions). From Eq.~(\ref{eq:eg}), we see that within $\Lambda$CDM+GR, $E_G \propto \Omega_m^{-0.45}$. It is also clear that combining RSD measurements of $f\sigma_8(z)$ with $E_G$ measurements can enormously help in disentangling $f(z)$ and $\sigma_8(z)$~\cite{Skara:2019usd}. This allows for better constraints on $\sigma_8$, which in turn can help arbitrate the $S_8$ discrepancy.

In this work, we make use of the $E_G$ measurements compiled in Table~7 of Ref.~\cite{Pinho:2018unz}, which we collectively refer to as $E_G$. This consists of 9 measurements of $E_G(z)$ in the range $0.09<z<0.48$. Of these 9 points, 4 have been obtained from a joint analysis of RCSLenS and CFHTLenS imaging and WiggleZ and BOSS spectroscopy~\cite{Blake:2015vea}; 2 from a joint analysis of CFHTLenS imaging and VIPERS spectroscopy~\cite{delaTorre:2016rxm}; and 3 from a joint analysis of KiDS-450 imaging and 2dFLenS, BOSS, and GAMA spectroscopy~\cite{Amon:2017lia}. These $E_G$ measurements probe scales in the range $3\,h^{-1}\,{\rm Mpc}<R<60\,h^{-1}\,{\rm Mpc}$ and well in the linear regime. We treat the 9 $E_G$ measurements as being statistically uncorrelated, thus approximating the likelihood as being a multivariate Gaussian in $E_G$ with diagonal covariance matrix: Refs.~\cite{Blake:2015vea,delaTorre:2016rxm,Amon:2017lia} suggest that the covariance between $E_G$ measurements at two different redshift bins from the same analysis may be neglected, whereas to the best of our knowledge the covariance between measurements from different analyses has not been estimated in the literature.

\subsection{Other measurements}
\label{subsec:other}

In addition to $f\sigma_8$ (RSD) and $E_G$ measurements, we consider three additional geometrical measurements of distances and expansion rates, based on the use of standard rulers, standard candles, and standard clocks:
\begin{itemize}
\item Baryon Acoustic Oscillation (BAO) distance and expansion rate measurements from the 6dFGS~\cite{Beutler:2011hx}, SDSS-DR7 MGS~\cite{Ross:2014qpa}, BOSS DR12~\cite{Alam:2016hwk} galaxy surveys, as well as from eBOSS DR14 Lyman-$\alpha$ (Ly$\alpha$) absorption~\cite{Agathe:2019vsu} and Ly$\alpha$-quasars cross-correlation~\cite{Blomqvist:2019rah}. These consist of isotropic BAO measurements of $D_V(z)/r_d$ (with $D_V(z)$ and $r_d$ the spherically averaged volume distance, and sound horizon at baryon drag respectively) for 6dFGS and MGS, and anisotropic BAO measurements of $D_M(z)/r_d$ and $D_H(z)/r_d$ (with $D_M(z)$ the comoving angular diameter distance and $D_H(z)=c/H(z)$ the Hubble distance) for BOSS DR12, eBOSS DR14 Ly$\alpha$, and eBOSS DR14 Ly$\alpha$-quasars cross-correlation. At the time of writing, the covariance matrix for the legacy eBOSS BAO measurements~\cite{Alam:2020sor} was not publicly available, which is the reason why we instead opted for these older measurements. At any rate, we expect that adopting these newer measurements should not qualitatively affect our results.
\item Type Ia Supernovae (SNeIa) distance moduli measurements from the \textit{Pantheon} sample, consisting of 1048 SNeIa in the range $0.01<z<2.3$~\cite{Scolnic:2017caz}. These measurements constrain the uncalibrated luminosity distance $H_0d_L(z)$, or in other words the slope of the late-time expansion rate (which in turn constrains $\Omega_m$). We refer to this dataset as \textit{Pantheon}.
\item Cosmic chronometer measurements of $H(z)$. These consist of measurements of $H(z)$ from the differential age evolution of massive, early-time, passively evolving galaxies, which act as standard clocks~\cite{Jimenez:2001gg}. We make use of 31 CC measurements of $H(z)$ in the range $0.07<z<1.965$, compiled in Refs.~\cite{Jimenez:2003iv,Simon:2004tf,Stern:2009ep,Moresco:2012by,Zhang:2012mp,Moresco:2015cya,Moresco:2016mzx,Ratsimbazafy:2017vga}. We refer to this dataset as \textit{CC}.
\end{itemize}
We consider three different dataset combinations, all of which involve the \textit{RSD} dataset: \textit{RSD}+\textit{BAO}+\textit{Pantheon}, \textit{RSD}+\textit{BAO}+\textit{Pantheon}+\textit{CC}, and \textit{RSD}+$E_G$. Of the three, we consider the \textit{RSD}+\textit{BAO}+\textit{Pantheon} one to be the most robust one, and treat it as our baseline dataset combination. In particular, combining \textit{BAO} and \textit{Pantheon} measurements produces tight constraints on $\Omega_{m,0}$ which, once combined with the \textit{RSD} measurements, improves the constraints on $\sigma_8$. We note that, as we are assuming the validity of the $\Lambda$CDM model at high redshifts, we can compute the sound horizon $r_s$ given a Big Bang Nucleosynthesis (BBN) prior on $\omega_b$ (discussed in the paragraph below). Hence, the \textit{BAO}+\textit{Pantheon} combination corresponds to an \textit{inverse distance ladder} anchored to the early-Universe determination of $r_s$. In addition, we will occasionally also report the constraints we obtain from the \textit{RSD} dataset alone.

Model-wise, we consider a standard $\Lambda$CDM+GR model, spanned by the following 4 parameters: the Hubble constant $H_0$ or equivalently the reduced Hubble constant $h \equiv H_0/(100\,{\rm km}\,{\rm s}^{-1}\,{\rm Mpc}^{-1})$, the physical baryon density $\omega_b \equiv \Omega_bh^2$, the physical cold dark matter density $\omega_c \equiv \Omega_ch^2$, and $\sigma_8$. The matter density parameter today $\Omega_m$ is treated as a derived parameter, whose value is given by $\Omega_m = (\omega_b+\omega_c)/h^2$. Another important derived parameter is $S_8 \equiv \sigma_8\sqrt{\Omega_m/0.3}$.  To constrain the physical baryon density, we adopt a Gaussian prior on $\omega_b$ from Big Bang Nucleosynthesis (BBN): $100\omega_b =  2.233 \pm 0.036$~\cite{Mossa:2020gjc}. In what follows, the use of the BBN prior on $\omega_b$ will be implicitly assumed. With the exception of $\omega_b$, for which we adopt a Gaussian prior as discussed above, we adopt flat priors on all other cosmological parameters. At a later stage, we consider a 1-parameter extension of the previous model, where the dark energy equation of state (DE EoS) $w$ is allowed to vary. We refer to this extended model as $w$CDM.~\footnote{Although in principle interesting, we do not consider an extended cosmology involving spatial curvature $\Omega_K$ since it is known that, despite the apparent indication for a closed Universe from \textit{Planck} primary CMB measurements~\cite{Handley:2019tkm,DiValentino:2019qzk}, $\Omega_K$ is too well constrained close to spatial flatness by combining \textit{Planck} data with other datasets which break the geometrical degeneracy~\cite{Ryan:2018aif,Park:2018tgj,Efstathiou:2020wem,Chudaykin:2020ghx,Vagnozzi:2020zrh,Vagnozzi:2020dfn,Cao:2021ldv}. In addition, including $\Omega_K$ as a free parameter results in most cases in the $S_8$ discrepancy being considerably worsened (see for instance Ref.~\cite{DiValentino:2020hov}).}

We note, however, that our results might actually be seen applying more generally than just to $\Lambda$CDM. In fact, the evolution equation for $\delta_m$, Eq.~(\ref{eq:deltam}), which is our main equation as far as the interpretation of RSD measurements goes, really only assumes \textit{a)} the validity of GR, and that \textit{b)} DE does not cluster. The amount of matter is then constrained by the \textit{BAO}+\textit{Pantheon} dataset combination. While we will keep referring to the $\Lambda$CDM model in the remainder of our paper, the reader should keep in mind that the associated results are in fact more general than that.

We use Monte Carlo Markov Chain (MCMC) methods to sample the posterior distributions of the parameters considered. To generate our MCMC chains, we make use of the cosmological MCMC sampler \texttt{MontePython}~\cite{Blas:2011rf,Audren:2012wb,Brinckmann:2018cvx}, while theoretical predictions for the cosmological observables are computed through \texttt{CLASS}~\cite{Lesgourgues:2011re,Blas:2011rf}. We monitor the convergence of the generated MCMC chains via the Gelman-Rubin parameter $R-1$~\cite{Gelman:1992zz}, and require $R-1<0.001$ for the chains to be considered converged.

\subsection{Tension metrics}
\label{subsec:tension}

Once we have obtained constraints on the above cosmological parameters, and in particular the derived parameters $\Omega_m$ and $S_8$, our next goal is to quantify the level of concordance or discordance (if any) between the dataset combinations we have considered and the \textit{Planck} CMB measurements. Consider two datasets $i$ and $j$ for which the inferred values of $S_8$ are $S_{8,i} \pm \sigma_{S_{8,i}}$ and $S_{8,j} \pm \sigma_{S_{8,j}}$ respectively. Then, if one focuses solely on $S_8$, a na\"{i}ve 1D tension metric, which we refer to as $T_{S_8}$, can be constructed by the following:
\begin{equation}
T_{S_8} \equiv \frac{S_{8,i}-S_{8,j}}{\sqrt{\sigma^2_{S_{8,i}}+\sigma^2_{S_{8,j}}}}\,,
\label{eq:1D_estimator}
\end{equation}
where the value of $T_{S_8}$ can directly be interpreted as level of tension in equivalent Gaussian $\sigma$s. We note that $T_{S_8}$ was already used in a similar context by Refs.~\cite{Hildebrandt:2016iqg,Joudaki:2016kym}. While this tension metric is a good starting point, it can underestimate the level of tension due to its only focusing on one particular direction of parameter space. A more robust tension metric should instead take the whole parameter space into consideration, accounting for correlations between parameters.

To construct a more robust tension metric, we make use of the quadratic estimator proposed in Ref.~\cite{Addison:2015wyg}, which robustly assesses whether the differences between correlated parameters inferred from two different datasets are consistent with zero. Considering once more two independent datasets $i$ and $j$, we can assess the level of concordance or discordance between the two by considering the vector of differences of mean parameter values, treating it as being distributed according to a multivariate Gaussian distribution with zero mean and covariance given by the sum of the covariance matrices of the individual datasets. In practice, we construct the following test statistic:
\begin{eqnarray}
\chi^2 = (\mathbf{x}_i - \mathbf{x}_j)^{T}({\cal C}_i + {\cal C}_j)^{-1}(\mathbf{x}_i - \mathbf{x}_j)\,,
\label{eq:quadratic_estimator}
\end{eqnarray}
where $\mathbf{x}_i$ and $\mathbf{x}_j$ are the vectors containing the mean values for the cosmological parameters inferred from datasets $i$ and $j$ respectively, and similarly ${\cal C}_i$ and ${\cal C}_j$ are the covariance matrices for these datasets. It can easily be seen that Eq.~(\ref{eq:quadratic_estimator}) essentially corresponds to a generalized Mahalanobis distance between $\mathbf{x}_i$ and $\mathbf{x}_j$.

The significance of a given value of the test statistic $\chi^2$ is then converted to an equivalent Gaussian $\sigma$ level. We compute the test statistic in Eq.~(\ref{eq:quadratic_estimator}) over the whole 4-dimensional parameter space (5-dimensional when we also vary the DE EoS $w$), to fully account for correlations between the parameters. For each dataset, we estimate the parameter mean vector and covariance matrx directly from our MCMC chains. In closing, we also note that the same quadratic tension estimator was recently used by the ACT collaboration in Ref.~\cite{Aiola:2020azj} to quote the level of concordance with the \textit{Planck} measurements. For a selection of other tension metrics discussed in the recent literature, we refer the reader to e.g. Refs.~\cite{Karpenka:2014moa,MacCrann:2014wfa,Lin:2017ikq,Lin:2017bhs,Adhikari:2018wnk,Raveri:2018wln,Nicola:2018rcd,Handley:2019wlz,Handley:2019pqx,Garcia-Quintero:2019cgt,Lemos:2019txn,Raveri:2019gdp}.

\section{Results}
\label{sec:results}

\begin{figure*}
\begin{center}
\includegraphics[width=2.8in]{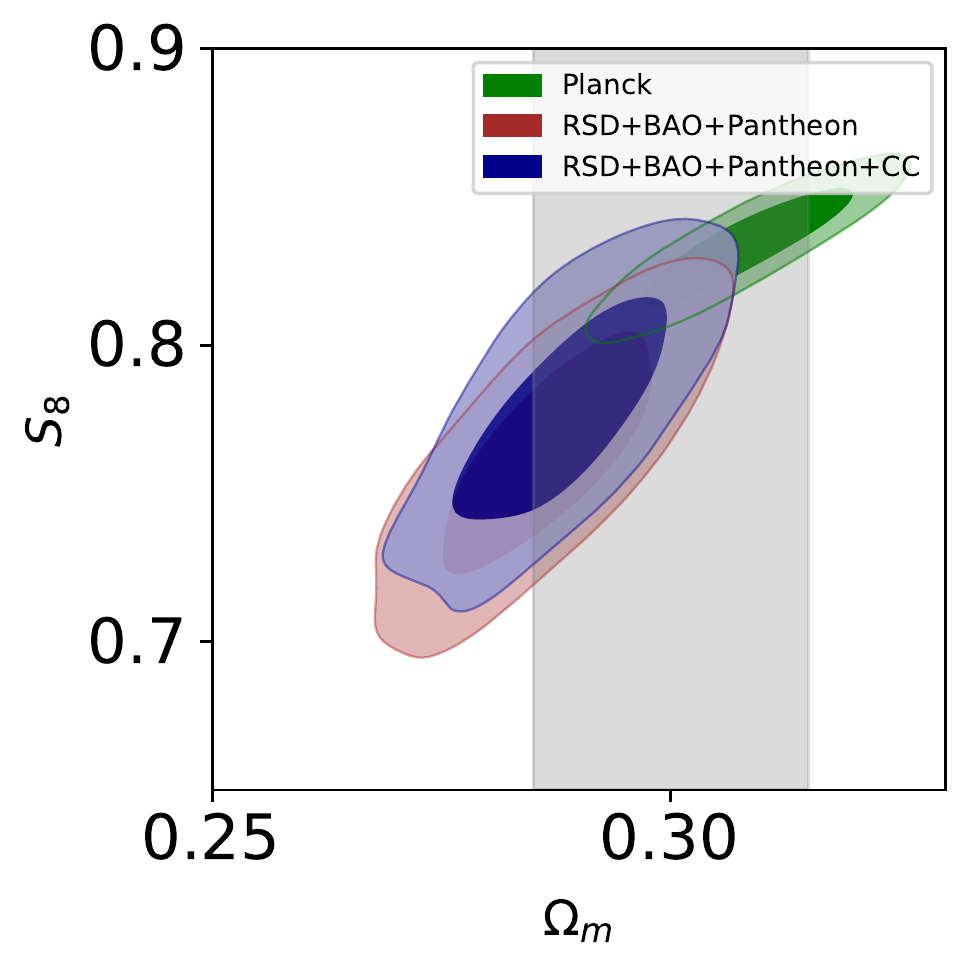} \,\,\,\,\,\,\,\, 
\includegraphics[width=2.8in]{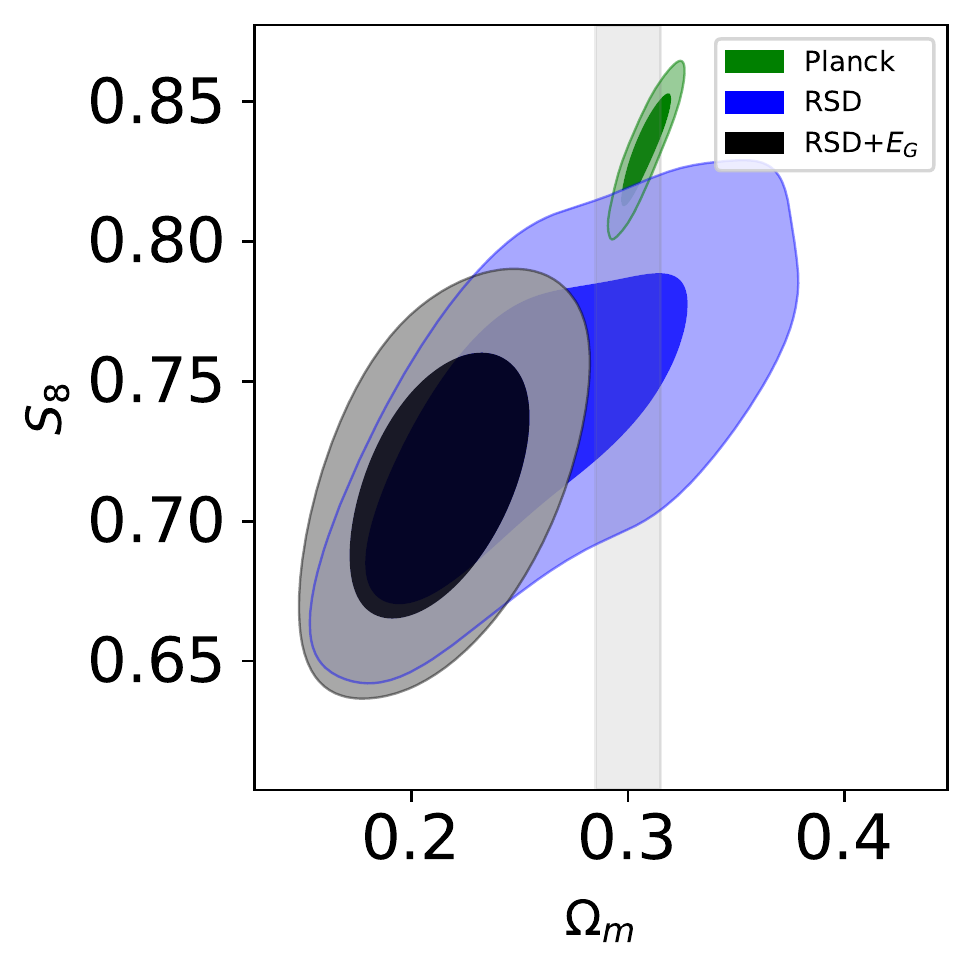}
\caption{\textit{Left panel}: 2D joint posterior distributions in the $S_8$-$\Omega_m$ plane, with the corresponding 68\%~C.L. and 95\%~C.L. contours, obtained from the following datasets/dataset combinations within the $\Lambda$CDM model: \textit{Planck} (green contours), \textit{RSD}+\textit{BAO}+\textit{Pantheon} (magenta contours), and \textit{RSD}+\textit{BAO}+\textit{Pantheon}+\textit{CC} (dark blue contours). \textit{Right panel}: as for the left panel, but considering the \textit{Planck} (green contours), \textit{RSD} (light blue contours), and \textit{RSD}+$E_G$ (black contours) dataset combinations respectively. In both the left and right panels, the vertical grey bands denote the $68\%$~C.L. interval on $\Omega_m=0.298 \pm 0.015$ obtained from \textit{BAO}+\textit{Pantheon}. The level of agreement or tension between \textit{Planck} and dataset combinations considered is quantified in the two rightmost columns of Tab.~\ref{tab:LCDM}. The \textit{RSD}+\textit{BAO}+\textit{Pantheon} and \textit{RSD}+\textit{BAO}+\textit{Pantheon}+\textit{CC} dataset combinations are the most robust ones, and should be considered as baseline dataset combinations.}
\label{fig:S8LCDM}
\end{center}
\end{figure*}

\begin{table*}
\centering
\scalebox{1.0}{
\begin{tabular}{|c||c|c|c||c|c|}
\hline
Dataset & $\Omega_m$ & $\sigma_8$ & $S_8$ & Tension [Eq.~(\ref{eq:1D_estimator})] & Tension [Eq.~(\ref{eq:quadratic_estimator})] \\ 
\hline
\textit{RSD}+\textit{BAO}+\textit{Pantheon} & $0.286 \pm 0.008$ & $0.781_{-0.019}^{+0.021}$ & $0.762^{+0.030}_{-0.025}$ & 2.1$\sigma$ & 2.2$\sigma$ \\ \hline
\textit{RSD}+\textit{BAO}+\textit{Pantheon}+\textit{CC} & $0.288 \pm 0.008$ & $0.793_{-0.020}^{+0.018}$ & $0.777^{+0.026}_{-0.027}$ & 1.8$\sigma$ & 2.1$\sigma$ \\ \hline
\textit{RSD}+$E_G$  & $0.200^{+0.020}_{-0.030}$ & $0.870_{-0.050}^{+0.039}$ & $0.710 \pm 0.029$ & 3.7$\sigma$ & 5.3$\sigma$ \\ \hline
\textit{RSD} & $0.254_{-0.058}^{+0.038}$ & $0.804_{-0.071}^{+0.048}$ & $0.739^{+0.036}_{-0.040}$ & 2.3$\sigma$ & 3.1$\sigma$ \\ \hline
\textit{BAO}+\textit{Pantheon} & $0.298 \pm 0.015$ & -- & -- & -- & -- \\ \hline 
\hline
\end{tabular}}
\caption{68\%~C.L. intervals on the matter density parameter $\Omega_m$, the present day linear theory amplitude of matter fluctuations averaged in spheres of radius $8\,h^{-1}{\rm Mpc}$ $\sigma_8$, and $S_8 \equiv \sigma_8\sqrt{\Omega_m/0.3}$, inferred from the datasets/dataset combinations given in the leftmost column, within the $\Lambda$CDM model. The two rightmost column quantify the level of agreement or tension between \textit{Planck} and the datasets in question, using either the 1D $T_{S_8}$ tension metric given by Eq.~(\ref{eq:1D_estimator}), or the more robust quadratic tension metric estimator given by Eq.~(\ref{eq:quadratic_estimator}). We encourage the use of the latter as reference value for the amount of tension. For the \textit{RSD}+\textit{BAO}+\textit{Pantheon} and \textit{RSD}+\textit{BAO}+\textit{Pantheon}+\textit{CC} dataset combinations, the level of agreement with \textit{Planck} is at the $\simeq 2\sigma$ level: at this level the mild disagreement, if any, is still consistent with a possible statistical fluctuation.}
\label{tab:LCDM}
\end{table*}

We first work within the context of the $\Lambda$CDM model. In a first instance, we consider our baseline dataset combination: \textit{RSD}+\textit{BAO}+\textit{Pantheon}. From this dataset combination we infer 68\% confidence level (C.L.) constraints of $\Omega_m=0.286 \pm 0.008$, $\sigma_8=0.7808_{-0.019}^{+0.021}$, and $S_8 = 0.762^{+0.030}_{-0.025}$. We note that \textit{BAO}+\textit{Pantheon} produce tight constraints on $\Omega_m=0.298 \pm 0.015$. Using the 1D $T_{S_8}$ tension metric given by Eq.~(\ref{eq:1D_estimator}), the value of $S_8$ is found to be in $1.8\sigma$ agreement with the \textit{Planck} determination, for which $S_8 = 0.834 \pm 0.016$. Adopting instead the more robust quadratic tension metric estimator given by Eq.~(\ref{eq:quadratic_estimator}), we find that the concordance between \textit{RSD}+\textit{BAO}+\textit{Pantheon} and \textit{Planck} decreases. However, we find that the two datasets are still in agreement at the $2.2\sigma$ level. The agreement between the two datasets is admittedly not perfect: there is clearly a mild disagreement between the two, with \textit{RSD}+\textit{BAO}+\textit{Pantheon} preferring lower values of $S_8$. However, we believe any reference to tensions is certainly premature, since a $\simeq 2\sigma$ agreement/disagreement could still be compatible with a statistical fluctuation.

Including the \textit{CC} dataset does not qualitatively alter the previous conclusions. In this case, we find $\Omega_m = 0.288 \pm 0.008$, $\sigma_8 = 0.7929_{-0.020}^{+0.018}$ and $S_8 = 0.777^{+0.026}_{-0.027}$. Again, using the 1D $T_{S_8}$ and quadratic tension metrics, we find that \textit{RSD}+\textit{BAO}+\textit{Pantheon}+\textit{CC} and \textit{Planck} are in agreement at the $1.8\sigma$ and $2.1\sigma$ level respectively. While again there is clearly a mild disagreement with \textit{Planck}, with \textit{RSD}+\textit{BAO}+\textit{Pantheon}+\textit{CC} preferring lower values of $S_8$, this disagreement is at a level which could be compatible with a statistical fluctuation.

Overall, the main message of the first part of our results therefore is: combining a wide range of RSD measurements of $f\sigma_8(z)$ with an inverse distance ladder constructed out of BAO and Hubble flow SNeIa and anchored to the high sound horizon value predicted within $\Lambda$CDM, while returning a slightly lower value of $S_8$, gives \textit{no strong evidence for the \textit{Planck} $\Lambda$CDM cosmology over-predicting the amplitude of matter fluctuations at $z \lesssim 1$}. In this sense, the \textit{RSD}+\textit{BAO}+\textit{Pantheon}(+\textit{CC}) dataset combination would suggest that it might be premature to invoke new physics to address the $S_8$ discrepancy, in qualitative agreement with the earlier results of Ref.~\cite{Efstathiou:2017rgv}.

\begin{figure*}
\begin{center}
\includegraphics[width=0.9\linewidth]{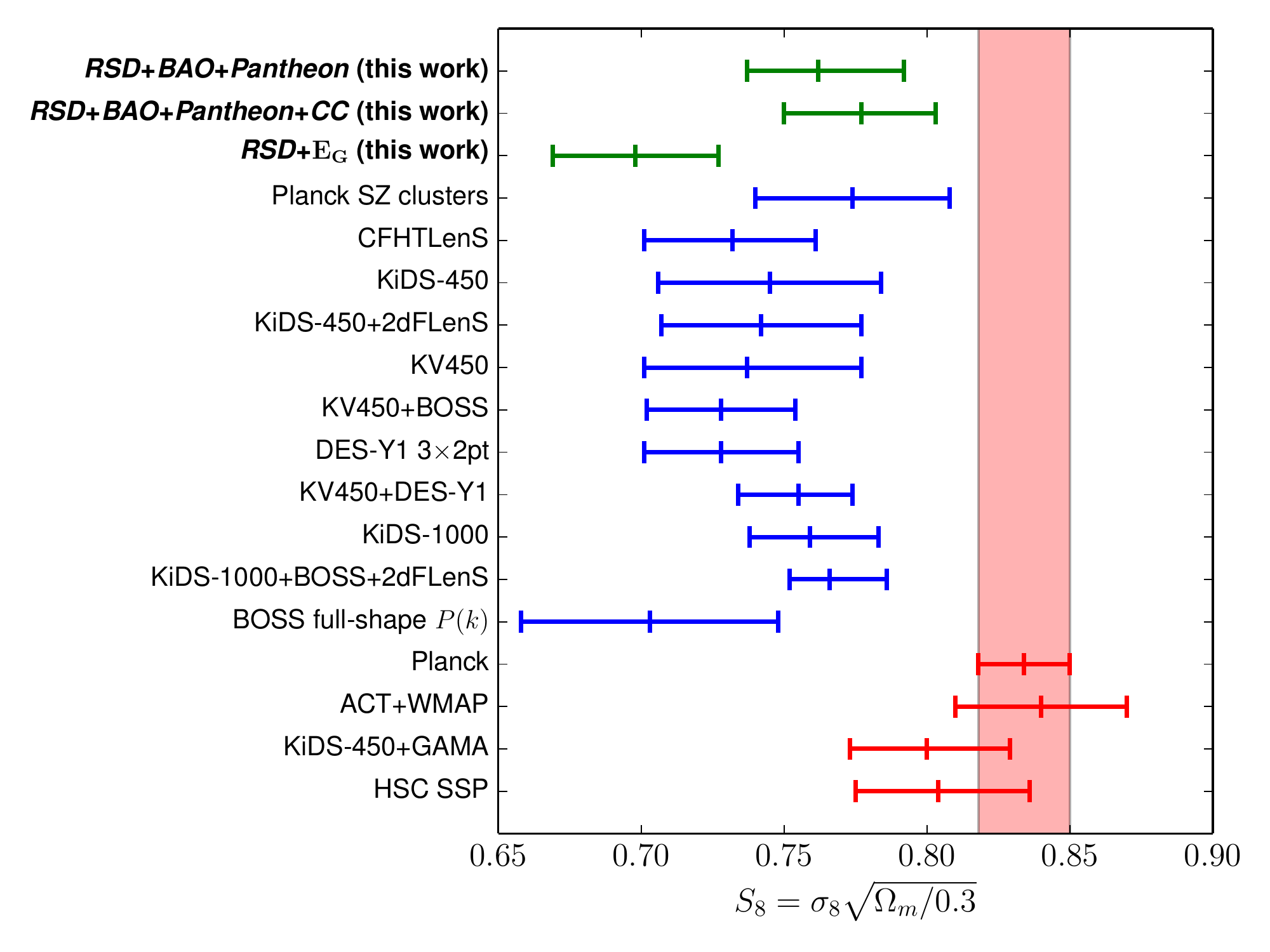}
\caption{Whisker plot displaying $68\%$~C.L. intervals on $S_8 \equiv \sigma_8\sqrt{\Omega_m/0.3}$, as inferred from a wide variety of measurements within the $\Lambda$CDM model. The color coding is such that green bars indicate our new results, blue bars indicate probes which infer an overall lower value of $S_8$ (mostly weak lensing surveys), and red bars indicate probes which infer an overall higher value of $S_8$. The red band denotes the 68\%~C.L. interval on $S_8=0.834 \pm 0.016$ determined by \textit{Planck} CMB measurements. From top to bottom, the reported measurements and surveys are: $S_8=0.762^{+0.030}_{-0.025}$ from \textit{RSD}+\textit{BAO}+\textit{Pantheon} (our work); $S_8=0.777^{+0.026}_{-0.027}$ from \textit{RSD}+\textit{BAO}+\textit{Pantheon}+\textit{CC} (our work); $S_8=0.710 \pm 0.029$ from \textit{RSD}+$E_G$ (our work); $S_8 = 0.774 \pm 0.034$ from \textit{Planck} Sunyaev-Zeldovich cluster counts~\cite{Ade:2015fva}; $S_8 = 0.732^{+0.029}_{-0.031}$ from CFHTLenS~\cite{Joudaki:2016mvz}; $S_8 = 0.745 \pm 0.039$ from KiDS-450~\cite{Joudaki:2016kym}; $S_8 = 0.742 \pm 0.035$ from KiDS-450+2dFLenS~\cite{Joudaki:2017zdt}; $S_8 = 0.737^{+0.040}_{-0.036}$ from KV450~\cite{Hildebrandt:2018yau};  $S_8 = 0.728 \pm 0.026$ from KV450+BOSS~\cite{Troster:2019ean}; $S_8 = 0.782 \pm 0.027$ from the DES-Y1 $3\times2$pt analysis~\cite{Troxel:2017xyo}; $S_8 = 0.755^{+0.019}_{-0.021}$ from KV450+DES-Y1~\cite{Joudaki:2019pmv,Asgari:2019fkq}; $S_8 = 0.759^{+0.024}_{-0.021}$ from KiDS-1000~\cite{Asgari:2020wuj}; $S_8 = 0.766^{+0.020}_{-0.014}$ from KiDS-1000+BOSS+2dFLenS~\cite{Heymans:2020ghw}; $S_8 = 0.703 \pm 0.045$ from a re-analysis of the BOSS galaxy power spectrum~\cite{Ivanov:2019pdj}; $S_8 = 0.834 \pm 0.016$ from \textit{Planck}~\cite{Aghanim:2018eyx}; $S_8 = 0.834 \pm 0.016$ from ACT+WMAP~\cite{Aiola:2020azj}; $S_8 = 0.800^{+0.029}_{-0.027}$ from KiDS-450+GAMA~\cite{vanUitert:2017ieu}; and $S_8 = 0.804^{+0.032}_{-0.029}$ from HSC SSP~\cite{Hamana:2019etx}.}
\label{fig:S8_summary}
\end{center}
\end{figure*}

We now consider the \textit{RSD}+$E_G$ dataset combination, which we find leads to rather unexpected results. In particular, we recover extremely low values for $\Omega_m = 0.200^{+0.020}_{-0.030}$ and $S_8 = 0.698 \pm 0.029$ respectively. The extremely low value of $\Omega_m$ is in strong tension with any independent probe of $\Omega_m$, e.g. BAO~\cite{Alam:2020sor} and SNeIa~\cite{Scolnic:2017caz}, including probes which by themselves already tend to favor low values of $\Omega_m$, such as cluster counts~\cite{Ade:2015fva,Sakr:2018new} (see also Ref.~\cite{Zubeldia:2019brr} for revised constraints). The recovered extremely low value of $S_8$ is also in mild tension with weak lensing measurements, which by themselves already prefer a lower value of $S_8$ as discussed in Section~\ref{sec:introduction}. Using the 1D $T_{S_8}$ and quadratic tension metrics, we find that \textit{RSD}+$E_G$ and \textit{Planck} are in tension at the $4.2\sigma$ and $5.3\sigma$ levels respectively. In this case it is very clear that focusing only on $S_8$ underestimates the level of the tension. It is worth noting that, following Refs.~\cite{Blake:2015vea,delaTorre:2016rxm,Amon:2017lia,Pinho:2018unz} we have treated the $E_G$ measurements as being independent, \textit{i.e.} neglecting the covariance between them. To the best of our knowledge, the covariance between all the available $E_G$ measurements has yet to be robustly quantified in the literature. We can generically expect that including the covariance between these measurements, if any, might reduce the significance of the tension, if only by virtue of enlarged error bars.

We note that these results are in qualitative agreement with those of Ref.~\cite{Skara:2019usd}, who also found that a similar dataset combination exacerbated the $S_8$ discrepancy at a similar level. We also note that the inferred low value of $S_8$ is in qualitative agreement with the value inferred from the re-analysis of the BOSS full-shape power spectrum of Ref.~\cite{Ivanov:2019pdj}, which finds $S_8=0.703 \pm 0.045$. If taken at face value, these results could indicate a weakening of gravity at low redshifts, as suggested in Ref.~\cite{Skara:2019usd}, where a model in which the lensing and growth effective Newton's constants $G_L$ and $G_{\rm eff}$ weaken was studied in this context. Similar hints were found in related works, including Refs.~\cite{Nesseris:2017vor,Kazantzidis:2018rnb,Perivolaropoulos:2019vkb,Kazantzidis:2019dvk}.

It is also worth noting that similar hints in $E_G$ data were found in Ref.~\cite{Pullen:2015vtb}, where combining \textit{Planck} 2015 CMB lensing maps and the galaxy velocity field reconstructed from the BOSS DR11 CMASS sample, an $E_G$ measurement of $E_G(z=0.57)=0.243 \pm 0.060 \pm 0.013$ was found, discrepant at the $2.6\sigma$ level from the GR expectation of $E_G(z=0.57) = 0.402 \pm 0.012$ given the \textit{Planck} and BOSS measurements. Possible systematic errors were studied in detail and found to be subdominant compared to the statistical error, and in any case unable to restore agreement with GR (see Fig.~11 of Ref.~\cite{Pullen:2015vtb}). It is worth noting that the later related work of Ref.~\cite{Alam:2016qcl} finds no evidence for these deviations.

We note that possible tensions between $E_G$ measurements and \textit{Planck} might be related to the ``\textit{lensing is low}'' (LIL) problem. This amounts to the observation that galaxy clustering measurements, together with standard galaxy-halo connection models, predict a galaxy-galaxy lensing signal which is higher by $\simeq 20$-$40\%$ compared to observations~\cite{Leauthaud:2016jdb}. Possible explanations for the LIL problem range from an incomplete/incorrect galaxy-halo connection model, baryonic physics, additional systematics, or new physics (see e.g. Refs.~\cite{Lange:2019nya,Yuan:2019fbi,Zu:2020uqo,Yuan:2020xlk,Lange:2020mnl}). However, none of the proposed scenarios have been fully able to address the problem. There is also some debate as to how much do uncertainties on photometric redshifts impact or bias the inferred $S_8$, and therefore the discrepancy with CMB measurements (see e.g. Refs.~\cite{Joudaki:2016mvz,Efstathiou:2017rgv}). We also note that measurements of the cross-correlation between CMB lensing and galaxy overdensities have systematically been reporting evidence of a deficit of power on large scales (see e.g. Refs.~\cite{Liu:2015xfa,Giannantonio:2015ahz,Kuntz:2015wza,Pullen:2015vtb,Giusarma:2018jei}). While this lack of power might be related to the LIL problem, it might also be at least partially due to contamination from the thermal Sunyaev-Zel'dovich effect (see e.g. Ref.~\cite{Darwish:2020fwf}).

In view of these possible problems with $E_G$ measurements, we caution the reader against over-interpreting the results obtained from the \textit{RSD}+$E_G$ dataset combination, and to consider our \textit{RSD}+\textit{BAO}+\textit{Pantheon}(+\textit{CC}) results as being the baseline ones. At the same time, it is worth noting that $E_G$ and weak lensing measurements are closely related -- in fact, all our $E_G$ measurements were obtained from analyses which made use of weak lensing data (from RSCLenS, CFHTLenS, and KiDS-450). In this sense, the \textit{RSD}+$E_G$ combination does not allow us to assess the status of the $S_8$ discrepancy in a way which is completely independent of weak lensing surveys. On the other hand, this can be achieved by the \textit{RSD}+\textit{BAO}+\textit{Pantheon}(+\textit{CC}) dataset combination(s), which is one of the reasons why we invite the reader to consider the results obtained from the latter as being our baseline ones.

Given these tensions, we also do not combine the \textit{RSD}+$E_G$ and \textit{BAO}+\textit{Pantheon}(+\textit{CC}) datasets. In closing we finally note that, while most independent analyses infer values of $\Omega_m$ in the ballpark of $\simeq 0.3$, a few analyses do infer rather low values of $\Omega_m$: these include a combination of DES cluster counts and weak lensing inferring $\Omega_m = 0.179^{+0.031}_{-0.038}$~\cite{Abbott:2020knk}, as well as the KiDS-450+2dFLenS and KiDS-450+2dFLenS+GAMA analyses, which infer $\Omega_m = 0.23^{+0.038}_{-0.038}$~\cite{Joudaki:2017zdt} and $\Omega_m = 0.25^{+0.03}_{-0.03}$~\cite{Amon:2017lia} respectively, all in extremely strong tension with \textit{Planck}. However, these studies themselves appear to suggest that the cause of these low values of $\Omega_m$ can be tracked back, at least partially, to systematics. These systematics are argued to most likely concern the modelling of the weak lensing signal rather than the cluster counts one, although adopting a higher richness threshold in the selection of clusters appears to reduce the tension with other probes~\cite{Abbott:2020knk}.

Finally, in order to investigate whether the tension between \textit{RSD}+$E_G$ and \textit{Planck} is entirely or mostly due to the $E_G$ dataset and the possible problems discussed previously, we also consider the \textit{RSD} dataset alone. In this case, we still find rather low values of $\Omega_m = 0.227^{+0.068}_{-0.033}$ and $S_8 = 0.734^{+0.036}_{-0.040}$.  Using the 1D $T_{S_8}$ and quadratic tension metrics, we find that \textit{RSD} and \textit{Planck} are in tension at the $2.8\sigma$ and $3.1\sigma$ levels respectively. These results are in qualitative agreement with earlier works in Refs.~\cite{Nesseris:2017vor,Kazantzidis:2018rnb,Perivolaropoulos:2019vkb,Kazantzidis:2019dvk}, which also identified tensions between RSD and \textit{Planck} measurements at the $2.5$-$3\sigma$ level which, if taken at face value, point towards a lack of gravitational power in structures on intermediate and small cosmological scales, which could indicate a time-dependent (weakening) gravitational constant.

Our results are summarized in Fig.~\ref{fig:S8LCDM}, where we show the joint $S_8$-$\Omega_m$ constraints obtained from the different dataset combinations considered, in the whisker plot of Fig.~\ref{fig:S8_summary}, where we compare our inferred values of $S_8$ to those inferred from a number of independent surveys (mentioned earlier in Section~\ref{sec:introduction}), and in Tab.~\ref{tab:LCDM}, where we summarize our constraints and level of concordance/discordance between the dataset combinations considered and \textit{Planck}. In particular, from Fig.~\ref{fig:S8_summary} we see that the values of $S_8$ we infer from our \textit{RSD}+\textit{BAO}+\textit{Pantheon} and \textit{RSD}+\textit{BAO}+\textit{Pantheon}+\textit{CC} dataset combinations, while in $\simeq 2\sigma$ agreement with \textit{Planck}, are in better agreement with the value inferred from various weak lensing surveys.

\begin{figure}
\begin{center}
\includegraphics[width=2.8in]{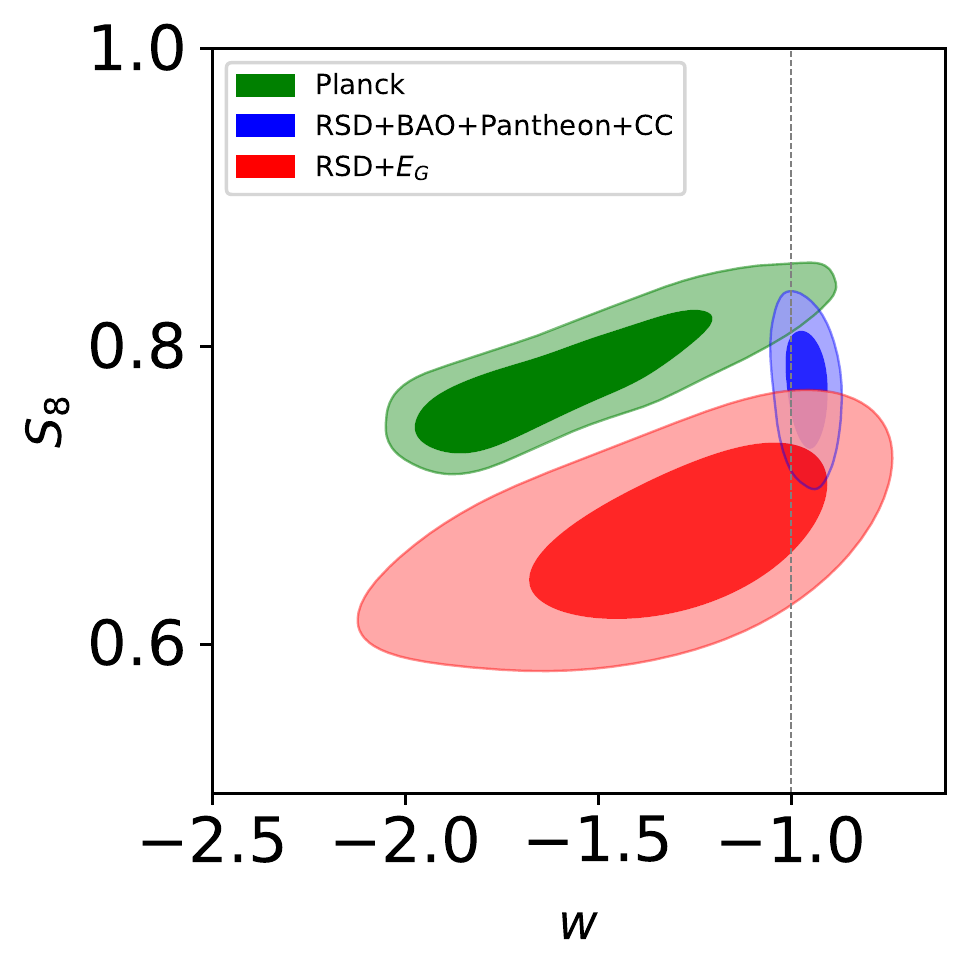} 
\caption{2D joint posterior distributions in the $S_8$-$w$ plane, with the corresponding 68\%~C.L. and 95\%~C.L. contours, obtained from the following datasets/dataset combinations within the $w$CDM model: \textit{Planck} (green contours), \textit{RSD}+\textit{BAO}+\textit{Pantheon}+\textit{CC} (blue contours), and \textit{RSD}+$E_G$ (red contours). Using the more robust quadratic tension metric estimator given by Eq.~(\ref{eq:quadratic_estimator}), we infer that \textit{Planck} and \textit{RSD}+\textit{BAO}+\textit{Pantheon}+\textit{CC} are in agreement at the $2.2\sigma$ level, while we infer that \textit{Planck} and \textit{RSD}+$E_G$ are in tension at the $3.5\sigma$ level.}
\label{fig:wCDM}
\end{center}
\end{figure}

\subsection{$w$CDM}
\label{subsec:wcdm}

Earlier we found the \textit{RSD}+\textit{BAO}+\textit{Pantheon}(+\textit{CC}) dataset combination to be in $\simeq 2\sigma$ agreement with \textit{Planck}. While this level of agreement does not call for new physics, it is worth noting that the value of $S_8$ we inferred is nonetheless lower than that of \textit{Planck} and moves in the direction of the value inferred from weak lensing measurements. In this sense, we believe it is still worth exploring whether extended models may improve the agreement between these two datasets. With this in mind, we repeat the previous analysis for the $w$CDM model, where the DE EoS $w$ is allowed to vary.

For the \textit{RSD}+\textit{BAO}+\textit{Pantheon}+\textit{CC} dataset combination, we infer 68\% C.L. constraints of $w=-0.96 \pm 0.04$, $\Omega_m=0.293^{+0.008}_{-0.009}$, $\sigma_8=0.781 \pm 0.021$, and $S_8=0.775^{+0.027}_{-0.030}$. In particular, we find the inferred value of $w$ to be in excellent agreement with the cosmological constant value $w=-1$. Moreover, using the quadratic estimator of Eq.~(\ref{eq:quadratic_estimator}), we find \textit{RSD}+\textit{BAO}+\textit{Pantheon}+\textit{CC} and \textit{Planck} to be in agreement at $2.2\sigma$ within the $w$CDM model. Therefore, the extension allowing for $w$ to vary has essentially left the level of concordance/discordance between these two probes unchanged compared to the value within the $\Lambda$CDM model discussed earlier. Dropping the \textit{CC} dataset leads to essentially identical results.

If we instead consider the \textit{RSD}+$E_G$ dataset combination, we infer 68\%~C.L. constraints of $w=-1.31^{+0.33}_{-0.15}$, $\Omega_m=0.209^{+0.017}_{-0.027}$, $\sigma_8=0.809^{+0.066}_{-0.047}$, and $S_8= 0.670^{+0.037}_{-0.036}$ in line with the earlier results within $\Lambda$CDM supporting a lower matter density. While the inferred value of $w$ is consistent with $w=-1$ within better than $1\sigma$, we notice a curious trend towards phantom values $w<-1$. This is directly related to the preference for lower values of $\Omega_m$, given the positive correlation between $w$ and $\Omega_m$. Using the quadratic estimator of Eq.~(\ref{eq:quadratic_estimator}), we find \textit{RSD}+$E_G$ and \textit{Planck} to be in $3.5\sigma$ tension within the $w$CDM model. While this figure is significantly lower than the $5.3\sigma$ obtained earlier within $\Lambda$CDM, mostly by virtue of the larger error bars, the amount of tension between the two probes remains significant.

As for our earlier results, given the possible issues with the $E_G$ measurements, we urge the reader to take our \textit{RSD}+\textit{BAO}+\textit{Pantheon}(+\textit{CC}) results as baseline. With this in mind, the main result of this Section is that our previous results obtained within the $\Lambda$CDM model, and in particular the inferred level of concordance between \textit{Planck} and a combination of RSD and inverse distance ladder measurements, is stable against a minimal parameter space extension where the DE EoS $w$ is allowed to vary. Freeing up $w$ does not improve the level of agreement between these two probes.

\section{Conclusions}
\label{sec:conclusions}

The $\Lambda$CDM model is, without question, an extremely successful one. Despite its many successes there are persisting hints, in the form of cosmological tensions, that this model might be about to break down. However, before claiming the definitive failure of an otherwise extremely successful, albeit phenomenological model, it is important to check whether these hints persist when viewed from a different perspective.

In this spirit, we have re-assessed the $S_8$ discrepancy between CMB and weak lensing probes of the amplitude of matter fluctuations. We have examined this tension from the point of view of Redshift Space Distortions (RSD) measurements of the growth rate, and more precisely of $f\sigma_8(z)$. A robust assessment of the RSD take on the $S_8$ discrepancy cannot afford to leave out geometrical data in the form of BAO and high-$z$ SNeIa measurements, given the importance of the these measurements in constraining $\Omega_m$.

The cosmological constraints we infer from our baseline combination of RSD, BAO, and \textit{Pantheon} SNeIa data (eventually including cosmic chronometer measurements), and in particular the inferred value of $S_8$, are somewhat intermediate between the weak lensing and \textit{Planck} CMB results (although tending towards the former), as is visually shown by the 2 uppermost bars in Fig.~\ref{fig:S8_summary}. Using the tension metric defined in Eq.~(\ref{eq:quadratic_estimator}), we find the \textit{RSD}+\textit{BAO}+\textit{Pantheon} combination to be in agreement with \textit{Planck} at the $2.2\sigma$ level. From this perspective the hints for a $S_8$ discrepancy from growth rate data, if any, could be ascribable to a statistical fluctuation. These results, though obtained adopting a more up-to-date set of RSD measurements, agree qualitatively with the earlier results of Ref.~\cite{Efstathiou:2017rgv}.

We have also combined RSD measurements with measurements of the $E_G$ statistic, which measures a combination of gravitational lensing, galaxy clustering, and redshift-space distortions~\cite{Zhang:2007nk}. We have found the \textit{RSD}+$E_G$ combination to be in $5.3\sigma$ tension with \textit{Planck} (see the third bar from the top of Fig.~\ref{fig:S8_summary}), ultimately due to the extremely low inferred value of $\Omega_m$. We caution the reader against over-interpreting the results arising from the \textit{RSD}+$E_G$ combination, and to take the results coming from the \textit{RSD}+\textit{BAO}+\textit{Pantheon} (+\textit{CC}) dataset combination as being our baseline ones, for reasons discussed in more depth in Section~\ref{sec:results}. Finally, we have examined the stability of our results against a minimal parameter space extension where we free the dark energy equation of state $w$, and have found our results to be qualitatively unchanged.

Our initial goal was to answer the question: ``\textit{Is there evidence from data other than weak lensing measurements for the \textit{Planck} $\Lambda$CDM cosmology over-predicting the amplitude of matter fluctuations at $z \lesssim 1$}?'' From the perspective of growth rate measurements, the answer is that there are hints at the $\approx 2\sigma$ level, but no definitive evidence of a tension: in this sense, we believe it might still be too early to claim evidence for new physics in light of the $S_8$ discrepancy. We also note that new physics models constructed to alleviate the $S_8$ discrepancy should not do so at the expense of worsening the $H_0$ tension (and viceversa). It is noteworthy that many proposed models fail in doing so (see e.g. the discussion in Ref.~\cite{Alestas:2021xes}), suggesting that if the $S_8$ discrepancy does indeed call for new physics, a joint solution to the $S_8$ and $H_0$ tensions will likely involve a rather non-trivial physical scenario. Future more precise measurements from the CMB~\cite{Abazajian:2016yjj,Ade:2018sbj,Abitbol:2019nhf}, growth rate~\cite{Ivezic:2008fe,Aghamousa:2016zmz,Bull:2018lat}, and weak lensing~\cite{Ivezic:2008fe,Amendola:2012ys,Aghamousa:2016zmz} sides will certainly shed more light on the issue, and will either confirm or disprove whether new physics is needed in this context.

\begin{acknowledgments}
\noindent S.V. thanks George Efstathiou for useful discussions on the role of RSD measurements in arbitrating the $S_8$ discrepancy (if any). R.C.N. acknowledges financial support from the Funda\c{c}\~{a}o de Amparo \`{a} Pesquisa do Estado de S\~{a}o Paulo (FAPESP, S\~{a}o Paulo Research Foundation) under the project No. 2018/18036-5. S.V. is supported by the Isaac Newton Trust and the Kavli Foundation through a Newton-Kavli Fellowship, and by a grant from the Foundation Blanceflor Boncompagni Ludovisi, n\'{e}e Bildt. S.V. acknowledges a College Research Associateship at Homerton College, University of Cambridge.
\end{acknowledgments}

\bibliography{S8_PRD}

\begin{thebibliography}{199}%
\makeatletter
\providecommand \@ifxundefined [1]{%
 \@ifx{#1\undefined}
}%
\providecommand \@ifnum [1]{%
 \ifnum #1\expandafter \@firstoftwo
 \else \expandafter \@secondoftwo
 \fi
}%
\providecommand \@ifx [1]{%
 \ifx #1\expandafter \@firstoftwo
 \else \expandafter \@secondoftwo
 \fi
}%
\providecommand \natexlab [1]{#1}%
\providecommand \enquote  [1]{``#1''}%
\providecommand \bibnamefont  [1]{#1}%
\providecommand \bibfnamefont [1]{#1}%
\providecommand \citenamefont [1]{#1}%
\providecommand \href@noop [0]{\@secondoftwo}%
\providecommand \href [0]{\begingroup \@sanitize@url \@href}%
\providecommand \@href[1]{\@@startlink{#1}\@@href}%
\providecommand \@@href[1]{\endgroup#1\@@endlink}%
\providecommand \@sanitize@url [0]{\catcode `\\12\catcode `\$12\catcode
  `\&12\catcode `\#12\catcode `\^12\catcode `\_12\catcode `\%12\relax}%
\providecommand \@@startlink[1]{}%
\providecommand \@@endlink[0]{}%
\providecommand \url  [0]{\begingroup\@sanitize@url \@url }%
\providecommand \@url [1]{\endgroup\@href {#1}{\urlprefix }}%
\providecommand \urlprefix  [0]{URL }%
\providecommand \Eprint [0]{\href }%
\providecommand \doibase [0]{http://dx.doi.org/}%
\providecommand \selectlanguage [0]{\@gobble}%
\providecommand \bibinfo  [0]{\@secondoftwo}%
\providecommand \bibfield  [0]{\@secondoftwo}%
\providecommand \translation [1]{[#1]}%
\providecommand \BibitemOpen [0]{}%
\providecommand \bibitemStop [0]{}%
\providecommand \bibitemNoStop [0]{.\EOS\space}%
\providecommand \EOS [0]{\spacefactor3000\relax}%
\providecommand \BibitemShut  [1]{\csname bibitem#1\endcsname}%
\let\auto@bib@innerbib\@empty
\bibitem [{\citenamefont {Riess}\ \emph {et~al.}(1998)\citenamefont {Riess}
  \emph {et~al.}}]{Riess:1998cb}%
  \BibitemOpen
  \bibfield  {author} {\bibinfo {author} {\bibfnamefont {A.~G.}\ \bibnamefont
  {Riess}} \emph {et~al.} (\bibinfo {collaboration} {Supernova Search Team}),\
  }\href {\doibase 10.1086/300499} {\bibfield  {journal} {\bibinfo  {journal}
  {Astron. J.}\ }\textbf {\bibinfo {volume} {116}},\ \bibinfo {pages} {1009}
  (\bibinfo {year} {1998})},\ \Eprint {http://arxiv.org/abs/astro-ph/9805201}
  {arXiv:astro-ph/9805201} \BibitemShut {NoStop}%
\bibitem [{\citenamefont {Perlmutter}\ \emph {et~al.}(1999)\citenamefont
  {Perlmutter} \emph {et~al.}}]{Perlmutter:1998np}%
  \BibitemOpen
  \bibfield  {author} {\bibinfo {author} {\bibfnamefont {S.}~\bibnamefont
  {Perlmutter}} \emph {et~al.} (\bibinfo {collaboration} {Supernova Cosmology
  Project}),\ }\href {\doibase 10.1086/307221} {\bibfield  {journal} {\bibinfo
  {journal} {Astrophys. J.}\ }\textbf {\bibinfo {volume} {517}},\ \bibinfo
  {pages} {565} (\bibinfo {year} {1999})},\ \Eprint
  {http://arxiv.org/abs/astro-ph/9812133} {arXiv:astro-ph/9812133} \BibitemShut
  {NoStop}%
\bibitem [{\citenamefont {Aghanim}\ \emph {et~al.}(2020)\citenamefont {Aghanim}
  \emph {et~al.}}]{Aghanim:2018eyx}%
  \BibitemOpen
  \bibfield  {author} {\bibinfo {author} {\bibfnamefont {N.}~\bibnamefont
  {Aghanim}} \emph {et~al.} (\bibinfo {collaboration} {Planck}),\ }\href
  {\doibase 10.1051/0004-6361/201833910} {\bibfield  {journal} {\bibinfo
  {journal} {Astron. Astrophys.}\ }\textbf {\bibinfo {volume} {641}},\ \bibinfo
  {pages} {A6} (\bibinfo {year} {2020})},\ \Eprint
  {http://arxiv.org/abs/1807.06209} {arXiv:1807.06209 [astro-ph.CO]}
  \BibitemShut {NoStop}%
\bibitem [{\citenamefont {Aiola}\ \emph {et~al.}(2020)\citenamefont {Aiola}
  \emph {et~al.}}]{Aiola:2020azj}%
  \BibitemOpen
  \bibfield  {author} {\bibinfo {author} {\bibfnamefont {S.}~\bibnamefont
  {Aiola}} \emph {et~al.} (\bibinfo {collaboration} {ACT}),\ }\href {\doibase
  10.1088/1475-7516/2020/12/047} {\bibfield  {journal} {\bibinfo  {journal}
  {JCAP}\ }\textbf {\bibinfo {volume} {12}},\ \bibinfo {pages} {047} (\bibinfo
  {year} {2020})},\ \Eprint {http://arxiv.org/abs/2007.07288} {arXiv:2007.07288
  [astro-ph.CO]} \BibitemShut {NoStop}%
\bibitem [{\citenamefont {Alam}\ \emph {et~al.}(2020)\citenamefont {Alam} \emph
  {et~al.}}]{Alam:2020sor}%
  \BibitemOpen
  \bibfield  {author} {\bibinfo {author} {\bibfnamefont {S.}~\bibnamefont
  {Alam}} \emph {et~al.} (\bibinfo {collaboration} {eBOSS}),\ }\href@noop {} {\
   (\bibinfo {year} {2020})},\ \Eprint {http://arxiv.org/abs/2007.08991}
  {arXiv:2007.08991 [astro-ph.CO]} \BibitemShut {NoStop}%
\bibitem [{\citenamefont {Riess}\ \emph {et~al.}(2019)\citenamefont {Riess},
  \citenamefont {Casertano}, \citenamefont {Yuan}, \citenamefont {Macri},\ and\
  \citenamefont {Scolnic}}]{Riess:2019cxk}%
  \BibitemOpen
  \bibfield  {author} {\bibinfo {author} {\bibfnamefont {A.~G.}\ \bibnamefont
  {Riess}}, \bibinfo {author} {\bibfnamefont {S.}~\bibnamefont {Casertano}},
  \bibinfo {author} {\bibfnamefont {W.}~\bibnamefont {Yuan}}, \bibinfo {author}
  {\bibfnamefont {L.~M.}\ \bibnamefont {Macri}}, \ and\ \bibinfo {author}
  {\bibfnamefont {D.}~\bibnamefont {Scolnic}},\ }\href {\doibase
  10.3847/1538-4357/ab1422} {\bibfield  {journal} {\bibinfo  {journal}
  {Astrophys. J.}\ }\textbf {\bibinfo {volume} {876}},\ \bibinfo {pages} {85}
  (\bibinfo {year} {2019})},\ \Eprint {http://arxiv.org/abs/1903.07603}
  {arXiv:1903.07603 [astro-ph.CO]} \BibitemShut {NoStop}%
\bibitem [{\citenamefont {Wong}\ \emph {et~al.}(2020)\citenamefont {Wong} \emph
  {et~al.}}]{Wong:2019kwg}%
  \BibitemOpen
  \bibfield  {author} {\bibinfo {author} {\bibfnamefont {K.~C.}\ \bibnamefont
  {Wong}} \emph {et~al.},\ }\href {\doibase 10.1093/mnras/stz3094} {\bibfield
  {journal} {\bibinfo  {journal} {Mon. Not. Roy. Astron. Soc.}\ }\textbf
  {\bibinfo {volume} {498}},\ \bibinfo {pages} {1420} (\bibinfo {year}
  {2020})},\ \Eprint {http://arxiv.org/abs/1907.04869} {arXiv:1907.04869
  [astro-ph.CO]} \BibitemShut {NoStop}%
\bibitem [{\citenamefont {Freedman}\ \emph {et~al.}(2019)\citenamefont
  {Freedman} \emph {et~al.}}]{Freedman:2019jwv}%
  \BibitemOpen
  \bibfield  {author} {\bibinfo {author} {\bibfnamefont {W.~L.}\ \bibnamefont
  {Freedman}} \emph {et~al.},\ }\href {\doibase 10.3847/1538-4357/ab2f73}
  {\bibfield  {journal} {\bibinfo  {journal} {Astrophys. J.}\ }\textbf
  {\bibinfo {volume} {882}},\ \bibinfo {pages} {34} (\bibinfo {year} {2019})},\
  \Eprint {http://arxiv.org/abs/1907.05922} {arXiv:1907.05922 [astro-ph.CO]}
  \BibitemShut {NoStop}%
\bibitem [{\citenamefont {Bernal}\ \emph {et~al.}(2016)\citenamefont {Bernal},
  \citenamefont {Verde},\ and\ \citenamefont {Riess}}]{Bernal:2016gxb}%
  \BibitemOpen
  \bibfield  {author} {\bibinfo {author} {\bibfnamefont {J.~L.}\ \bibnamefont
  {Bernal}}, \bibinfo {author} {\bibfnamefont {L.}~\bibnamefont {Verde}}, \
  and\ \bibinfo {author} {\bibfnamefont {A.~G.}\ \bibnamefont {Riess}},\ }\href
  {\doibase 10.1088/1475-7516/2016/10/019} {\bibfield  {journal} {\bibinfo
  {journal} {JCAP}\ }\textbf {\bibinfo {volume} {10}},\ \bibinfo {pages} {019}
  (\bibinfo {year} {2016})},\ \Eprint {http://arxiv.org/abs/1607.05617}
  {arXiv:1607.05617 [astro-ph.CO]} \BibitemShut {NoStop}%
\bibitem [{\citenamefont {M\"ortsell}\ and\ \citenamefont
  {Dhawan}(2018)}]{Mortsell:2018mfj}%
  \BibitemOpen
  \bibfield  {author} {\bibinfo {author} {\bibfnamefont {E.}~\bibnamefont
  {M\"ortsell}}\ and\ \bibinfo {author} {\bibfnamefont {S.}~\bibnamefont
  {Dhawan}},\ }\href {\doibase 10.1088/1475-7516/2018/09/025} {\bibfield
  {journal} {\bibinfo  {journal} {JCAP}\ }\textbf {\bibinfo {volume} {09}},\
  \bibinfo {pages} {025} (\bibinfo {year} {2018})},\ \Eprint
  {http://arxiv.org/abs/1801.07260} {arXiv:1801.07260 [astro-ph.CO]}
  \BibitemShut {NoStop}%
\bibitem [{\citenamefont {Poulin}\ \emph {et~al.}(2019)\citenamefont {Poulin},
  \citenamefont {Smith}, \citenamefont {Karwal},\ and\ \citenamefont
  {Kamionkowski}}]{Poulin:2018cxd}%
  \BibitemOpen
  \bibfield  {author} {\bibinfo {author} {\bibfnamefont {V.}~\bibnamefont
  {Poulin}}, \bibinfo {author} {\bibfnamefont {T.~L.}\ \bibnamefont {Smith}},
  \bibinfo {author} {\bibfnamefont {T.}~\bibnamefont {Karwal}}, \ and\ \bibinfo
  {author} {\bibfnamefont {M.}~\bibnamefont {Kamionkowski}},\ }\href {\doibase
  10.1103/PhysRevLett.122.221301} {\bibfield  {journal} {\bibinfo  {journal}
  {Phys. Rev. Lett.}\ }\textbf {\bibinfo {volume} {122}},\ \bibinfo {pages}
  {221301} (\bibinfo {year} {2019})},\ \Eprint
  {http://arxiv.org/abs/1811.04083} {arXiv:1811.04083 [astro-ph.CO]}
  \BibitemShut {NoStop}%
\bibitem [{\citenamefont {Kreisch}\ \emph {et~al.}(2020)\citenamefont
  {Kreisch}, \citenamefont {Cyr-Racine},\ and\ \citenamefont
  {Dor\'e}}]{Kreisch:2019yzn}%
  \BibitemOpen
  \bibfield  {author} {\bibinfo {author} {\bibfnamefont {C.~D.}\ \bibnamefont
  {Kreisch}}, \bibinfo {author} {\bibfnamefont {F.-Y.}\ \bibnamefont
  {Cyr-Racine}}, \ and\ \bibinfo {author} {\bibfnamefont {O.}~\bibnamefont
  {Dor\'e}},\ }\href {\doibase 10.1103/PhysRevD.101.123505} {\bibfield
  {journal} {\bibinfo  {journal} {Phys. Rev. D}\ }\textbf {\bibinfo {volume}
  {101}},\ \bibinfo {pages} {123505} (\bibinfo {year} {2020})},\ \Eprint
  {http://arxiv.org/abs/1902.00534} {arXiv:1902.00534 [astro-ph.CO]}
  \BibitemShut {NoStop}%
\bibitem [{\citenamefont {Vagnozzi}(2020)}]{Vagnozzi:2019ezj}%
  \BibitemOpen
  \bibfield  {author} {\bibinfo {author} {\bibfnamefont {S.}~\bibnamefont
  {Vagnozzi}},\ }\href {\doibase 10.1103/PhysRevD.102.023518} {\bibfield
  {journal} {\bibinfo  {journal} {Phys. Rev. D}\ }\textbf {\bibinfo {volume}
  {102}},\ \bibinfo {pages} {023518} (\bibinfo {year} {2020})},\ \Eprint
  {http://arxiv.org/abs/1907.07569} {arXiv:1907.07569 [astro-ph.CO]}
  \BibitemShut {NoStop}%
\bibitem [{\citenamefont {Visinelli}\ \emph {et~al.}(2019)\citenamefont
  {Visinelli}, \citenamefont {Vagnozzi},\ and\ \citenamefont
  {Danielsson}}]{Visinelli:2019qqu}%
  \BibitemOpen
  \bibfield  {author} {\bibinfo {author} {\bibfnamefont {L.}~\bibnamefont
  {Visinelli}}, \bibinfo {author} {\bibfnamefont {S.}~\bibnamefont {Vagnozzi}},
  \ and\ \bibinfo {author} {\bibfnamefont {U.}~\bibnamefont {Danielsson}},\
  }\href {\doibase 10.3390/sym11081035} {\bibfield  {journal} {\bibinfo
  {journal} {Symmetry}\ }\textbf {\bibinfo {volume} {11}},\ \bibinfo {pages}
  {1035} (\bibinfo {year} {2019})},\ \Eprint {http://arxiv.org/abs/1907.07953}
  {arXiv:1907.07953 [astro-ph.CO]} \BibitemShut {NoStop}%
\bibitem [{\citenamefont {Sakstein}\ and\ \citenamefont
  {Trodden}(2020)}]{Sakstein:2019fmf}%
  \BibitemOpen
  \bibfield  {author} {\bibinfo {author} {\bibfnamefont {J.}~\bibnamefont
  {Sakstein}}\ and\ \bibinfo {author} {\bibfnamefont {M.}~\bibnamefont
  {Trodden}},\ }\href {\doibase 10.1103/PhysRevLett.124.161301} {\bibfield
  {journal} {\bibinfo  {journal} {Phys. Rev. Lett.}\ }\textbf {\bibinfo
  {volume} {124}},\ \bibinfo {pages} {161301} (\bibinfo {year} {2020})},\
  \Eprint {http://arxiv.org/abs/1911.11760} {arXiv:1911.11760 [astro-ph.CO]}
  \BibitemShut {NoStop}%
\bibitem [{\citenamefont {Hill}\ \emph {et~al.}(2020)\citenamefont {Hill},
  \citenamefont {McDonough}, \citenamefont {Toomey},\ and\ \citenamefont
  {Alexander}}]{Hill:2020osr}%
  \BibitemOpen
  \bibfield  {author} {\bibinfo {author} {\bibfnamefont {J.~C.}\ \bibnamefont
  {Hill}}, \bibinfo {author} {\bibfnamefont {E.}~\bibnamefont {McDonough}},
  \bibinfo {author} {\bibfnamefont {M.~W.}\ \bibnamefont {Toomey}}, \ and\
  \bibinfo {author} {\bibfnamefont {S.}~\bibnamefont {Alexander}},\ }\href
  {\doibase 10.1103/PhysRevD.102.043507} {\bibfield  {journal} {\bibinfo
  {journal} {Phys. Rev. D}\ }\textbf {\bibinfo {volume} {102}},\ \bibinfo
  {pages} {043507} (\bibinfo {year} {2020})},\ \Eprint
  {http://arxiv.org/abs/2003.07355} {arXiv:2003.07355 [astro-ph.CO]}
  \BibitemShut {NoStop}%
\bibitem [{\citenamefont {Ballesteros}\ \emph {et~al.}(2020)\citenamefont
  {Ballesteros}, \citenamefont {Notari},\ and\ \citenamefont
  {Rompineve}}]{Ballesteros:2020sik}%
  \BibitemOpen
  \bibfield  {author} {\bibinfo {author} {\bibfnamefont {G.}~\bibnamefont
  {Ballesteros}}, \bibinfo {author} {\bibfnamefont {A.}~\bibnamefont {Notari}},
  \ and\ \bibinfo {author} {\bibfnamefont {F.}~\bibnamefont {Rompineve}},\
  }\href {\doibase 10.1088/1475-7516/2020/11/024} {\bibfield  {journal}
  {\bibinfo  {journal} {JCAP}\ }\textbf {\bibinfo {volume} {11}},\ \bibinfo
  {pages} {024} (\bibinfo {year} {2020})},\ \Eprint
  {http://arxiv.org/abs/2004.05049} {arXiv:2004.05049 [astro-ph.CO]}
  \BibitemShut {NoStop}%
\bibitem [{\citenamefont {Braglia}\ \emph {et~al.}(2020)\citenamefont
  {Braglia}, \citenamefont {Ballardini}, \citenamefont {Emond}, \citenamefont
  {Finelli}, \citenamefont {Gumrukcuoglu}, \citenamefont {Koyama},\ and\
  \citenamefont {Paoletti}}]{Braglia:2020iik}%
  \BibitemOpen
  \bibfield  {author} {\bibinfo {author} {\bibfnamefont {M.}~\bibnamefont
  {Braglia}}, \bibinfo {author} {\bibfnamefont {M.}~\bibnamefont {Ballardini}},
  \bibinfo {author} {\bibfnamefont {W.~T.}\ \bibnamefont {Emond}}, \bibinfo
  {author} {\bibfnamefont {F.}~\bibnamefont {Finelli}}, \bibinfo {author}
  {\bibfnamefont {A.~E.}\ \bibnamefont {Gumrukcuoglu}}, \bibinfo {author}
  {\bibfnamefont {K.}~\bibnamefont {Koyama}}, \ and\ \bibinfo {author}
  {\bibfnamefont {D.}~\bibnamefont {Paoletti}},\ }\href {\doibase
  10.1103/PhysRevD.102.023529} {\bibfield  {journal} {\bibinfo  {journal}
  {Phys. Rev. D}\ }\textbf {\bibinfo {volume} {102}},\ \bibinfo {pages}
  {023529} (\bibinfo {year} {2020})},\ \Eprint
  {http://arxiv.org/abs/2004.11161} {arXiv:2004.11161 [astro-ph.CO]}
  \BibitemShut {NoStop}%
\bibitem [{\citenamefont {Efstathiou}(2020)}]{Efstathiou:2020wxn}%
  \BibitemOpen
  \bibfield  {author} {\bibinfo {author} {\bibfnamefont {G.}~\bibnamefont
  {Efstathiou}},\ }\href@noop {} {\  (\bibinfo {year} {2020})},\ \Eprint
  {http://arxiv.org/abs/2007.10716} {arXiv:2007.10716 [astro-ph.CO]}
  \BibitemShut {NoStop}%
\bibitem [{\citenamefont {Das}\ and\ \citenamefont
  {Ghosh}(2020)}]{Das:2020xke}%
  \BibitemOpen
  \bibfield  {author} {\bibinfo {author} {\bibfnamefont {A.}~\bibnamefont
  {Das}}\ and\ \bibinfo {author} {\bibfnamefont {S.}~\bibnamefont {Ghosh}},\
  }\href@noop {} {\  (\bibinfo {year} {2020})},\ \Eprint
  {http://arxiv.org/abs/2011.12315} {arXiv:2011.12315 [hep-ph]} \BibitemShut
  {NoStop}%
\bibitem [{\citenamefont {Roy~Choudhury}\ \emph {et~al.}(2020)\citenamefont
  {Roy~Choudhury}, \citenamefont {Hannestad},\ and\ \citenamefont
  {Tram}}]{Choudhury:2020tka}%
  \BibitemOpen
  \bibfield  {author} {\bibinfo {author} {\bibfnamefont {S.}~\bibnamefont
  {Roy~Choudhury}}, \bibinfo {author} {\bibfnamefont {S.}~\bibnamefont
  {Hannestad}}, \ and\ \bibinfo {author} {\bibfnamefont {T.}~\bibnamefont
  {Tram}},\ }\href@noop {} {\  (\bibinfo {year} {2020})},\ \Eprint
  {http://arxiv.org/abs/2012.07519} {arXiv:2012.07519 [astro-ph.CO]}
  \BibitemShut {NoStop}%
\bibitem [{\citenamefont {Brinckmann}\ \emph {et~al.}(2020)\citenamefont
  {Brinckmann}, \citenamefont {Chang},\ and\ \citenamefont
  {LoVerde}}]{Brinckmann:2020bcn}%
  \BibitemOpen
  \bibfield  {author} {\bibinfo {author} {\bibfnamefont {T.}~\bibnamefont
  {Brinckmann}}, \bibinfo {author} {\bibfnamefont {J.~H.}\ \bibnamefont
  {Chang}}, \ and\ \bibinfo {author} {\bibfnamefont {M.}~\bibnamefont
  {LoVerde}},\ }\href@noop {} {\  (\bibinfo {year} {2020})},\ \Eprint
  {http://arxiv.org/abs/2012.11830} {arXiv:2012.11830 [astro-ph.CO]}
  \BibitemShut {NoStop}%
\bibitem [{\citenamefont {Efstathiou}(2021)}]{Efstathiou:2021ocp}%
  \BibitemOpen
  \bibfield  {author} {\bibinfo {author} {\bibfnamefont {G.}~\bibnamefont
  {Efstathiou}},\ }\href@noop {} {\  (\bibinfo {year} {2021})},\ \Eprint
  {http://arxiv.org/abs/2103.08723} {arXiv:2103.08723 [astro-ph.CO]}
  \BibitemShut {NoStop}%
\bibitem [{\citenamefont {De~Felice}\ \emph {et~al.}(2021)\citenamefont
  {De~Felice}, \citenamefont {Mukohyama},\ and\ \citenamefont
  {Pookkillath}}]{De_Felice_2021}%
  \BibitemOpen
  \bibfield  {author} {\bibinfo {author} {\bibfnamefont {A.}~\bibnamefont
  {De~Felice}}, \bibinfo {author} {\bibfnamefont {S.}~\bibnamefont
  {Mukohyama}}, \ and\ \bibinfo {author} {\bibfnamefont {M.~C.}\ \bibnamefont
  {Pookkillath}},\ }\href {\doibase 10.1016/j.physletb.2021.136201} {\bibfield
  {journal} {\bibinfo  {journal} {Physics Letters B}\ }\textbf {\bibinfo
  {volume} {816}},\ \bibinfo {pages} {136201} (\bibinfo {year}
  {2021})}\BibitemShut {NoStop}%
\bibitem [{\citenamefont {Verde}\ \emph {et~al.}(2019)\citenamefont {Verde},
  \citenamefont {Treu},\ and\ \citenamefont {Riess}}]{Verde:2019ivm}%
  \BibitemOpen
  \bibfield  {author} {\bibinfo {author} {\bibfnamefont {L.}~\bibnamefont
  {Verde}}, \bibinfo {author} {\bibfnamefont {T.}~\bibnamefont {Treu}}, \ and\
  \bibinfo {author} {\bibfnamefont {A.~G.}\ \bibnamefont {Riess}},\ }\href
  {\doibase 10.1038/s41550-019-0902-0} {\bibfield  {journal} {\bibinfo
  {journal} {Nature Astron.}\ }\textbf {\bibinfo {volume} {3}},\ \bibinfo
  {pages} {891} (\bibinfo {year} {2019})},\ \Eprint
  {http://arxiv.org/abs/1907.10625} {arXiv:1907.10625 [astro-ph.CO]}
  \BibitemShut {NoStop}%
\bibitem [{\citenamefont {Di~Valentino}\ \emph
  {et~al.}(2020{\natexlab{a}})\citenamefont {Di~Valentino} \emph
  {et~al.}}]{DiValentino:2020zio}%
  \BibitemOpen
  \bibfield  {author} {\bibinfo {author} {\bibfnamefont {E.}~\bibnamefont
  {Di~Valentino}} \emph {et~al.},\ }\href@noop {} {\  (\bibinfo {year}
  {2020}{\natexlab{a}})},\ \Eprint {http://arxiv.org/abs/2008.11284}
  {arXiv:2008.11284 [astro-ph.CO]} \BibitemShut {NoStop}%
\bibitem [{\citenamefont {Di~Valentino}\ \emph
  {et~al.}(2021{\natexlab{a}})\citenamefont {Di~Valentino}, \citenamefont
  {Mena}, \citenamefont {Pan}, \citenamefont {Visinelli}, \citenamefont {Yang},
  \citenamefont {Melchiorri}, \citenamefont {Mota}, \citenamefont {Riess},\
  and\ \citenamefont {Silk}}]{DiValentino:2021izs}%
  \BibitemOpen
  \bibfield  {author} {\bibinfo {author} {\bibfnamefont {E.}~\bibnamefont
  {Di~Valentino}}, \bibinfo {author} {\bibfnamefont {O.}~\bibnamefont {Mena}},
  \bibinfo {author} {\bibfnamefont {S.}~\bibnamefont {Pan}}, \bibinfo {author}
  {\bibfnamefont {L.}~\bibnamefont {Visinelli}}, \bibinfo {author}
  {\bibfnamefont {W.}~\bibnamefont {Yang}}, \bibinfo {author} {\bibfnamefont
  {A.}~\bibnamefont {Melchiorri}}, \bibinfo {author} {\bibfnamefont {D.~F.}\
  \bibnamefont {Mota}}, \bibinfo {author} {\bibfnamefont {A.~G.}\ \bibnamefont
  {Riess}}, \ and\ \bibinfo {author} {\bibfnamefont {J.}~\bibnamefont {Silk}},\
  }\href@noop {} {\  (\bibinfo {year} {2021}{\natexlab{a}})},\ \Eprint
  {http://arxiv.org/abs/2103.01183} {arXiv:2103.01183 [astro-ph.CO]}
  \BibitemShut {NoStop}%
\bibitem [{\citenamefont {Joudaki}\ \emph
  {et~al.}(2017{\natexlab{a}})\citenamefont {Joudaki} \emph
  {et~al.}}]{Joudaki:2016mvz}%
  \BibitemOpen
  \bibfield  {author} {\bibinfo {author} {\bibfnamefont {S.}~\bibnamefont
  {Joudaki}} \emph {et~al.},\ }\href {\doibase 10.1093/mnras/stw2665}
  {\bibfield  {journal} {\bibinfo  {journal} {Mon. Not. Roy. Astron. Soc.}\
  }\textbf {\bibinfo {volume} {465}},\ \bibinfo {pages} {2033} (\bibinfo {year}
  {2017}{\natexlab{a}})},\ \Eprint {http://arxiv.org/abs/1601.05786}
  {arXiv:1601.05786 [astro-ph.CO]} \BibitemShut {NoStop}%
\bibitem [{\citenamefont {Joudaki}\ \emph
  {et~al.}(2017{\natexlab{b}})\citenamefont {Joudaki} \emph
  {et~al.}}]{Joudaki:2016kym}%
  \BibitemOpen
  \bibfield  {author} {\bibinfo {author} {\bibfnamefont {S.}~\bibnamefont
  {Joudaki}} \emph {et~al.},\ }\href {\doibase 10.1093/mnras/stx998} {\bibfield
   {journal} {\bibinfo  {journal} {Mon. Not. Roy. Astron. Soc.}\ }\textbf
  {\bibinfo {volume} {471}},\ \bibinfo {pages} {1259} (\bibinfo {year}
  {2017}{\natexlab{b}})},\ \Eprint {http://arxiv.org/abs/1610.04606}
  {arXiv:1610.04606 [astro-ph.CO]} \BibitemShut {NoStop}%
\bibitem [{\citenamefont {Joudaki}\ \emph {et~al.}(2018)\citenamefont {Joudaki}
  \emph {et~al.}}]{Joudaki:2017zdt}%
  \BibitemOpen
  \bibfield  {author} {\bibinfo {author} {\bibfnamefont {S.}~\bibnamefont
  {Joudaki}} \emph {et~al.},\ }\href {\doibase 10.1093/mnras/stx2820}
  {\bibfield  {journal} {\bibinfo  {journal} {Mon. Not. Roy. Astron. Soc.}\
  }\textbf {\bibinfo {volume} {474}},\ \bibinfo {pages} {4894} (\bibinfo {year}
  {2018})},\ \Eprint {http://arxiv.org/abs/1707.06627} {arXiv:1707.06627
  [astro-ph.CO]} \BibitemShut {NoStop}%
\bibitem [{\citenamefont {Hildebrandt}\ \emph {et~al.}(2020)\citenamefont
  {Hildebrandt} \emph {et~al.}}]{Hildebrandt:2018yau}%
  \BibitemOpen
  \bibfield  {author} {\bibinfo {author} {\bibfnamefont {H.}~\bibnamefont
  {Hildebrandt}} \emph {et~al.},\ }\href {\doibase 10.1051/0004-6361/201834878}
  {\bibfield  {journal} {\bibinfo  {journal} {Astron. Astrophys.}\ }\textbf
  {\bibinfo {volume} {633}},\ \bibinfo {pages} {A69} (\bibinfo {year}
  {2020})},\ \Eprint {http://arxiv.org/abs/1812.06076} {arXiv:1812.06076
  [astro-ph.CO]} \BibitemShut {NoStop}%
\bibitem [{\citenamefont {Troxel}\ \emph {et~al.}(2018)\citenamefont {Troxel}
  \emph {et~al.}}]{Troxel:2017xyo}%
  \BibitemOpen
  \bibfield  {author} {\bibinfo {author} {\bibfnamefont {M.~A.}\ \bibnamefont
  {Troxel}} \emph {et~al.} (\bibinfo {collaboration} {DES}),\ }\href {\doibase
  10.1103/PhysRevD.98.043528} {\bibfield  {journal} {\bibinfo  {journal} {Phys.
  Rev. D}\ }\textbf {\bibinfo {volume} {98}},\ \bibinfo {pages} {043528}
  (\bibinfo {year} {2018})},\ \Eprint {http://arxiv.org/abs/1708.01538}
  {arXiv:1708.01538 [astro-ph.CO]} \BibitemShut {NoStop}%
\bibitem [{\citenamefont {Tr\"oster}\ \emph {et~al.}(2020)\citenamefont
  {Tr\"oster} \emph {et~al.}}]{Troster:2019ean}%
  \BibitemOpen
  \bibfield  {author} {\bibinfo {author} {\bibfnamefont {T.}~\bibnamefont
  {Tr\"oster}} \emph {et~al.},\ }\href {\doibase 10.1051/0004-6361/201936772}
  {\bibfield  {journal} {\bibinfo  {journal} {Astron. Astrophys.}\ }\textbf
  {\bibinfo {volume} {633}},\ \bibinfo {pages} {L10} (\bibinfo {year}
  {2020})},\ \Eprint {http://arxiv.org/abs/1909.11006} {arXiv:1909.11006
  [astro-ph.CO]} \BibitemShut {NoStop}%
\bibitem [{\citenamefont {Joudaki}\ \emph {et~al.}(2020)\citenamefont {Joudaki}
  \emph {et~al.}}]{Joudaki:2019pmv}%
  \BibitemOpen
  \bibfield  {author} {\bibinfo {author} {\bibfnamefont {S.}~\bibnamefont
  {Joudaki}} \emph {et~al.},\ }\href {\doibase 10.1051/0004-6361/201936154}
  {\bibfield  {journal} {\bibinfo  {journal} {Astron. Astrophys.}\ }\textbf
  {\bibinfo {volume} {638}},\ \bibinfo {pages} {L1} (\bibinfo {year} {2020})},\
  \Eprint {http://arxiv.org/abs/1906.09262} {arXiv:1906.09262 [astro-ph.CO]}
  \BibitemShut {NoStop}%
\bibitem [{\citenamefont {Asgari}\ \emph {et~al.}(2020)\citenamefont {Asgari}
  \emph {et~al.}}]{Asgari:2019fkq}%
  \BibitemOpen
  \bibfield  {author} {\bibinfo {author} {\bibfnamefont {M.}~\bibnamefont
  {Asgari}} \emph {et~al.},\ }\href {\doibase 10.1051/0004-6361/201936512}
  {\bibfield  {journal} {\bibinfo  {journal} {Astron. Astrophys.}\ }\textbf
  {\bibinfo {volume} {634}},\ \bibinfo {pages} {A127} (\bibinfo {year}
  {2020})},\ \Eprint {http://arxiv.org/abs/1910.05336} {arXiv:1910.05336
  [astro-ph.CO]} \BibitemShut {NoStop}%
\bibitem [{\citenamefont {Ivanov}\ \emph {et~al.}(2020)\citenamefont {Ivanov},
  \citenamefont {Simonovi\'c},\ and\ \citenamefont
  {Zaldarriaga}}]{Ivanov:2019pdj}%
  \BibitemOpen
  \bibfield  {author} {\bibinfo {author} {\bibfnamefont {M.~M.}\ \bibnamefont
  {Ivanov}}, \bibinfo {author} {\bibfnamefont {M.}~\bibnamefont {Simonovi\'c}},
  \ and\ \bibinfo {author} {\bibfnamefont {M.}~\bibnamefont {Zaldarriaga}},\
  }\href {\doibase 10.1088/1475-7516/2020/05/042} {\bibfield  {journal}
  {\bibinfo  {journal} {JCAP}\ }\textbf {\bibinfo {volume} {05}},\ \bibinfo
  {pages} {042} (\bibinfo {year} {2020})},\ \Eprint
  {http://arxiv.org/abs/1909.05277} {arXiv:1909.05277 [astro-ph.CO]}
  \BibitemShut {NoStop}%
\bibitem [{\citenamefont {Asgari}\ \emph {et~al.}(2021)\citenamefont {Asgari}
  \emph {et~al.}}]{Asgari:2020wuj}%
  \BibitemOpen
  \bibfield  {author} {\bibinfo {author} {\bibfnamefont {M.}~\bibnamefont
  {Asgari}} \emph {et~al.} (\bibinfo {collaboration} {KiDS}),\ }\href {\doibase
  10.1051/0004-6361/202039070} {\bibfield  {journal} {\bibinfo  {journal}
  {Astron. Astrophys.}\ }\textbf {\bibinfo {volume} {645}},\ \bibinfo {pages}
  {A104} (\bibinfo {year} {2021})},\ \Eprint {http://arxiv.org/abs/2007.15633}
  {arXiv:2007.15633 [astro-ph.CO]} \BibitemShut {NoStop}%
\bibitem [{\citenamefont {Heymans}\ \emph {et~al.}(2020)\citenamefont {Heymans}
  \emph {et~al.}}]{Heymans:2020ghw}%
  \BibitemOpen
  \bibfield  {author} {\bibinfo {author} {\bibfnamefont {C.}~\bibnamefont
  {Heymans}} \emph {et~al.},\ }\href@noop {} {\  (\bibinfo {year} {2020})},\
  \Eprint {http://arxiv.org/abs/2007.15632} {arXiv:2007.15632 [astro-ph.CO]}
  \BibitemShut {NoStop}%
\bibitem [{\citenamefont {Ade}\ \emph {et~al.}(2016)\citenamefont {Ade} \emph
  {et~al.}}]{Ade:2015fva}%
  \BibitemOpen
  \bibfield  {author} {\bibinfo {author} {\bibfnamefont {P.~A.~R.}\
  \bibnamefont {Ade}} \emph {et~al.} (\bibinfo {collaboration} {Planck}),\
  }\href {\doibase 10.1051/0004-6361/201525833} {\bibfield  {journal} {\bibinfo
   {journal} {Astron. Astrophys.}\ }\textbf {\bibinfo {volume} {594}},\
  \bibinfo {pages} {A24} (\bibinfo {year} {2016})},\ \Eprint
  {http://arxiv.org/abs/1502.01597} {arXiv:1502.01597 [astro-ph.CO]}
  \BibitemShut {NoStop}%
\bibitem [{\citenamefont {van Uitert}\ \emph {et~al.}(2018)\citenamefont {van
  Uitert} \emph {et~al.}}]{vanUitert:2017ieu}%
  \BibitemOpen
  \bibfield  {author} {\bibinfo {author} {\bibfnamefont {E.}~\bibnamefont {van
  Uitert}} \emph {et~al.},\ }\href {\doibase 10.1093/mnras/sty551} {\bibfield
  {journal} {\bibinfo  {journal} {Mon. Not. Roy. Astron. Soc.}\ }\textbf
  {\bibinfo {volume} {476}},\ \bibinfo {pages} {4662} (\bibinfo {year}
  {2018})},\ \Eprint {http://arxiv.org/abs/1706.05004} {arXiv:1706.05004
  [astro-ph.CO]} \BibitemShut {NoStop}%
\bibitem [{\citenamefont {Hamana}\ \emph {et~al.}(2020)\citenamefont {Hamana}
  \emph {et~al.}}]{Hamana:2019etx}%
  \BibitemOpen
  \bibfield  {author} {\bibinfo {author} {\bibfnamefont {T.}~\bibnamefont
  {Hamana}} \emph {et~al.},\ }\href {\doibase 10.1093/pasj/psz138} {\bibfield
  {journal} {\bibinfo  {journal} {Publ. Astron. Soc. Jap.}\ }\textbf {\bibinfo
  {volume} {72}},\ \bibinfo {pages} {16} (\bibinfo {year} {2020})},\ \Eprint
  {http://arxiv.org/abs/1906.06041} {arXiv:1906.06041 [astro-ph.CO]}
  \BibitemShut {NoStop}%
\bibitem [{\citenamefont {Di~Valentino}\ \emph
  {et~al.}(2020{\natexlab{b}})\citenamefont {Di~Valentino} \emph
  {et~al.}}]{DiValentino:2020vvd}%
  \BibitemOpen
  \bibfield  {author} {\bibinfo {author} {\bibfnamefont {E.}~\bibnamefont
  {Di~Valentino}} \emph {et~al.},\ }\href@noop {} {\  (\bibinfo {year}
  {2020}{\natexlab{b}})},\ \Eprint {http://arxiv.org/abs/2008.11285}
  {arXiv:2008.11285 [astro-ph.CO]} \BibitemShut {NoStop}%
\bibitem [{\citenamefont {Battye}\ and\ \citenamefont
  {Moss}(2014)}]{Battye:2013xqa}%
  \BibitemOpen
  \bibfield  {author} {\bibinfo {author} {\bibfnamefont {R.~A.}\ \bibnamefont
  {Battye}}\ and\ \bibinfo {author} {\bibfnamefont {A.}~\bibnamefont {Moss}},\
  }\href {\doibase 10.1103/PhysRevLett.112.051303} {\bibfield  {journal}
  {\bibinfo  {journal} {Phys. Rev. Lett.}\ }\textbf {\bibinfo {volume} {112}},\
  \bibinfo {pages} {051303} (\bibinfo {year} {2014})},\ \Eprint
  {http://arxiv.org/abs/1308.5870} {arXiv:1308.5870 [astro-ph.CO]} \BibitemShut
  {NoStop}%
\bibitem [{\citenamefont {MacCrann}\ \emph {et~al.}(2015)\citenamefont
  {MacCrann}, \citenamefont {Zuntz}, \citenamefont {Bridle}, \citenamefont
  {Jain},\ and\ \citenamefont {Becker}}]{MacCrann:2014wfa}%
  \BibitemOpen
  \bibfield  {author} {\bibinfo {author} {\bibfnamefont {N.}~\bibnamefont
  {MacCrann}}, \bibinfo {author} {\bibfnamefont {J.}~\bibnamefont {Zuntz}},
  \bibinfo {author} {\bibfnamefont {S.}~\bibnamefont {Bridle}}, \bibinfo
  {author} {\bibfnamefont {B.}~\bibnamefont {Jain}}, \ and\ \bibinfo {author}
  {\bibfnamefont {M.~R.}\ \bibnamefont {Becker}},\ }\href {\doibase
  10.1093/mnras/stv1154} {\bibfield  {journal} {\bibinfo  {journal} {Mon. Not.
  Roy. Astron. Soc.}\ }\textbf {\bibinfo {volume} {451}},\ \bibinfo {pages}
  {2877} (\bibinfo {year} {2015})},\ \Eprint {http://arxiv.org/abs/1408.4742}
  {arXiv:1408.4742 [astro-ph.CO]} \BibitemShut {NoStop}%
\bibitem [{\citenamefont {Vagnozzi}\ \emph {et~al.}(2017)\citenamefont
  {Vagnozzi}, \citenamefont {Giusarma}, \citenamefont {Mena}, \citenamefont
  {Freese}, \citenamefont {Gerbino}, \citenamefont {Ho},\ and\ \citenamefont
  {Lattanzi}}]{Vagnozzi:2017ovm}%
  \BibitemOpen
  \bibfield  {author} {\bibinfo {author} {\bibfnamefont {S.}~\bibnamefont
  {Vagnozzi}}, \bibinfo {author} {\bibfnamefont {E.}~\bibnamefont {Giusarma}},
  \bibinfo {author} {\bibfnamefont {O.}~\bibnamefont {Mena}}, \bibinfo {author}
  {\bibfnamefont {K.}~\bibnamefont {Freese}}, \bibinfo {author} {\bibfnamefont
  {M.}~\bibnamefont {Gerbino}}, \bibinfo {author} {\bibfnamefont
  {S.}~\bibnamefont {Ho}}, \ and\ \bibinfo {author} {\bibfnamefont
  {M.}~\bibnamefont {Lattanzi}},\ }\href {\doibase 10.1103/PhysRevD.96.123503}
  {\bibfield  {journal} {\bibinfo  {journal} {Phys. Rev. D}\ }\textbf {\bibinfo
  {volume} {96}},\ \bibinfo {pages} {123503} (\bibinfo {year} {2017})},\
  \Eprint {http://arxiv.org/abs/1701.08172} {arXiv:1701.08172 [astro-ph.CO]}
  \BibitemShut {NoStop}%
\bibitem [{\citenamefont {Mccarthy}\ \emph {et~al.}(2018)\citenamefont
  {Mccarthy}, \citenamefont {Bird}, \citenamefont {Schaye}, \citenamefont
  {Harnois-Deraps}, \citenamefont {Font},\ and\ \citenamefont
  {Van~Waerbeke}}]{McCarthy:2017csu}%
  \BibitemOpen
  \bibfield  {author} {\bibinfo {author} {\bibfnamefont {I.~G.}\ \bibnamefont
  {Mccarthy}}, \bibinfo {author} {\bibfnamefont {S.}~\bibnamefont {Bird}},
  \bibinfo {author} {\bibfnamefont {J.}~\bibnamefont {Schaye}}, \bibinfo
  {author} {\bibfnamefont {J.}~\bibnamefont {Harnois-Deraps}}, \bibinfo
  {author} {\bibfnamefont {A.~S.}\ \bibnamefont {Font}}, \ and\ \bibinfo
  {author} {\bibfnamefont {L.}~\bibnamefont {Van~Waerbeke}},\ }\href {\doibase
  10.1093/mnras/sty377} {\bibfield  {journal} {\bibinfo  {journal} {Mon. Not.
  Roy. Astron. Soc.}\ }\textbf {\bibinfo {volume} {476}},\ \bibinfo {pages}
  {2999} (\bibinfo {year} {2018})},\ \Eprint {http://arxiv.org/abs/1712.02411}
  {arXiv:1712.02411 [astro-ph.CO]} \BibitemShut {NoStop}%
\bibitem [{\citenamefont {Feng}\ \emph {et~al.}(2017)\citenamefont {Feng},
  \citenamefont {Zhang},\ and\ \citenamefont {Zhang}}]{Feng:2017nss}%
  \BibitemOpen
  \bibfield  {author} {\bibinfo {author} {\bibfnamefont {L.}~\bibnamefont
  {Feng}}, \bibinfo {author} {\bibfnamefont {J.-F.}\ \bibnamefont {Zhang}}, \
  and\ \bibinfo {author} {\bibfnamefont {X.}~\bibnamefont {Zhang}},\ }\href
  {\doibase 10.1140/epjc/s10052-017-4986-3} {\bibfield  {journal} {\bibinfo
  {journal} {Eur. Phys. J. C}\ }\textbf {\bibinfo {volume} {77}},\ \bibinfo
  {pages} {418} (\bibinfo {year} {2017})},\ \Eprint
  {http://arxiv.org/abs/1703.04884} {arXiv:1703.04884 [astro-ph.CO]}
  \BibitemShut {NoStop}%
\bibitem [{\citenamefont {Hlozek}\ \emph {et~al.}(2015)\citenamefont {Hlozek},
  \citenamefont {Grin}, \citenamefont {Marsh},\ and\ \citenamefont
  {Ferreira}}]{Hlozek:2014lca}%
  \BibitemOpen
  \bibfield  {author} {\bibinfo {author} {\bibfnamefont {R.}~\bibnamefont
  {Hlozek}}, \bibinfo {author} {\bibfnamefont {D.}~\bibnamefont {Grin}},
  \bibinfo {author} {\bibfnamefont {D.~J.~E.}\ \bibnamefont {Marsh}}, \ and\
  \bibinfo {author} {\bibfnamefont {P.~G.}\ \bibnamefont {Ferreira}},\ }\href
  {\doibase 10.1103/PhysRevD.91.103512} {\bibfield  {journal} {\bibinfo
  {journal} {Phys. Rev. D}\ }\textbf {\bibinfo {volume} {91}},\ \bibinfo
  {pages} {103512} (\bibinfo {year} {2015})},\ \Eprint
  {http://arxiv.org/abs/1410.2896} {arXiv:1410.2896 [astro-ph.CO]} \BibitemShut
  {NoStop}%
\bibitem [{\citenamefont {Enqvist}\ \emph {et~al.}(2015)\citenamefont
  {Enqvist}, \citenamefont {Nadathur}, \citenamefont {Sekiguchi},\ and\
  \citenamefont {Takahashi}}]{Enqvist:2015ara}%
  \BibitemOpen
  \bibfield  {author} {\bibinfo {author} {\bibfnamefont {K.}~\bibnamefont
  {Enqvist}}, \bibinfo {author} {\bibfnamefont {S.}~\bibnamefont {Nadathur}},
  \bibinfo {author} {\bibfnamefont {T.}~\bibnamefont {Sekiguchi}}, \ and\
  \bibinfo {author} {\bibfnamefont {T.}~\bibnamefont {Takahashi}},\ }\href
  {\doibase 10.1088/1475-7516/2015/09/067} {\bibfield  {journal} {\bibinfo
  {journal} {JCAP}\ }\textbf {\bibinfo {volume} {09}},\ \bibinfo {pages} {067}
  (\bibinfo {year} {2015})},\ \Eprint {http://arxiv.org/abs/1505.05511}
  {arXiv:1505.05511 [astro-ph.CO]} \BibitemShut {NoStop}%
\bibitem [{\citenamefont {Di~Valentino}\ \emph {et~al.}(2018)\citenamefont
  {Di~Valentino}, \citenamefont {B\o{}ehm}, \citenamefont {Hivon},\ and\
  \citenamefont {Bouchet}}]{DiValentino:2017oaw}%
  \BibitemOpen
  \bibfield  {author} {\bibinfo {author} {\bibfnamefont {E.}~\bibnamefont
  {Di~Valentino}}, \bibinfo {author} {\bibfnamefont {C.}~\bibnamefont
  {B\o{}ehm}}, \bibinfo {author} {\bibfnamefont {E.}~\bibnamefont {Hivon}}, \
  and\ \bibinfo {author} {\bibfnamefont {F.~R.}\ \bibnamefont {Bouchet}},\
  }\href {\doibase 10.1103/PhysRevD.97.043513} {\bibfield  {journal} {\bibinfo
  {journal} {Phys. Rev. D}\ }\textbf {\bibinfo {volume} {97}},\ \bibinfo
  {pages} {043513} (\bibinfo {year} {2018})},\ \Eprint
  {http://arxiv.org/abs/1710.02559} {arXiv:1710.02559 [astro-ph.CO]}
  \BibitemShut {NoStop}%
\bibitem [{\citenamefont {Chudaykin}\ \emph {et~al.}(2018)\citenamefont
  {Chudaykin}, \citenamefont {Gorbunov},\ and\ \citenamefont
  {Tkachev}}]{Chudaykin:2017ptd}%
  \BibitemOpen
  \bibfield  {author} {\bibinfo {author} {\bibfnamefont {A.}~\bibnamefont
  {Chudaykin}}, \bibinfo {author} {\bibfnamefont {D.}~\bibnamefont {Gorbunov}},
  \ and\ \bibinfo {author} {\bibfnamefont {I.}~\bibnamefont {Tkachev}},\ }\href
  {\doibase 10.1103/PhysRevD.97.083508} {\bibfield  {journal} {\bibinfo
  {journal} {Phys. Rev. D}\ }\textbf {\bibinfo {volume} {97}},\ \bibinfo
  {pages} {083508} (\bibinfo {year} {2018})},\ \Eprint
  {http://arxiv.org/abs/1711.06738} {arXiv:1711.06738 [astro-ph.CO]}
  \BibitemShut {NoStop}%
\bibitem [{\citenamefont {Pandey}\ \emph {et~al.}(2020)\citenamefont {Pandey},
  \citenamefont {Karwal},\ and\ \citenamefont {Das}}]{Pandey:2019plg}%
  \BibitemOpen
  \bibfield  {author} {\bibinfo {author} {\bibfnamefont {K.~L.}\ \bibnamefont
  {Pandey}}, \bibinfo {author} {\bibfnamefont {T.}~\bibnamefont {Karwal}}, \
  and\ \bibinfo {author} {\bibfnamefont {S.}~\bibnamefont {Das}},\ }\href
  {\doibase 10.1088/1475-7516/2020/07/026} {\bibfield  {journal} {\bibinfo
  {journal} {JCAP}\ }\textbf {\bibinfo {volume} {07}},\ \bibinfo {pages} {026}
  (\bibinfo {year} {2020})},\ \Eprint {http://arxiv.org/abs/1902.10636}
  {arXiv:1902.10636 [astro-ph.CO]} \BibitemShut {NoStop}%
\bibitem [{\citenamefont {Xiao}\ \emph {et~al.}(2020)\citenamefont {Xiao},
  \citenamefont {Zhang}, \citenamefont {An}, \citenamefont {Feng},\ and\
  \citenamefont {Wang}}]{Xiao:2019ccl}%
  \BibitemOpen
  \bibfield  {author} {\bibinfo {author} {\bibfnamefont {L.}~\bibnamefont
  {Xiao}}, \bibinfo {author} {\bibfnamefont {L.}~\bibnamefont {Zhang}},
  \bibinfo {author} {\bibfnamefont {R.}~\bibnamefont {An}}, \bibinfo {author}
  {\bibfnamefont {C.}~\bibnamefont {Feng}}, \ and\ \bibinfo {author}
  {\bibfnamefont {B.}~\bibnamefont {Wang}},\ }\href {\doibase
  10.1088/1475-7516/2020/01/045} {\bibfield  {journal} {\bibinfo  {journal}
  {JCAP}\ }\textbf {\bibinfo {volume} {01}},\ \bibinfo {pages} {045} (\bibinfo
  {year} {2020})},\ \Eprint {http://arxiv.org/abs/1908.02668} {arXiv:1908.02668
  [astro-ph.CO]} \BibitemShut {NoStop}%
\bibitem [{\citenamefont {Abellan}\ \emph {et~al.}(2020)\citenamefont
  {Abellan}, \citenamefont {Murgia}, \citenamefont {Poulin},\ and\
  \citenamefont {Lavalle}}]{Abellan:2020pmw}%
  \BibitemOpen
  \bibfield  {author} {\bibinfo {author} {\bibfnamefont {G.~F.}\ \bibnamefont
  {Abellan}}, \bibinfo {author} {\bibfnamefont {R.}~\bibnamefont {Murgia}},
  \bibinfo {author} {\bibfnamefont {V.}~\bibnamefont {Poulin}}, \ and\ \bibinfo
  {author} {\bibfnamefont {J.}~\bibnamefont {Lavalle}},\ }\href@noop {} {\
  (\bibinfo {year} {2020})},\ \Eprint {http://arxiv.org/abs/2008.09615}
  {arXiv:2008.09615 [astro-ph.CO]} \BibitemShut {NoStop}%
\bibitem [{\citenamefont {Chen}\ \emph {et~al.}(2020)\citenamefont {Chen} \emph
  {et~al.}}]{Chen:2020iwm}%
  \BibitemOpen
  \bibfield  {author} {\bibinfo {author} {\bibfnamefont {A.}~\bibnamefont
  {Chen}} \emph {et~al.} (\bibinfo {collaboration} {DES}),\ }\href@noop {} {\
  (\bibinfo {year} {2020})},\ \Eprint {http://arxiv.org/abs/2011.04606}
  {arXiv:2011.04606 [astro-ph.CO]} \BibitemShut {NoStop}%
\bibitem [{\citenamefont {Abell\'an}\ \emph {et~al.}(2021)\citenamefont
  {Abell\'an}, \citenamefont {Murgia},\ and\ \citenamefont
  {Poulin}}]{Abellan:2021bpx}%
  \BibitemOpen
  \bibfield  {author} {\bibinfo {author} {\bibfnamefont {G.~F.}\ \bibnamefont
  {Abell\'an}}, \bibinfo {author} {\bibfnamefont {R.}~\bibnamefont {Murgia}}, \
  and\ \bibinfo {author} {\bibfnamefont {V.}~\bibnamefont {Poulin}},\
  }\href@noop {} {\  (\bibinfo {year} {2021})},\ \Eprint
  {http://arxiv.org/abs/2102.12498} {arXiv:2102.12498 [astro-ph.CO]}
  \BibitemShut {NoStop}%
\bibitem [{\citenamefont {Kunz}\ \emph {et~al.}(2015)\citenamefont {Kunz},
  \citenamefont {Nesseris},\ and\ \citenamefont {Sawicki}}]{Kunz:2015oqa}%
  \BibitemOpen
  \bibfield  {author} {\bibinfo {author} {\bibfnamefont {M.}~\bibnamefont
  {Kunz}}, \bibinfo {author} {\bibfnamefont {S.}~\bibnamefont {Nesseris}}, \
  and\ \bibinfo {author} {\bibfnamefont {I.}~\bibnamefont {Sawicki}},\ }\href
  {\doibase 10.1103/PhysRevD.92.063006} {\bibfield  {journal} {\bibinfo
  {journal} {Phys. Rev. D}\ }\textbf {\bibinfo {volume} {92}},\ \bibinfo
  {pages} {063006} (\bibinfo {year} {2015})},\ \Eprint
  {http://arxiv.org/abs/1507.01486} {arXiv:1507.01486 [astro-ph.CO]}
  \BibitemShut {NoStop}%
\bibitem [{\citenamefont {Pourtsidou}\ and\ \citenamefont
  {Tram}(2016)}]{Pourtsidou:2016ico}%
  \BibitemOpen
  \bibfield  {author} {\bibinfo {author} {\bibfnamefont {A.}~\bibnamefont
  {Pourtsidou}}\ and\ \bibinfo {author} {\bibfnamefont {T.}~\bibnamefont
  {Tram}},\ }\href {\doibase 10.1103/PhysRevD.94.043518} {\bibfield  {journal}
  {\bibinfo  {journal} {Phys. Rev. D}\ }\textbf {\bibinfo {volume} {94}},\
  \bibinfo {pages} {043518} (\bibinfo {year} {2016})},\ \Eprint
  {http://arxiv.org/abs/1604.04222} {arXiv:1604.04222 [astro-ph.CO]}
  \BibitemShut {NoStop}%
\bibitem [{\citenamefont {Kumar}\ and\ \citenamefont
  {Nunes}(2016)}]{Kumar:2016zpg}%
  \BibitemOpen
  \bibfield  {author} {\bibinfo {author} {\bibfnamefont {S.}~\bibnamefont
  {Kumar}}\ and\ \bibinfo {author} {\bibfnamefont {R.~C.}\ \bibnamefont
  {Nunes}},\ }\href {\doibase 10.1103/PhysRevD.94.123511} {\bibfield  {journal}
  {\bibinfo  {journal} {Phys. Rev. D}\ }\textbf {\bibinfo {volume} {94}},\
  \bibinfo {pages} {123511} (\bibinfo {year} {2016})},\ \Eprint
  {http://arxiv.org/abs/1608.02454} {arXiv:1608.02454 [astro-ph.CO]}
  \BibitemShut {NoStop}%
\bibitem [{\citenamefont {Gariazzo}\ \emph {et~al.}(2017)\citenamefont
  {Gariazzo}, \citenamefont {Escudero}, \citenamefont {Diamanti},\ and\
  \citenamefont {Mena}}]{Gariazzo:2017pzb}%
  \BibitemOpen
  \bibfield  {author} {\bibinfo {author} {\bibfnamefont {S.}~\bibnamefont
  {Gariazzo}}, \bibinfo {author} {\bibfnamefont {M.}~\bibnamefont {Escudero}},
  \bibinfo {author} {\bibfnamefont {R.}~\bibnamefont {Diamanti}}, \ and\
  \bibinfo {author} {\bibfnamefont {O.}~\bibnamefont {Mena}},\ }\href {\doibase
  10.1103/PhysRevD.96.043501} {\bibfield  {journal} {\bibinfo  {journal} {Phys.
  Rev. D}\ }\textbf {\bibinfo {volume} {96}},\ \bibinfo {pages} {043501}
  (\bibinfo {year} {2017})},\ \Eprint {http://arxiv.org/abs/1704.02991}
  {arXiv:1704.02991 [astro-ph.CO]} \BibitemShut {NoStop}%
\bibitem [{\citenamefont {Benetti}\ \emph {et~al.}(2018)\citenamefont
  {Benetti}, \citenamefont {Graef},\ and\ \citenamefont
  {Alcaniz}}]{Benetti:2017juy}%
  \BibitemOpen
  \bibfield  {author} {\bibinfo {author} {\bibfnamefont {M.}~\bibnamefont
  {Benetti}}, \bibinfo {author} {\bibfnamefont {L.~L.}\ \bibnamefont {Graef}},
  \ and\ \bibinfo {author} {\bibfnamefont {J.~S.}\ \bibnamefont {Alcaniz}},\
  }\href {\doibase 10.1088/1475-7516/2018/07/066} {\bibfield  {journal}
  {\bibinfo  {journal} {JCAP}\ }\textbf {\bibinfo {volume} {07}},\ \bibinfo
  {pages} {066} (\bibinfo {year} {2018})},\ \Eprint
  {http://arxiv.org/abs/1712.00677} {arXiv:1712.00677 [astro-ph.CO]}
  \BibitemShut {NoStop}%
\bibitem [{\citenamefont {Buen-Abad}\ \emph {et~al.}(2018)\citenamefont
  {Buen-Abad}, \citenamefont {Schmaltz}, \citenamefont {Lesgourgues},\ and\
  \citenamefont {Brinckmann}}]{Buen-Abad:2017gxg}%
  \BibitemOpen
  \bibfield  {author} {\bibinfo {author} {\bibfnamefont {M.~A.}\ \bibnamefont
  {Buen-Abad}}, \bibinfo {author} {\bibfnamefont {M.}~\bibnamefont {Schmaltz}},
  \bibinfo {author} {\bibfnamefont {J.}~\bibnamefont {Lesgourgues}}, \ and\
  \bibinfo {author} {\bibfnamefont {T.}~\bibnamefont {Brinckmann}},\ }\href
  {\doibase 10.1088/1475-7516/2018/01/008} {\bibfield  {journal} {\bibinfo
  {journal} {JCAP}\ }\textbf {\bibinfo {volume} {01}},\ \bibinfo {pages} {008}
  (\bibinfo {year} {2018})},\ \Eprint {http://arxiv.org/abs/1708.09406}
  {arXiv:1708.09406 [astro-ph.CO]} \BibitemShut {NoStop}%
\bibitem [{\citenamefont {Poulin}\ \emph {et~al.}(2018)\citenamefont {Poulin},
  \citenamefont {Boddy}, \citenamefont {Bird},\ and\ \citenamefont
  {Kamionkowski}}]{Poulin:2018zxs}%
  \BibitemOpen
  \bibfield  {author} {\bibinfo {author} {\bibfnamefont {V.}~\bibnamefont
  {Poulin}}, \bibinfo {author} {\bibfnamefont {K.~K.}\ \bibnamefont {Boddy}},
  \bibinfo {author} {\bibfnamefont {S.}~\bibnamefont {Bird}}, \ and\ \bibinfo
  {author} {\bibfnamefont {M.}~\bibnamefont {Kamionkowski}},\ }\href {\doibase
  10.1103/PhysRevD.97.123504} {\bibfield  {journal} {\bibinfo  {journal} {Phys.
  Rev. D}\ }\textbf {\bibinfo {volume} {97}},\ \bibinfo {pages} {123504}
  (\bibinfo {year} {2018})},\ \Eprint {http://arxiv.org/abs/1803.02474}
  {arXiv:1803.02474 [astro-ph.CO]} \BibitemShut {NoStop}%
\bibitem [{\citenamefont {Kumar}\ \emph {et~al.}(2018)\citenamefont {Kumar},
  \citenamefont {Nunes},\ and\ \citenamefont {Yadav}}]{Kumar:2018yhh}%
  \BibitemOpen
  \bibfield  {author} {\bibinfo {author} {\bibfnamefont {S.}~\bibnamefont
  {Kumar}}, \bibinfo {author} {\bibfnamefont {R.~C.}\ \bibnamefont {Nunes}}, \
  and\ \bibinfo {author} {\bibfnamefont {S.~K.}\ \bibnamefont {Yadav}},\ }\href
  {\doibase 10.1103/PhysRevD.98.043521} {\bibfield  {journal} {\bibinfo
  {journal} {Phys. Rev. D}\ }\textbf {\bibinfo {volume} {98}},\ \bibinfo
  {pages} {043521} (\bibinfo {year} {2018})},\ \Eprint
  {http://arxiv.org/abs/1803.10229} {arXiv:1803.10229 [astro-ph.CO]}
  \BibitemShut {NoStop}%
\bibitem [{\citenamefont {Lambiase}\ \emph {et~al.}(2019)\citenamefont
  {Lambiase}, \citenamefont {Mohanty}, \citenamefont {Narang},\ and\
  \citenamefont {Parashari}}]{Lambiase:2018ows}%
  \BibitemOpen
  \bibfield  {author} {\bibinfo {author} {\bibfnamefont {G.}~\bibnamefont
  {Lambiase}}, \bibinfo {author} {\bibfnamefont {S.}~\bibnamefont {Mohanty}},
  \bibinfo {author} {\bibfnamefont {A.}~\bibnamefont {Narang}}, \ and\ \bibinfo
  {author} {\bibfnamefont {P.}~\bibnamefont {Parashari}},\ }\href {\doibase
  10.1140/epjc/s10052-019-6634-6} {\bibfield  {journal} {\bibinfo  {journal}
  {Eur. Phys. J. C}\ }\textbf {\bibinfo {volume} {79}},\ \bibinfo {pages} {141}
  (\bibinfo {year} {2019})},\ \Eprint {http://arxiv.org/abs/1804.07154}
  {arXiv:1804.07154 [astro-ph.CO]} \BibitemShut {NoStop}%
\bibitem [{\citenamefont {Dutta}\ \emph {et~al.}(2020)\citenamefont {Dutta},
  \citenamefont {Ruchika}, \citenamefont {Roy}, \citenamefont {Sen},\ and\
  \citenamefont {Sheikh-Jabbari}}]{Dutta:2018vmq}%
  \BibitemOpen
  \bibfield  {author} {\bibinfo {author} {\bibfnamefont {K.}~\bibnamefont
  {Dutta}}, \bibinfo {author} {\bibnamefont {Ruchika}}, \bibinfo {author}
  {\bibfnamefont {A.}~\bibnamefont {Roy}}, \bibinfo {author} {\bibfnamefont
  {A.~A.}\ \bibnamefont {Sen}}, \ and\ \bibinfo {author} {\bibfnamefont
  {M.~M.}\ \bibnamefont {Sheikh-Jabbari}},\ }\href {\doibase
  10.1007/s10714-020-2665-4} {\bibfield  {journal} {\bibinfo  {journal} {Gen.
  Rel. Grav.}\ }\textbf {\bibinfo {volume} {52}},\ \bibinfo {pages} {15}
  (\bibinfo {year} {2020})},\ \Eprint {http://arxiv.org/abs/1808.06623}
  {arXiv:1808.06623 [astro-ph.CO]} \BibitemShut {NoStop}%
\bibitem [{\citenamefont {Kumar}\ \emph
  {et~al.}(2019{\natexlab{a}})\citenamefont {Kumar}, \citenamefont {Nunes},\
  and\ \citenamefont {Yadav}}]{Kumar:2019gfl}%
  \BibitemOpen
  \bibfield  {author} {\bibinfo {author} {\bibfnamefont {S.}~\bibnamefont
  {Kumar}}, \bibinfo {author} {\bibfnamefont {R.~C.}\ \bibnamefont {Nunes}}, \
  and\ \bibinfo {author} {\bibfnamefont {S.~K.}\ \bibnamefont {Yadav}},\ }\href
  {\doibase 10.1093/mnras/stz2676} {\bibfield  {journal} {\bibinfo  {journal}
  {Mon. Not. Roy. Astron. Soc.}\ }\textbf {\bibinfo {volume} {490}},\ \bibinfo
  {pages} {1406} (\bibinfo {year} {2019}{\natexlab{a}})},\ \Eprint
  {http://arxiv.org/abs/1901.07549} {arXiv:1901.07549 [astro-ph.CO]}
  \BibitemShut {NoStop}%
\bibitem [{\citenamefont {Kumar}\ \emph
  {et~al.}(2019{\natexlab{b}})\citenamefont {Kumar}, \citenamefont {Nunes},\
  and\ \citenamefont {Yadav}}]{Kumar:2019wfs}%
  \BibitemOpen
  \bibfield  {author} {\bibinfo {author} {\bibfnamefont {S.}~\bibnamefont
  {Kumar}}, \bibinfo {author} {\bibfnamefont {R.~C.}\ \bibnamefont {Nunes}}, \
  and\ \bibinfo {author} {\bibfnamefont {S.~K.}\ \bibnamefont {Yadav}},\ }\href
  {\doibase 10.1140/epjc/s10052-019-7087-7} {\bibfield  {journal} {\bibinfo
  {journal} {Eur. Phys. J. C}\ }\textbf {\bibinfo {volume} {79}},\ \bibinfo
  {pages} {576} (\bibinfo {year} {2019}{\natexlab{b}})},\ \Eprint
  {http://arxiv.org/abs/1903.04865} {arXiv:1903.04865 [astro-ph.CO]}
  \BibitemShut {NoStop}%
\bibitem [{\citenamefont {Archidiacono}\ \emph {et~al.}(2019)\citenamefont
  {Archidiacono}, \citenamefont {Hooper}, \citenamefont {Murgia}, \citenamefont
  {Bohr}, \citenamefont {Lesgourgues},\ and\ \citenamefont
  {Viel}}]{Archidiacono:2019wdp}%
  \BibitemOpen
  \bibfield  {author} {\bibinfo {author} {\bibfnamefont {M.}~\bibnamefont
  {Archidiacono}}, \bibinfo {author} {\bibfnamefont {D.~C.}\ \bibnamefont
  {Hooper}}, \bibinfo {author} {\bibfnamefont {R.}~\bibnamefont {Murgia}},
  \bibinfo {author} {\bibfnamefont {S.}~\bibnamefont {Bohr}}, \bibinfo {author}
  {\bibfnamefont {J.}~\bibnamefont {Lesgourgues}}, \ and\ \bibinfo {author}
  {\bibfnamefont {M.}~\bibnamefont {Viel}},\ }\href {\doibase
  10.1088/1475-7516/2019/10/055} {\bibfield  {journal} {\bibinfo  {journal}
  {JCAP}\ }\textbf {\bibinfo {volume} {10}},\ \bibinfo {pages} {055} (\bibinfo
  {year} {2019})},\ \Eprint {http://arxiv.org/abs/1907.01496} {arXiv:1907.01496
  [astro-ph.CO]} \BibitemShut {NoStop}%
\bibitem [{\citenamefont {Di~Valentino}\ \emph
  {et~al.}(2020{\natexlab{c}})\citenamefont {Di~Valentino}, \citenamefont
  {Melchiorri}, \citenamefont {Mena},\ and\ \citenamefont
  {Vagnozzi}}]{DiValentino:2019ffd}%
  \BibitemOpen
  \bibfield  {author} {\bibinfo {author} {\bibfnamefont {E.}~\bibnamefont
  {Di~Valentino}}, \bibinfo {author} {\bibfnamefont {A.}~\bibnamefont
  {Melchiorri}}, \bibinfo {author} {\bibfnamefont {O.}~\bibnamefont {Mena}}, \
  and\ \bibinfo {author} {\bibfnamefont {S.}~\bibnamefont {Vagnozzi}},\ }\href
  {\doibase 10.1016/j.dark.2020.100666} {\bibfield  {journal} {\bibinfo
  {journal} {Phys. Dark Univ.}\ }\textbf {\bibinfo {volume} {30}},\ \bibinfo
  {pages} {100666} (\bibinfo {year} {2020}{\natexlab{c}})},\ \Eprint
  {http://arxiv.org/abs/1908.04281} {arXiv:1908.04281 [astro-ph.CO]}
  \BibitemShut {NoStop}%
\bibitem [{\citenamefont {Di~Valentino}\ \emph
  {et~al.}(2020{\natexlab{d}})\citenamefont {Di~Valentino}, \citenamefont
  {Melchiorri}, \citenamefont {Mena},\ and\ \citenamefont
  {Vagnozzi}}]{DiValentino:2019jae}%
  \BibitemOpen
  \bibfield  {author} {\bibinfo {author} {\bibfnamefont {E.}~\bibnamefont
  {Di~Valentino}}, \bibinfo {author} {\bibfnamefont {A.}~\bibnamefont
  {Melchiorri}}, \bibinfo {author} {\bibfnamefont {O.}~\bibnamefont {Mena}}, \
  and\ \bibinfo {author} {\bibfnamefont {S.}~\bibnamefont {Vagnozzi}},\ }\href
  {\doibase 10.1103/PhysRevD.101.063502} {\bibfield  {journal} {\bibinfo
  {journal} {Phys. Rev. D}\ }\textbf {\bibinfo {volume} {101}},\ \bibinfo
  {pages} {063502} (\bibinfo {year} {2020}{\natexlab{d}})},\ \Eprint
  {http://arxiv.org/abs/1910.09853} {arXiv:1910.09853 [astro-ph.CO]}
  \BibitemShut {NoStop}%
\bibitem [{\citenamefont {Vagnozzi}\ \emph
  {et~al.}(2020{\natexlab{a}})\citenamefont {Vagnozzi}, \citenamefont
  {Visinelli}, \citenamefont {Mena},\ and\ \citenamefont
  {Mota}}]{Vagnozzi:2019kvw}%
  \BibitemOpen
  \bibfield  {author} {\bibinfo {author} {\bibfnamefont {S.}~\bibnamefont
  {Vagnozzi}}, \bibinfo {author} {\bibfnamefont {L.}~\bibnamefont {Visinelli}},
  \bibinfo {author} {\bibfnamefont {O.}~\bibnamefont {Mena}}, \ and\ \bibinfo
  {author} {\bibfnamefont {D.~F.}\ \bibnamefont {Mota}},\ }\href {\doibase
  10.1093/mnras/staa311} {\bibfield  {journal} {\bibinfo  {journal} {Mon. Not.
  Roy. Astron. Soc.}\ }\textbf {\bibinfo {volume} {493}},\ \bibinfo {pages}
  {1139} (\bibinfo {year} {2020}{\natexlab{a}})},\ \Eprint
  {http://arxiv.org/abs/1911.12374} {arXiv:1911.12374 [gr-qc]} \BibitemShut
  {NoStop}%
\bibitem [{\citenamefont {Chamings}\ \emph {et~al.}(2020)\citenamefont
  {Chamings}, \citenamefont {Avgoustidis}, \citenamefont {Copeland},
  \citenamefont {Green},\ and\ \citenamefont {Pourtsidou}}]{Chamings:2019kcl}%
  \BibitemOpen
  \bibfield  {author} {\bibinfo {author} {\bibfnamefont {F.~N.}\ \bibnamefont
  {Chamings}}, \bibinfo {author} {\bibfnamefont {A.}~\bibnamefont
  {Avgoustidis}}, \bibinfo {author} {\bibfnamefont {E.~J.}\ \bibnamefont
  {Copeland}}, \bibinfo {author} {\bibfnamefont {A.~M.}\ \bibnamefont {Green}},
  \ and\ \bibinfo {author} {\bibfnamefont {A.}~\bibnamefont {Pourtsidou}},\
  }\href {\doibase 10.1103/PhysRevD.101.043531} {\bibfield  {journal} {\bibinfo
   {journal} {Phys. Rev. D}\ }\textbf {\bibinfo {volume} {101}},\ \bibinfo
  {pages} {043531} (\bibinfo {year} {2020})},\ \Eprint
  {http://arxiv.org/abs/1912.09858} {arXiv:1912.09858 [astro-ph.CO]}
  \BibitemShut {NoStop}%
\bibitem [{\citenamefont {Jim\'enez}\ \emph {et~al.}(2020)\citenamefont
  {Jim\'enez}, \citenamefont {Bettoni}, \citenamefont {Figueruelo},\ and\
  \citenamefont {Teppa~Pannia}}]{Jimenez:2020ysu}%
  \BibitemOpen
  \bibfield  {author} {\bibinfo {author} {\bibfnamefont {J.~B.}\ \bibnamefont
  {Jim\'enez}}, \bibinfo {author} {\bibfnamefont {D.}~\bibnamefont {Bettoni}},
  \bibinfo {author} {\bibfnamefont {D.}~\bibnamefont {Figueruelo}}, \ and\
  \bibinfo {author} {\bibfnamefont {F.~A.}\ \bibnamefont {Teppa~Pannia}},\
  }\href {\doibase 10.1088/1475-7516/2020/08/020} {\bibfield  {journal}
  {\bibinfo  {journal} {JCAP}\ }\textbf {\bibinfo {volume} {08}},\ \bibinfo
  {pages} {020} (\bibinfo {year} {2020})},\ \Eprint
  {http://arxiv.org/abs/2004.14661} {arXiv:2004.14661 [astro-ph.CO]}
  \BibitemShut {NoStop}%
\bibitem [{\citenamefont {Heimersheim}\ \emph {et~al.}(2020)\citenamefont
  {Heimersheim}, \citenamefont {Sch\"oneberg}, \citenamefont {Hooper},\ and\
  \citenamefont {Lesgourgues}}]{Heimersheim:2020aoc}%
  \BibitemOpen
  \bibfield  {author} {\bibinfo {author} {\bibfnamefont {S.}~\bibnamefont
  {Heimersheim}}, \bibinfo {author} {\bibfnamefont {N.}~\bibnamefont
  {Sch\"oneberg}}, \bibinfo {author} {\bibfnamefont {D.~C.}\ \bibnamefont
  {Hooper}}, \ and\ \bibinfo {author} {\bibfnamefont {J.}~\bibnamefont
  {Lesgourgues}},\ }\href {\doibase 10.1088/1475-7516/2020/12/016} {\bibfield
  {journal} {\bibinfo  {journal} {JCAP}\ }\textbf {\bibinfo {volume} {12}},\
  \bibinfo {pages} {016} (\bibinfo {year} {2020})},\ \Eprint
  {http://arxiv.org/abs/2008.08486} {arXiv:2008.08486 [astro-ph.CO]}
  \BibitemShut {NoStop}%
\bibitem [{\citenamefont {Choi}\ \emph {et~al.}(2021)\citenamefont {Choi},
  \citenamefont {Yanagida},\ and\ \citenamefont {Yokozaki}}]{Choi:2020pyy}%
  \BibitemOpen
  \bibfield  {author} {\bibinfo {author} {\bibfnamefont {G.}~\bibnamefont
  {Choi}}, \bibinfo {author} {\bibfnamefont {T.~T.}\ \bibnamefont {Yanagida}},
  \ and\ \bibinfo {author} {\bibfnamefont {N.}~\bibnamefont {Yokozaki}},\
  }\href {\doibase 10.1007/JHEP01(2021)127} {\bibfield  {journal} {\bibinfo
  {journal} {JHEP}\ }\textbf {\bibinfo {volume} {01}},\ \bibinfo {pages} {127}
  (\bibinfo {year} {2021})},\ \Eprint {http://arxiv.org/abs/2010.06892}
  {arXiv:2010.06892 [hep-ph]} \BibitemShut {NoStop}%
\bibitem [{\citenamefont {Camera}\ \emph {et~al.}(2019)\citenamefont {Camera},
  \citenamefont {Martinelli},\ and\ \citenamefont {Bertacca}}]{Camera:2017tws}%
  \BibitemOpen
  \bibfield  {author} {\bibinfo {author} {\bibfnamefont {S.}~\bibnamefont
  {Camera}}, \bibinfo {author} {\bibfnamefont {M.}~\bibnamefont {Martinelli}},
  \ and\ \bibinfo {author} {\bibfnamefont {D.}~\bibnamefont {Bertacca}},\
  }\href {\doibase 10.1016/j.dark.2018.11.008} {\bibfield  {journal} {\bibinfo
  {journal} {Phys. Dark Univ.}\ }\textbf {\bibinfo {volume} {23}},\ \bibinfo
  {pages} {100247} (\bibinfo {year} {2019})},\ \Eprint
  {http://arxiv.org/abs/1704.06277} {arXiv:1704.06277 [astro-ph.CO]}
  \BibitemShut {NoStop}%
\bibitem [{\citenamefont {De~Felice}\ and\ \citenamefont
  {Mukohyama}(2017)}]{De_Felice_2017}%
  \BibitemOpen
  \bibfield  {author} {\bibinfo {author} {\bibfnamefont {A.}~\bibnamefont
  {De~Felice}}\ and\ \bibinfo {author} {\bibfnamefont {S.}~\bibnamefont
  {Mukohyama}},\ }\href {\doibase 10.1103/physrevlett.118.091104} {\bibfield
  {journal} {\bibinfo  {journal} {Physical Review Letters}\ }\textbf {\bibinfo
  {volume} {118}} (\bibinfo {year} {2017}),\
  10.1103/physrevlett.118.091104}\BibitemShut {NoStop}%
\bibitem [{\citenamefont {Dossett}\ \emph {et~al.}(2015)\citenamefont
  {Dossett}, \citenamefont {Ishak}, \citenamefont {Parkinson},\ and\
  \citenamefont {Davis}}]{Dossett:2015nda}%
  \BibitemOpen
  \bibfield  {author} {\bibinfo {author} {\bibfnamefont {J.~N.}\ \bibnamefont
  {Dossett}}, \bibinfo {author} {\bibfnamefont {M.}~\bibnamefont {Ishak}},
  \bibinfo {author} {\bibfnamefont {D.}~\bibnamefont {Parkinson}}, \ and\
  \bibinfo {author} {\bibfnamefont {T.}~\bibnamefont {Davis}},\ }\href
  {\doibase 10.1103/PhysRevD.92.023003} {\bibfield  {journal} {\bibinfo
  {journal} {Phys. Rev. D}\ }\textbf {\bibinfo {volume} {92}},\ \bibinfo
  {pages} {023003} (\bibinfo {year} {2015})},\ \Eprint
  {http://arxiv.org/abs/1501.03119} {arXiv:1501.03119 [astro-ph.CO]}
  \BibitemShut {NoStop}%
\bibitem [{\citenamefont {Nesseris}\ \emph {et~al.}(2017)\citenamefont
  {Nesseris}, \citenamefont {Pantazis},\ and\ \citenamefont
  {Perivolaropoulos}}]{Nesseris:2017vor}%
  \BibitemOpen
  \bibfield  {author} {\bibinfo {author} {\bibfnamefont {S.}~\bibnamefont
  {Nesseris}}, \bibinfo {author} {\bibfnamefont {G.}~\bibnamefont {Pantazis}},
  \ and\ \bibinfo {author} {\bibfnamefont {L.}~\bibnamefont
  {Perivolaropoulos}},\ }\href {\doibase 10.1103/PhysRevD.96.023542} {\bibfield
   {journal} {\bibinfo  {journal} {Phys. Rev. D}\ }\textbf {\bibinfo {volume}
  {96}},\ \bibinfo {pages} {023542} (\bibinfo {year} {2017})},\ \Eprint
  {http://arxiv.org/abs/1703.10538} {arXiv:1703.10538 [astro-ph.CO]}
  \BibitemShut {NoStop}%
\bibitem [{\citenamefont {Kazantzidis}\ and\ \citenamefont
  {Perivolaropoulos}(2018)}]{Kazantzidis:2018rnb}%
  \BibitemOpen
  \bibfield  {author} {\bibinfo {author} {\bibfnamefont {L.}~\bibnamefont
  {Kazantzidis}}\ and\ \bibinfo {author} {\bibfnamefont {L.}~\bibnamefont
  {Perivolaropoulos}},\ }\href {\doibase 10.1103/PhysRevD.97.103503} {\bibfield
   {journal} {\bibinfo  {journal} {Phys. Rev. D}\ }\textbf {\bibinfo {volume}
  {97}},\ \bibinfo {pages} {103503} (\bibinfo {year} {2018})},\ \Eprint
  {http://arxiv.org/abs/1803.01337} {arXiv:1803.01337 [astro-ph.CO]}
  \BibitemShut {NoStop}%
\bibitem [{\citenamefont {Kazantzidis}\ and\ \citenamefont
  {Perivolaropoulos}(2019)}]{Kazantzidis:2019dvk}%
  \BibitemOpen
  \bibfield  {author} {\bibinfo {author} {\bibfnamefont {L.}~\bibnamefont
  {Kazantzidis}}\ and\ \bibinfo {author} {\bibfnamefont {L.}~\bibnamefont
  {Perivolaropoulos}},\ }\href@noop {} {\  (\bibinfo {year} {2019})},\ \Eprint
  {http://arxiv.org/abs/1907.03176} {arXiv:1907.03176 [astro-ph.CO]}
  \BibitemShut {NoStop}%
\bibitem [{\citenamefont {Skara}\ and\ \citenamefont
  {Perivolaropoulos}(2020)}]{Skara:2019usd}%
  \BibitemOpen
  \bibfield  {author} {\bibinfo {author} {\bibfnamefont {F.}~\bibnamefont
  {Skara}}\ and\ \bibinfo {author} {\bibfnamefont {L.}~\bibnamefont
  {Perivolaropoulos}},\ }\href {\doibase 10.1103/PhysRevD.101.063521}
  {\bibfield  {journal} {\bibinfo  {journal} {Phys. Rev. D}\ }\textbf {\bibinfo
  {volume} {101}},\ \bibinfo {pages} {063521} (\bibinfo {year} {2020})},\
  \Eprint {http://arxiv.org/abs/1911.10609} {arXiv:1911.10609 [astro-ph.CO]}
  \BibitemShut {NoStop}%
\bibitem [{\citenamefont {Zumalacarregui}(2020)}]{Zumalacarregui:2020cjh}%
  \BibitemOpen
  \bibfield  {author} {\bibinfo {author} {\bibfnamefont {M.}~\bibnamefont
  {Zumalacarregui}},\ }\href {\doibase 10.1103/PhysRevD.102.023523} {\bibfield
  {journal} {\bibinfo  {journal} {Phys. Rev. D}\ }\textbf {\bibinfo {volume}
  {102}},\ \bibinfo {pages} {023523} (\bibinfo {year} {2020})},\ \Eprint
  {http://arxiv.org/abs/2003.06396} {arXiv:2003.06396 [astro-ph.CO]}
  \BibitemShut {NoStop}%
\bibitem [{\citenamefont {Barros}\ \emph {et~al.}(2020)\citenamefont {Barros},
  \citenamefont {Barreiro}, \citenamefont {Koivisto},\ and\ \citenamefont
  {Nunes}}]{Barros:2020bgg}%
  \BibitemOpen
  \bibfield  {author} {\bibinfo {author} {\bibfnamefont {B.~J.}\ \bibnamefont
  {Barros}}, \bibinfo {author} {\bibfnamefont {T.}~\bibnamefont {Barreiro}},
  \bibinfo {author} {\bibfnamefont {T.}~\bibnamefont {Koivisto}}, \ and\
  \bibinfo {author} {\bibfnamefont {N.~J.}\ \bibnamefont {Nunes}},\ }\href
  {\doibase 10.1016/j.dark.2020.100616} {\bibfield  {journal} {\bibinfo
  {journal} {Phys. Dark Univ.}\ }\textbf {\bibinfo {volume} {30}},\ \bibinfo
  {pages} {100616} (\bibinfo {year} {2020})},\ \Eprint
  {http://arxiv.org/abs/2004.07867} {arXiv:2004.07867 [gr-qc]} \BibitemShut
  {NoStop}%
\bibitem [{\citenamefont {Marra}\ and\ \citenamefont
  {Perivolaropoulos}(2021)}]{Marra:2021fvf}%
  \BibitemOpen
  \bibfield  {author} {\bibinfo {author} {\bibfnamefont {V.}~\bibnamefont
  {Marra}}\ and\ \bibinfo {author} {\bibfnamefont {L.}~\bibnamefont
  {Perivolaropoulos}},\ }\href@noop {} {\  (\bibinfo {year} {2021})},\ \Eprint
  {http://arxiv.org/abs/2102.06012} {arXiv:2102.06012 [astro-ph.CO]}
  \BibitemShut {NoStop}%
\bibitem [{\citenamefont {De~Felice}\ \emph {et~al.}(2020)\citenamefont
  {De~Felice}, \citenamefont {Nakamura},\ and\ \citenamefont
  {Tsujikawa}}]{De_Felice_2020}%
  \BibitemOpen
  \bibfield  {author} {\bibinfo {author} {\bibfnamefont {A.}~\bibnamefont
  {De~Felice}}, \bibinfo {author} {\bibfnamefont {S.}~\bibnamefont {Nakamura}},
  \ and\ \bibinfo {author} {\bibfnamefont {S.}~\bibnamefont {Tsujikawa}},\
  }\href {\doibase 10.1103/physrevd.102.063531} {\bibfield  {journal} {\bibinfo
   {journal} {Physical Review D}\ }\textbf {\bibinfo {volume} {102}} (\bibinfo
  {year} {2020}),\ 10.1103/physrevd.102.063531}\BibitemShut {NoStop}%
\bibitem [{\citenamefont {Di~Valentino}\ and\ \citenamefont
  {Bridle}(2018)}]{DiValentino:2018gcu}%
  \BibitemOpen
  \bibfield  {author} {\bibinfo {author} {\bibfnamefont {E.}~\bibnamefont
  {Di~Valentino}}\ and\ \bibinfo {author} {\bibfnamefont {S.}~\bibnamefont
  {Bridle}},\ }\href {\doibase 10.3390/sym10110585} {\bibfield  {journal}
  {\bibinfo  {journal} {Symmetry}\ }\textbf {\bibinfo {volume} {10}},\ \bibinfo
  {pages} {585} (\bibinfo {year} {2018})}\BibitemShut {NoStop}%
\bibitem [{\citenamefont {Di~Valentino}\ \emph
  {et~al.}(2020{\natexlab{e}})\citenamefont {Di~Valentino}, \citenamefont
  {Melchiorri},\ and\ \citenamefont {Silk}}]{DiValentino:2019dzu}%
  \BibitemOpen
  \bibfield  {author} {\bibinfo {author} {\bibfnamefont {E.}~\bibnamefont
  {Di~Valentino}}, \bibinfo {author} {\bibfnamefont {A.}~\bibnamefont
  {Melchiorri}}, \ and\ \bibinfo {author} {\bibfnamefont {J.}~\bibnamefont
  {Silk}},\ }\href {\doibase 10.1088/1475-7516/2020/01/013} {\bibfield
  {journal} {\bibinfo  {journal} {JCAP}\ }\textbf {\bibinfo {volume} {01}},\
  \bibinfo {pages} {013} (\bibinfo {year} {2020}{\natexlab{e}})},\ \Eprint
  {http://arxiv.org/abs/1908.01391} {arXiv:1908.01391 [astro-ph.CO]}
  \BibitemShut {NoStop}%
\bibitem [{\citenamefont {Vagnozzi}\ \emph {et~al.}(2018)\citenamefont
  {Vagnozzi}, \citenamefont {Dhawan}, \citenamefont {Gerbino}, \citenamefont
  {Freese}, \citenamefont {Goobar},\ and\ \citenamefont
  {Mena}}]{Vagnozzi:2018jhn}%
  \BibitemOpen
  \bibfield  {author} {\bibinfo {author} {\bibfnamefont {S.}~\bibnamefont
  {Vagnozzi}}, \bibinfo {author} {\bibfnamefont {S.}~\bibnamefont {Dhawan}},
  \bibinfo {author} {\bibfnamefont {M.}~\bibnamefont {Gerbino}}, \bibinfo
  {author} {\bibfnamefont {K.}~\bibnamefont {Freese}}, \bibinfo {author}
  {\bibfnamefont {A.}~\bibnamefont {Goobar}}, \ and\ \bibinfo {author}
  {\bibfnamefont {O.}~\bibnamefont {Mena}},\ }\href {\doibase
  10.1103/PhysRevD.98.083501} {\bibfield  {journal} {\bibinfo  {journal} {Phys.
  Rev. D}\ }\textbf {\bibinfo {volume} {98}},\ \bibinfo {pages} {083501}
  (\bibinfo {year} {2018})},\ \Eprint {http://arxiv.org/abs/1801.08553}
  {arXiv:1801.08553 [astro-ph.CO]} \BibitemShut {NoStop}%
\bibitem [{\citenamefont {Alestas}\ and\ \citenamefont
  {Perivolaropoulos}(2021)}]{Alestas:2021xes}%
  \BibitemOpen
  \bibfield  {author} {\bibinfo {author} {\bibfnamefont {G.}~\bibnamefont
  {Alestas}}\ and\ \bibinfo {author} {\bibfnamefont {L.}~\bibnamefont
  {Perivolaropoulos}},\ }\href@noop {} {\  (\bibinfo {year} {2021})},\ \Eprint
  {http://arxiv.org/abs/2103.04045} {arXiv:2103.04045 [astro-ph.CO]}
  \BibitemShut {NoStop}%
\bibitem [{\citenamefont {Di~Valentino}\ \emph
  {et~al.}(2020{\natexlab{f}})\citenamefont {Di~Valentino}, \citenamefont
  {Linder},\ and\ \citenamefont {Melchiorri}}]{DiValentino:2020kha}%
  \BibitemOpen
  \bibfield  {author} {\bibinfo {author} {\bibfnamefont {E.}~\bibnamefont
  {Di~Valentino}}, \bibinfo {author} {\bibfnamefont {E.~V.}\ \bibnamefont
  {Linder}}, \ and\ \bibinfo {author} {\bibfnamefont {A.}~\bibnamefont
  {Melchiorri}},\ }\href {\doibase 10.1016/j.dark.2020.100733} {\bibfield
  {journal} {\bibinfo  {journal} {Phys. Dark Univ.}\ }\textbf {\bibinfo
  {volume} {30}},\ \bibinfo {pages} {100733} (\bibinfo {year}
  {2020}{\natexlab{f}})},\ \Eprint {http://arxiv.org/abs/2006.16291}
  {arXiv:2006.16291 [astro-ph.CO]} \BibitemShut {NoStop}%
\bibitem [{\citenamefont {Efstathiou}\ and\ \citenamefont
  {Lemos}(2018)}]{Efstathiou:2017rgv}%
  \BibitemOpen
  \bibfield  {author} {\bibinfo {author} {\bibfnamefont {G.}~\bibnamefont
  {Efstathiou}}\ and\ \bibinfo {author} {\bibfnamefont {P.}~\bibnamefont
  {Lemos}},\ }\href {\doibase 10.1093/mnras/sty099} {\bibfield  {journal}
  {\bibinfo  {journal} {Mon. Not. Roy. Astron. Soc.}\ }\textbf {\bibinfo
  {volume} {476}},\ \bibinfo {pages} {151} (\bibinfo {year} {2018})},\ \Eprint
  {http://arxiv.org/abs/1707.00483} {arXiv:1707.00483 [astro-ph.CO]}
  \BibitemShut {NoStop}%
\bibitem [{\citenamefont {Lemos}\ \emph {et~al.}(2019)\citenamefont {Lemos},
  \citenamefont {Lee}, \citenamefont {Efstathiou},\ and\ \citenamefont
  {Gratton}}]{Lemos:2018smw}%
  \BibitemOpen
  \bibfield  {author} {\bibinfo {author} {\bibfnamefont {P.}~\bibnamefont
  {Lemos}}, \bibinfo {author} {\bibfnamefont {E.}~\bibnamefont {Lee}}, \bibinfo
  {author} {\bibfnamefont {G.}~\bibnamefont {Efstathiou}}, \ and\ \bibinfo
  {author} {\bibfnamefont {S.}~\bibnamefont {Gratton}},\ }\href {\doibase
  10.1093/mnras/sty3082} {\bibfield  {journal} {\bibinfo  {journal} {Mon. Not.
  Roy. Astron. Soc.}\ }\textbf {\bibinfo {volume} {483}},\ \bibinfo {pages}
  {4803} (\bibinfo {year} {2019})},\ \Eprint {http://arxiv.org/abs/1806.06781}
  {arXiv:1806.06781 [astro-ph.CO]} \BibitemShut {NoStop}%
\bibitem [{\citenamefont {Aylor}\ \emph {et~al.}(2019)\citenamefont {Aylor},
  \citenamefont {Joy}, \citenamefont {Knox}, \citenamefont {Millea},
  \citenamefont {Raghunathan},\ and\ \citenamefont {Wu}}]{Aylor:2018drw}%
  \BibitemOpen
  \bibfield  {author} {\bibinfo {author} {\bibfnamefont {K.}~\bibnamefont
  {Aylor}}, \bibinfo {author} {\bibfnamefont {M.}~\bibnamefont {Joy}}, \bibinfo
  {author} {\bibfnamefont {L.}~\bibnamefont {Knox}}, \bibinfo {author}
  {\bibfnamefont {M.}~\bibnamefont {Millea}}, \bibinfo {author} {\bibfnamefont
  {S.}~\bibnamefont {Raghunathan}}, \ and\ \bibinfo {author} {\bibfnamefont
  {W.~L.~K.}\ \bibnamefont {Wu}},\ }\href {\doibase 10.3847/1538-4357/ab0898}
  {\bibfield  {journal} {\bibinfo  {journal} {Astrophys. J.}\ }\textbf
  {\bibinfo {volume} {874}},\ \bibinfo {pages} {4} (\bibinfo {year} {2019})},\
  \Eprint {http://arxiv.org/abs/1811.00537} {arXiv:1811.00537 [astro-ph.CO]}
  \BibitemShut {NoStop}%
\bibitem [{\citenamefont {Sch\"oneberg}\ \emph {et~al.}(2019)\citenamefont
  {Sch\"oneberg}, \citenamefont {Lesgourgues},\ and\ \citenamefont
  {Hooper}}]{Schoneberg:2019wmt}%
  \BibitemOpen
  \bibfield  {author} {\bibinfo {author} {\bibfnamefont {N.}~\bibnamefont
  {Sch\"oneberg}}, \bibinfo {author} {\bibfnamefont {J.}~\bibnamefont
  {Lesgourgues}}, \ and\ \bibinfo {author} {\bibfnamefont {D.~C.}\ \bibnamefont
  {Hooper}},\ }\href {\doibase 10.1088/1475-7516/2019/10/029} {\bibfield
  {journal} {\bibinfo  {journal} {JCAP}\ }\textbf {\bibinfo {volume} {10}},\
  \bibinfo {pages} {029} (\bibinfo {year} {2019})},\ \Eprint
  {http://arxiv.org/abs/1907.11594} {arXiv:1907.11594 [astro-ph.CO]}
  \BibitemShut {NoStop}%
\bibitem [{\citenamefont {Knox}\ and\ \citenamefont
  {Millea}(2020)}]{Knox:2019rjx}%
  \BibitemOpen
  \bibfield  {author} {\bibinfo {author} {\bibfnamefont {L.}~\bibnamefont
  {Knox}}\ and\ \bibinfo {author} {\bibfnamefont {M.}~\bibnamefont {Millea}},\
  }\href {\doibase 10.1103/PhysRevD.101.043533} {\bibfield  {journal} {\bibinfo
   {journal} {Phys. Rev. D}\ }\textbf {\bibinfo {volume} {101}},\ \bibinfo
  {pages} {043533} (\bibinfo {year} {2020})},\ \Eprint
  {http://arxiv.org/abs/1908.03663} {arXiv:1908.03663 [astro-ph.CO]}
  \BibitemShut {NoStop}%
\bibitem [{\citenamefont {Kaiser}(1987)}]{Kaiser:1987qv}%
  \BibitemOpen
  \bibfield  {author} {\bibinfo {author} {\bibfnamefont {N.}~\bibnamefont
  {Kaiser}},\ }\href@noop {} {\bibfield  {journal} {\bibinfo  {journal} {Mon.
  Not. Roy. Astron. Soc.}\ }\textbf {\bibinfo {volume} {227}},\ \bibinfo
  {pages} {1} (\bibinfo {year} {1987})}\BibitemShut {NoStop}%
\bibitem [{\citenamefont {Zhang}\ \emph {et~al.}(2007)\citenamefont {Zhang},
  \citenamefont {Liguori}, \citenamefont {Bean},\ and\ \citenamefont
  {Dodelson}}]{Zhang:2007nk}%
  \BibitemOpen
  \bibfield  {author} {\bibinfo {author} {\bibfnamefont {P.}~\bibnamefont
  {Zhang}}, \bibinfo {author} {\bibfnamefont {M.}~\bibnamefont {Liguori}},
  \bibinfo {author} {\bibfnamefont {R.}~\bibnamefont {Bean}}, \ and\ \bibinfo
  {author} {\bibfnamefont {S.}~\bibnamefont {Dodelson}},\ }\href {\doibase
  10.1103/PhysRevLett.99.141302} {\bibfield  {journal} {\bibinfo  {journal}
  {Phys. Rev. Lett.}\ }\textbf {\bibinfo {volume} {99}},\ \bibinfo {pages}
  {141302} (\bibinfo {year} {2007})},\ \Eprint {http://arxiv.org/abs/0704.1932}
  {arXiv:0704.1932 [astro-ph]} \BibitemShut {NoStop}%
\bibitem [{\citenamefont {Quelle}\ and\ \citenamefont
  {Maroto}(2020)}]{Quelle:2019vam}%
  \BibitemOpen
  \bibfield  {author} {\bibinfo {author} {\bibfnamefont {A.}~\bibnamefont
  {Quelle}}\ and\ \bibinfo {author} {\bibfnamefont {A.~L.}\ \bibnamefont
  {Maroto}},\ }\href {\doibase 10.1140/epjc/s10052-020-7941-7} {\bibfield
  {journal} {\bibinfo  {journal} {Eur. Phys. J. C}\ }\textbf {\bibinfo {volume}
  {80}},\ \bibinfo {pages} {369} (\bibinfo {year} {2020})},\ \Eprint
  {http://arxiv.org/abs/1908.00900} {arXiv:1908.00900 [astro-ph.CO]}
  \BibitemShut {NoStop}%
\bibitem [{\citenamefont {Li}\ \emph {et~al.}(2021)\citenamefont {Li},
  \citenamefont {Du}, \citenamefont {Zhou}, \citenamefont {Zhang},\ and\
  \citenamefont {Xu}}]{Li:2019nux}%
  \BibitemOpen
  \bibfield  {author} {\bibinfo {author} {\bibfnamefont {E.-K.}\ \bibnamefont
  {Li}}, \bibinfo {author} {\bibfnamefont {M.}~\bibnamefont {Du}}, \bibinfo
  {author} {\bibfnamefont {Z.-H.}\ \bibnamefont {Zhou}}, \bibinfo {author}
  {\bibfnamefont {H.}~\bibnamefont {Zhang}}, \ and\ \bibinfo {author}
  {\bibfnamefont {L.}~\bibnamefont {Xu}},\ }\href {\doibase
  10.1093/mnras/staa3894} {\bibfield  {journal} {\bibinfo  {journal} {Mon. Not.
  Roy. Astron. Soc.}\ }\textbf {\bibinfo {volume} {501}},\ \bibinfo {pages}
  {4452} (\bibinfo {year} {2021})},\ \Eprint {http://arxiv.org/abs/1911.12076}
  {arXiv:1911.12076 [astro-ph.CO]} \BibitemShut {NoStop}%
\bibitem [{\citenamefont {Benisty}(2021)}]{Benisty:2020kdt}%
  \BibitemOpen
  \bibfield  {author} {\bibinfo {author} {\bibfnamefont {D.}~\bibnamefont
  {Benisty}},\ }\href {\doibase 10.1016/j.dark.2020.100766} {\bibfield
  {journal} {\bibinfo  {journal} {Phys. Dark Univ.}\ }\textbf {\bibinfo
  {volume} {31}},\ \bibinfo {pages} {100766} (\bibinfo {year} {2021})},\
  \Eprint {http://arxiv.org/abs/2005.03751} {arXiv:2005.03751 [astro-ph.CO]}
  \BibitemShut {NoStop}%
\bibitem [{\citenamefont {Garcia-Quintero}\ \emph {et~al.}(2020)\citenamefont
  {Garcia-Quintero}, \citenamefont {Ishak},\ and\ \citenamefont
  {Ning}}]{Garcia-Quintero:2020bac}%
  \BibitemOpen
  \bibfield  {author} {\bibinfo {author} {\bibfnamefont {C.}~\bibnamefont
  {Garcia-Quintero}}, \bibinfo {author} {\bibfnamefont {M.}~\bibnamefont
  {Ishak}}, \ and\ \bibinfo {author} {\bibfnamefont {O.}~\bibnamefont {Ning}},\
  }\href {\doibase 10.1088/1475-7516/2020/12/018} {\bibfield  {journal}
  {\bibinfo  {journal} {JCAP}\ }\textbf {\bibinfo {volume} {12}},\ \bibinfo
  {pages} {018} (\bibinfo {year} {2020})},\ \Eprint
  {http://arxiv.org/abs/2010.12519} {arXiv:2010.12519 [astro-ph.CO]}
  \BibitemShut {NoStop}%
\bibitem [{\citenamefont {Lahav}\ \emph {et~al.}(1991)\citenamefont {Lahav},
  \citenamefont {Lilje}, \citenamefont {Primack},\ and\ \citenamefont
  {Rees}}]{Lahav:1991wc}%
  \BibitemOpen
  \bibfield  {author} {\bibinfo {author} {\bibfnamefont {O.}~\bibnamefont
  {Lahav}}, \bibinfo {author} {\bibfnamefont {P.~B.}\ \bibnamefont {Lilje}},
  \bibinfo {author} {\bibfnamefont {J.~R.}\ \bibnamefont {Primack}}, \ and\
  \bibinfo {author} {\bibfnamefont {M.~J.}\ \bibnamefont {Rees}},\ }\href@noop
  {} {\bibfield  {journal} {\bibinfo  {journal} {Mon. Not. Roy. Astron. Soc.}\
  }\textbf {\bibinfo {volume} {251}},\ \bibinfo {pages} {128} (\bibinfo {year}
  {1991})}\BibitemShut {NoStop}%
\bibitem [{\citenamefont {Sagredo}\ \emph {et~al.}(2018)\citenamefont
  {Sagredo}, \citenamefont {Nesseris},\ and\ \citenamefont
  {Sapone}}]{Sagredo:2018ahx}%
  \BibitemOpen
  \bibfield  {author} {\bibinfo {author} {\bibfnamefont {B.}~\bibnamefont
  {Sagredo}}, \bibinfo {author} {\bibfnamefont {S.}~\bibnamefont {Nesseris}}, \
  and\ \bibinfo {author} {\bibfnamefont {D.}~\bibnamefont {Sapone}},\ }\href
  {\doibase 10.1103/PhysRevD.98.083543} {\bibfield  {journal} {\bibinfo
  {journal} {Phys. Rev. D}\ }\textbf {\bibinfo {volume} {98}},\ \bibinfo
  {pages} {083543} (\bibinfo {year} {2018})},\ \Eprint
  {http://arxiv.org/abs/1806.10822} {arXiv:1806.10822 [astro-ph.CO]}
  \BibitemShut {NoStop}%
\bibitem [{\citenamefont {Song}\ and\ \citenamefont
  {Percival}(2009)}]{Song:2008qt}%
  \BibitemOpen
  \bibfield  {author} {\bibinfo {author} {\bibfnamefont {Y.-S.}\ \bibnamefont
  {Song}}\ and\ \bibinfo {author} {\bibfnamefont {W.~J.}\ \bibnamefont
  {Percival}},\ }\href {\doibase 10.1088/1475-7516/2009/10/004} {\bibfield
  {journal} {\bibinfo  {journal} {JCAP}\ }\textbf {\bibinfo {volume} {10}},\
  \bibinfo {pages} {004} (\bibinfo {year} {2009})},\ \Eprint
  {http://arxiv.org/abs/0807.0810} {arXiv:0807.0810 [astro-ph]} \BibitemShut
  {NoStop}%
\bibitem [{\citenamefont {Davis}\ \emph {et~al.}(2011)\citenamefont {Davis},
  \citenamefont {Nusser}, \citenamefont {Masters}, \citenamefont {Springob},
  \citenamefont {Huchra},\ and\ \citenamefont {Lemson}}]{Davis:2010sw}%
  \BibitemOpen
  \bibfield  {author} {\bibinfo {author} {\bibfnamefont {M.}~\bibnamefont
  {Davis}}, \bibinfo {author} {\bibfnamefont {A.}~\bibnamefont {Nusser}},
  \bibinfo {author} {\bibfnamefont {K.}~\bibnamefont {Masters}}, \bibinfo
  {author} {\bibfnamefont {C.}~\bibnamefont {Springob}}, \bibinfo {author}
  {\bibfnamefont {J.~P.}\ \bibnamefont {Huchra}}, \ and\ \bibinfo {author}
  {\bibfnamefont {G.}~\bibnamefont {Lemson}},\ }\href {\doibase
  10.1111/j.1365-2966.2011.18362.x} {\bibfield  {journal} {\bibinfo  {journal}
  {Mon. Not. Roy. Astron. Soc.}\ }\textbf {\bibinfo {volume} {413}},\ \bibinfo
  {pages} {2906} (\bibinfo {year} {2011})},\ \Eprint
  {http://arxiv.org/abs/1011.3114} {arXiv:1011.3114 [astro-ph.CO]} \BibitemShut
  {NoStop}%
\bibitem [{\citenamefont {Samushia}\ \emph {et~al.}(2012)\citenamefont
  {Samushia}, \citenamefont {Percival},\ and\ \citenamefont
  {Raccanelli}}]{Samushia:2011cs}%
  \BibitemOpen
  \bibfield  {author} {\bibinfo {author} {\bibfnamefont {L.}~\bibnamefont
  {Samushia}}, \bibinfo {author} {\bibfnamefont {W.~J.}\ \bibnamefont
  {Percival}}, \ and\ \bibinfo {author} {\bibfnamefont {A.}~\bibnamefont
  {Raccanelli}},\ }\href {\doibase 10.1111/j.1365-2966.2011.20169.x} {\bibfield
   {journal} {\bibinfo  {journal} {Mon. Not. Roy. Astron. Soc.}\ }\textbf
  {\bibinfo {volume} {420}},\ \bibinfo {pages} {2102} (\bibinfo {year}
  {2012})},\ \Eprint {http://arxiv.org/abs/1102.1014} {arXiv:1102.1014
  [astro-ph.CO]} \BibitemShut {NoStop}%
\bibitem [{\citenamefont {Turnbull}\ \emph {et~al.}(2012)\citenamefont
  {Turnbull}, \citenamefont {Hudson}, \citenamefont {Feldman}, \citenamefont
  {Hicken}, \citenamefont {Kirshner},\ and\ \citenamefont
  {Watkins}}]{Turnbull:2011ty}%
  \BibitemOpen
  \bibfield  {author} {\bibinfo {author} {\bibfnamefont {S.~J.}\ \bibnamefont
  {Turnbull}}, \bibinfo {author} {\bibfnamefont {M.~J.}\ \bibnamefont
  {Hudson}}, \bibinfo {author} {\bibfnamefont {H.~A.}\ \bibnamefont {Feldman}},
  \bibinfo {author} {\bibfnamefont {M.}~\bibnamefont {Hicken}}, \bibinfo
  {author} {\bibfnamefont {R.~P.}\ \bibnamefont {Kirshner}}, \ and\ \bibinfo
  {author} {\bibfnamefont {R.}~\bibnamefont {Watkins}},\ }\href {\doibase
  10.1111/j.1365-2966.2011.20050.x} {\bibfield  {journal} {\bibinfo  {journal}
  {Mon. Not. Roy. Astron. Soc.}\ }\textbf {\bibinfo {volume} {420}},\ \bibinfo
  {pages} {447} (\bibinfo {year} {2012})},\ \Eprint
  {http://arxiv.org/abs/1111.0631} {arXiv:1111.0631 [astro-ph.CO]} \BibitemShut
  {NoStop}%
\bibitem [{\citenamefont {Hudson}\ and\ \citenamefont
  {Turnbull}(2013)}]{Hudson:2012gt}%
  \BibitemOpen
  \bibfield  {author} {\bibinfo {author} {\bibfnamefont {M.~J.}\ \bibnamefont
  {Hudson}}\ and\ \bibinfo {author} {\bibfnamefont {S.~J.}\ \bibnamefont
  {Turnbull}},\ }\href {\doibase 10.1088/2041-8205/751/2/L30} {\bibfield
  {journal} {\bibinfo  {journal} {Astrophys. J. Lett.}\ }\textbf {\bibinfo
  {volume} {751}},\ \bibinfo {pages} {L30} (\bibinfo {year} {2013})},\ \Eprint
  {http://arxiv.org/abs/1203.4814} {arXiv:1203.4814 [astro-ph.CO]} \BibitemShut
  {NoStop}%
\bibitem [{\citenamefont {Blake}\ \emph {et~al.}(2012)\citenamefont {Blake}
  \emph {et~al.}}]{Blake:2012pj}%
  \BibitemOpen
  \bibfield  {author} {\bibinfo {author} {\bibfnamefont {C.}~\bibnamefont
  {Blake}} \emph {et~al.},\ }\href {\doibase 10.1111/j.1365-2966.2012.21473.x}
  {\bibfield  {journal} {\bibinfo  {journal} {Mon. Not. Roy. Astron. Soc.}\
  }\textbf {\bibinfo {volume} {425}},\ \bibinfo {pages} {405} (\bibinfo {year}
  {2012})},\ \Eprint {http://arxiv.org/abs/1204.3674} {arXiv:1204.3674
  [astro-ph.CO]} \BibitemShut {NoStop}%
\bibitem [{\citenamefont {Blake}\ \emph {et~al.}(2013)\citenamefont {Blake}
  \emph {et~al.}}]{Blake:2013nif}%
  \BibitemOpen
  \bibfield  {author} {\bibinfo {author} {\bibfnamefont {C.}~\bibnamefont
  {Blake}} \emph {et~al.},\ }\href {\doibase 10.1093/mnras/stt1791} {\bibfield
  {journal} {\bibinfo  {journal} {Mon. Not. Roy. Astron. Soc.}\ }\textbf
  {\bibinfo {volume} {436}},\ \bibinfo {pages} {3089} (\bibinfo {year}
  {2013})},\ \Eprint {http://arxiv.org/abs/1309.5556} {arXiv:1309.5556
  [astro-ph.CO]} \BibitemShut {NoStop}%
\bibitem [{\citenamefont {Sanchez}\ \emph {et~al.}(2014)\citenamefont {Sanchez}
  \emph {et~al.}}]{Sanchez:2013tga}%
  \BibitemOpen
  \bibfield  {author} {\bibinfo {author} {\bibfnamefont {A.~G.}\ \bibnamefont
  {Sanchez}} \emph {et~al.},\ }\href {\doibase 10.1093/mnras/stu342} {\bibfield
   {journal} {\bibinfo  {journal} {Mon. Not. Roy. Astron. Soc.}\ }\textbf
  {\bibinfo {volume} {440}},\ \bibinfo {pages} {2692} (\bibinfo {year}
  {2014})},\ \Eprint {http://arxiv.org/abs/1312.4854} {arXiv:1312.4854
  [astro-ph.CO]} \BibitemShut {NoStop}%
\bibitem [{\citenamefont {Chuang}\ \emph {et~al.}(2016)\citenamefont {Chuang}
  \emph {et~al.}}]{Chuang:2013wga}%
  \BibitemOpen
  \bibfield  {author} {\bibinfo {author} {\bibfnamefont {C.-H.}\ \bibnamefont
  {Chuang}} \emph {et~al.},\ }\href {\doibase 10.1093/mnras/stw1535} {\bibfield
   {journal} {\bibinfo  {journal} {Mon. Not. Roy. Astron. Soc.}\ }\textbf
  {\bibinfo {volume} {461}},\ \bibinfo {pages} {3781} (\bibinfo {year}
  {2016})},\ \Eprint {http://arxiv.org/abs/1312.4889} {arXiv:1312.4889
  [astro-ph.CO]} \BibitemShut {NoStop}%
\bibitem [{\citenamefont {Howlett}\ \emph {et~al.}(2015)\citenamefont
  {Howlett}, \citenamefont {Ross}, \citenamefont {Samushia}, \citenamefont
  {Percival},\ and\ \citenamefont {Manera}}]{Howlett:2014opa}%
  \BibitemOpen
  \bibfield  {author} {\bibinfo {author} {\bibfnamefont {C.}~\bibnamefont
  {Howlett}}, \bibinfo {author} {\bibfnamefont {A.}~\bibnamefont {Ross}},
  \bibinfo {author} {\bibfnamefont {L.}~\bibnamefont {Samushia}}, \bibinfo
  {author} {\bibfnamefont {W.}~\bibnamefont {Percival}}, \ and\ \bibinfo
  {author} {\bibfnamefont {M.}~\bibnamefont {Manera}},\ }\href {\doibase
  10.1093/mnras/stu2693} {\bibfield  {journal} {\bibinfo  {journal} {Mon. Not.
  Roy. Astron. Soc.}\ }\textbf {\bibinfo {volume} {449}},\ \bibinfo {pages}
  {848} (\bibinfo {year} {2015})},\ \Eprint {http://arxiv.org/abs/1409.3238}
  {arXiv:1409.3238 [astro-ph.CO]} \BibitemShut {NoStop}%
\bibitem [{\citenamefont {Feix}\ \emph {et~al.}(2015)\citenamefont {Feix},
  \citenamefont {Nusser},\ and\ \citenamefont {Branchini}}]{Feix:2015dla}%
  \BibitemOpen
  \bibfield  {author} {\bibinfo {author} {\bibfnamefont {M.}~\bibnamefont
  {Feix}}, \bibinfo {author} {\bibfnamefont {A.}~\bibnamefont {Nusser}}, \ and\
  \bibinfo {author} {\bibfnamefont {E.}~\bibnamefont {Branchini}},\ }\href
  {\doibase 10.1103/PhysRevLett.115.011301} {\bibfield  {journal} {\bibinfo
  {journal} {Phys. Rev. Lett.}\ }\textbf {\bibinfo {volume} {115}},\ \bibinfo
  {pages} {011301} (\bibinfo {year} {2015})},\ \Eprint
  {http://arxiv.org/abs/1503.05945} {arXiv:1503.05945 [astro-ph.CO]}
  \BibitemShut {NoStop}%
\bibitem [{\citenamefont {Okumura}\ \emph {et~al.}(2016)\citenamefont {Okumura}
  \emph {et~al.}}]{Okumura:2015lvp}%
  \BibitemOpen
  \bibfield  {author} {\bibinfo {author} {\bibfnamefont {T.}~\bibnamefont
  {Okumura}} \emph {et~al.},\ }\href {\doibase 10.1093/pasj/psw029} {\bibfield
  {journal} {\bibinfo  {journal} {Publ. Astron. Soc. Jap.}\ }\textbf {\bibinfo
  {volume} {68}},\ \bibinfo {pages} {38} (\bibinfo {year} {2016})},\ \Eprint
  {http://arxiv.org/abs/1511.08083} {arXiv:1511.08083 [astro-ph.CO]}
  \BibitemShut {NoStop}%
\bibitem [{\citenamefont {Huterer}\ \emph {et~al.}(2017)\citenamefont
  {Huterer}, \citenamefont {Shafer}, \citenamefont {Scolnic},\ and\
  \citenamefont {Schmidt}}]{Huterer:2016uyq}%
  \BibitemOpen
  \bibfield  {author} {\bibinfo {author} {\bibfnamefont {D.}~\bibnamefont
  {Huterer}}, \bibinfo {author} {\bibfnamefont {D.}~\bibnamefont {Shafer}},
  \bibinfo {author} {\bibfnamefont {D.}~\bibnamefont {Scolnic}}, \ and\
  \bibinfo {author} {\bibfnamefont {F.}~\bibnamefont {Schmidt}},\ }\href
  {\doibase 10.1088/1475-7516/2017/05/015} {\bibfield  {journal} {\bibinfo
  {journal} {JCAP}\ }\textbf {\bibinfo {volume} {05}},\ \bibinfo {pages} {015}
  (\bibinfo {year} {2017})},\ \Eprint {http://arxiv.org/abs/1611.09862}
  {arXiv:1611.09862 [astro-ph.CO]} \BibitemShut {NoStop}%
\bibitem [{\citenamefont {Pezzotta}\ \emph {et~al.}(2017)\citenamefont
  {Pezzotta} \emph {et~al.}}]{Pezzotta:2016gbo}%
  \BibitemOpen
  \bibfield  {author} {\bibinfo {author} {\bibfnamefont {A.}~\bibnamefont
  {Pezzotta}} \emph {et~al.},\ }\href {\doibase 10.1051/0004-6361/201630295}
  {\bibfield  {journal} {\bibinfo  {journal} {Astron. Astrophys.}\ }\textbf
  {\bibinfo {volume} {604}},\ \bibinfo {pages} {A33} (\bibinfo {year}
  {2017})},\ \Eprint {http://arxiv.org/abs/1612.05645} {arXiv:1612.05645
  [astro-ph.CO]} \BibitemShut {NoStop}%
\bibitem [{\citenamefont {Zhao}\ \emph {et~al.}(2019)\citenamefont {Zhao} \emph
  {et~al.}}]{Zhao:2018gvb}%
  \BibitemOpen
  \bibfield  {author} {\bibinfo {author} {\bibfnamefont {G.-B.}\ \bibnamefont
  {Zhao}} \emph {et~al.},\ }\href {\doibase 10.1093/mnras/sty2845} {\bibfield
  {journal} {\bibinfo  {journal} {Mon. Not. Roy. Astron. Soc.}\ }\textbf
  {\bibinfo {volume} {482}},\ \bibinfo {pages} {3497} (\bibinfo {year}
  {2019})},\ \Eprint {http://arxiv.org/abs/1801.03043} {arXiv:1801.03043
  [astro-ph.CO]} \BibitemShut {NoStop}%
\bibitem [{\citenamefont {Reyes}\ \emph {et~al.}(2010)\citenamefont {Reyes},
  \citenamefont {Mandelbaum}, \citenamefont {Seljak}, \citenamefont {Baldauf},
  \citenamefont {Gunn}, \citenamefont {Lombriser},\ and\ \citenamefont
  {Smith}}]{Reyes:2010tr}%
  \BibitemOpen
  \bibfield  {author} {\bibinfo {author} {\bibfnamefont {R.}~\bibnamefont
  {Reyes}}, \bibinfo {author} {\bibfnamefont {R.}~\bibnamefont {Mandelbaum}},
  \bibinfo {author} {\bibfnamefont {U.}~\bibnamefont {Seljak}}, \bibinfo
  {author} {\bibfnamefont {T.}~\bibnamefont {Baldauf}}, \bibinfo {author}
  {\bibfnamefont {J.~E.}\ \bibnamefont {Gunn}}, \bibinfo {author}
  {\bibfnamefont {L.}~\bibnamefont {Lombriser}}, \ and\ \bibinfo {author}
  {\bibfnamefont {R.~E.}\ \bibnamefont {Smith}},\ }\href {\doibase
  10.1038/nature08857} {\bibfield  {journal} {\bibinfo  {journal} {Nature}\
  }\textbf {\bibinfo {volume} {464}},\ \bibinfo {pages} {256} (\bibinfo {year}
  {2010})},\ \Eprint {http://arxiv.org/abs/1003.2185} {arXiv:1003.2185
  [astro-ph.CO]} \BibitemShut {NoStop}%
\bibitem [{\citenamefont {Amendola}\ \emph
  {et~al.}(2013{\natexlab{a}})\citenamefont {Amendola}, \citenamefont {Kunz},
  \citenamefont {Motta}, \citenamefont {Saltas},\ and\ \citenamefont
  {Sawicki}}]{Amendola:2012ky}%
  \BibitemOpen
  \bibfield  {author} {\bibinfo {author} {\bibfnamefont {L.}~\bibnamefont
  {Amendola}}, \bibinfo {author} {\bibfnamefont {M.}~\bibnamefont {Kunz}},
  \bibinfo {author} {\bibfnamefont {M.}~\bibnamefont {Motta}}, \bibinfo
  {author} {\bibfnamefont {I.~D.}\ \bibnamefont {Saltas}}, \ and\ \bibinfo
  {author} {\bibfnamefont {I.}~\bibnamefont {Sawicki}},\ }\href {\doibase
  10.1103/PhysRevD.87.023501} {\bibfield  {journal} {\bibinfo  {journal} {Phys.
  Rev. D}\ }\textbf {\bibinfo {volume} {87}},\ \bibinfo {pages} {023501}
  (\bibinfo {year} {2013}{\natexlab{a}})},\ \Eprint
  {http://arxiv.org/abs/1210.0439} {arXiv:1210.0439 [astro-ph.CO]} \BibitemShut
  {NoStop}%
\bibitem [{\citenamefont {Pullen}\ \emph {et~al.}(2015)\citenamefont {Pullen},
  \citenamefont {Alam},\ and\ \citenamefont {Ho}}]{Pullen:2014fva}%
  \BibitemOpen
  \bibfield  {author} {\bibinfo {author} {\bibfnamefont {A.~R.}\ \bibnamefont
  {Pullen}}, \bibinfo {author} {\bibfnamefont {S.}~\bibnamefont {Alam}}, \ and\
  \bibinfo {author} {\bibfnamefont {S.}~\bibnamefont {Ho}},\ }\href {\doibase
  10.1093/mnras/stv554} {\bibfield  {journal} {\bibinfo  {journal} {Mon. Not.
  Roy. Astron. Soc.}\ }\textbf {\bibinfo {volume} {449}},\ \bibinfo {pages}
  {4326} (\bibinfo {year} {2015})},\ \Eprint {http://arxiv.org/abs/1412.4454}
  {arXiv:1412.4454 [astro-ph.CO]} \BibitemShut {NoStop}%
\bibitem [{\citenamefont {Blake}\ \emph {et~al.}(2016)\citenamefont {Blake}
  \emph {et~al.}}]{Blake:2015vea}%
  \BibitemOpen
  \bibfield  {author} {\bibinfo {author} {\bibfnamefont {C.}~\bibnamefont
  {Blake}} \emph {et~al.},\ }\href {\doibase 10.1093/mnras/stv2875} {\bibfield
  {journal} {\bibinfo  {journal} {Mon. Not. Roy. Astron. Soc.}\ }\textbf
  {\bibinfo {volume} {456}},\ \bibinfo {pages} {2806} (\bibinfo {year}
  {2016})},\ \Eprint {http://arxiv.org/abs/1507.03086} {arXiv:1507.03086
  [astro-ph.CO]} \BibitemShut {NoStop}%
\bibitem [{\citenamefont {Leonard}\ \emph {et~al.}(2015)\citenamefont
  {Leonard}, \citenamefont {Ferreira},\ and\ \citenamefont
  {Heymans}}]{Leonard:2015cba}%
  \BibitemOpen
  \bibfield  {author} {\bibinfo {author} {\bibfnamefont {C.~D.}\ \bibnamefont
  {Leonard}}, \bibinfo {author} {\bibfnamefont {P.~G.}\ \bibnamefont
  {Ferreira}}, \ and\ \bibinfo {author} {\bibfnamefont {C.}~\bibnamefont
  {Heymans}},\ }\href {\doibase 10.1088/1475-7516/2015/12/051} {\bibfield
  {journal} {\bibinfo  {journal} {JCAP}\ }\textbf {\bibinfo {volume} {12}},\
  \bibinfo {pages} {051} (\bibinfo {year} {2015})},\ \Eprint
  {http://arxiv.org/abs/1510.04287} {arXiv:1510.04287 [astro-ph.CO]}
  \BibitemShut {NoStop}%
\bibitem [{\citenamefont {Pullen}\ \emph {et~al.}(2016)\citenamefont {Pullen},
  \citenamefont {Alam}, \citenamefont {He},\ and\ \citenamefont
  {Ho}}]{Pullen:2015vtb}%
  \BibitemOpen
  \bibfield  {author} {\bibinfo {author} {\bibfnamefont {A.~R.}\ \bibnamefont
  {Pullen}}, \bibinfo {author} {\bibfnamefont {S.}~\bibnamefont {Alam}},
  \bibinfo {author} {\bibfnamefont {S.}~\bibnamefont {He}}, \ and\ \bibinfo
  {author} {\bibfnamefont {S.}~\bibnamefont {Ho}},\ }\href {\doibase
  10.1093/mnras/stw1249} {\bibfield  {journal} {\bibinfo  {journal} {Mon. Not.
  Roy. Astron. Soc.}\ }\textbf {\bibinfo {volume} {460}},\ \bibinfo {pages}
  {4098} (\bibinfo {year} {2016})},\ \Eprint {http://arxiv.org/abs/1511.04457}
  {arXiv:1511.04457 [astro-ph.CO]} \BibitemShut {NoStop}%
\bibitem [{\citenamefont {Alam}\ \emph
  {et~al.}(2017{\natexlab{a}})\citenamefont {Alam}, \citenamefont {Miyatake},
  \citenamefont {More}, \citenamefont {Ho},\ and\ \citenamefont
  {Mandelbaum}}]{Alam:2016qcl}%
  \BibitemOpen
  \bibfield  {author} {\bibinfo {author} {\bibfnamefont {S.}~\bibnamefont
  {Alam}}, \bibinfo {author} {\bibfnamefont {H.}~\bibnamefont {Miyatake}},
  \bibinfo {author} {\bibfnamefont {S.}~\bibnamefont {More}}, \bibinfo {author}
  {\bibfnamefont {S.}~\bibnamefont {Ho}}, \ and\ \bibinfo {author}
  {\bibfnamefont {R.}~\bibnamefont {Mandelbaum}},\ }\href {\doibase
  10.1093/mnras/stw3056} {\bibfield  {journal} {\bibinfo  {journal} {Mon. Not.
  Roy. Astron. Soc.}\ }\textbf {\bibinfo {volume} {465}},\ \bibinfo {pages}
  {4853} (\bibinfo {year} {2017}{\natexlab{a}})},\ \Eprint
  {http://arxiv.org/abs/1610.09410} {arXiv:1610.09410 [astro-ph.CO]}
  \BibitemShut {NoStop}%
\bibitem [{\citenamefont {de~la Torre}\ \emph {et~al.}(2017)\citenamefont
  {de~la Torre} \emph {et~al.}}]{delaTorre:2016rxm}%
  \BibitemOpen
  \bibfield  {author} {\bibinfo {author} {\bibfnamefont {S.}~\bibnamefont
  {de~la Torre}} \emph {et~al.},\ }\href {\doibase 10.1051/0004-6361/201630276}
  {\bibfield  {journal} {\bibinfo  {journal} {Astron. Astrophys.}\ }\textbf
  {\bibinfo {volume} {608}},\ \bibinfo {pages} {A44} (\bibinfo {year}
  {2017})},\ \Eprint {http://arxiv.org/abs/1612.05647} {arXiv:1612.05647
  [astro-ph.CO]} \BibitemShut {NoStop}%
\bibitem [{\citenamefont {Amon}\ \emph {et~al.}(2018)\citenamefont {Amon} \emph
  {et~al.}}]{Amon:2017lia}%
  \BibitemOpen
  \bibfield  {author} {\bibinfo {author} {\bibfnamefont {A.}~\bibnamefont
  {Amon}} \emph {et~al.},\ }\href {\doibase 10.1093/mnras/sty1624} {\bibfield
  {journal} {\bibinfo  {journal} {Mon. Not. Roy. Astron. Soc.}\ }\textbf
  {\bibinfo {volume} {479}},\ \bibinfo {pages} {3422} (\bibinfo {year}
  {2018})},\ \Eprint {http://arxiv.org/abs/1711.10999} {arXiv:1711.10999
  [astro-ph.CO]} \BibitemShut {NoStop}%
\bibitem [{\citenamefont {Singh}\ \emph {et~al.}(2019)\citenamefont {Singh},
  \citenamefont {Alam}, \citenamefont {Mandelbaum}, \citenamefont {Seljak},
  \citenamefont {Rodriguez-Torres},\ and\ \citenamefont {Ho}}]{Singh:2018flu}%
  \BibitemOpen
  \bibfield  {author} {\bibinfo {author} {\bibfnamefont {S.}~\bibnamefont
  {Singh}}, \bibinfo {author} {\bibfnamefont {S.}~\bibnamefont {Alam}},
  \bibinfo {author} {\bibfnamefont {R.}~\bibnamefont {Mandelbaum}}, \bibinfo
  {author} {\bibfnamefont {U.}~\bibnamefont {Seljak}}, \bibinfo {author}
  {\bibfnamefont {S.}~\bibnamefont {Rodriguez-Torres}}, \ and\ \bibinfo
  {author} {\bibfnamefont {S.}~\bibnamefont {Ho}},\ }\href {\doibase
  10.1093/mnras/sty2681} {\bibfield  {journal} {\bibinfo  {journal} {Mon. Not.
  Roy. Astron. Soc.}\ }\textbf {\bibinfo {volume} {482}},\ \bibinfo {pages}
  {785} (\bibinfo {year} {2019})},\ \Eprint {http://arxiv.org/abs/1803.08915}
  {arXiv:1803.08915 [astro-ph.CO]} \BibitemShut {NoStop}%
\bibitem [{\citenamefont {Blake}\ \emph {et~al.}(2020)\citenamefont {Blake}
  \emph {et~al.}}]{Blake:2020mzy}%
  \BibitemOpen
  \bibfield  {author} {\bibinfo {author} {\bibfnamefont {C.}~\bibnamefont
  {Blake}} \emph {et~al.},\ }\href {\doibase 10.1051/0004-6361/202038505}
  {\bibfield  {journal} {\bibinfo  {journal} {Astron. Astrophys.}\ }\textbf
  {\bibinfo {volume} {642}},\ \bibinfo {pages} {A158} (\bibinfo {year}
  {2020})},\ \Eprint {http://arxiv.org/abs/2005.14351} {arXiv:2005.14351
  [astro-ph.CO]} \BibitemShut {NoStop}%
\bibitem [{\citenamefont {Zhang}\ \emph {et~al.}(2021)\citenamefont {Zhang}
  \emph {et~al.}}]{Zhang:2020vru}%
  \BibitemOpen
  \bibfield  {author} {\bibinfo {author} {\bibfnamefont {Y.}~\bibnamefont
  {Zhang}} \emph {et~al.},\ }\href {\doibase 10.1093/mnras/staa3672} {\bibfield
   {journal} {\bibinfo  {journal} {Mon. Not. Roy. Astron. Soc.}\ }\textbf
  {\bibinfo {volume} {501}},\ \bibinfo {pages} {1013} (\bibinfo {year}
  {2021})},\ \Eprint {http://arxiv.org/abs/2007.12607} {arXiv:2007.12607
  [astro-ph.CO]} \BibitemShut {NoStop}%
\bibitem [{\citenamefont {Pinho}\ \emph {et~al.}(2018)\citenamefont {Pinho},
  \citenamefont {Casas},\ and\ \citenamefont {Amendola}}]{Pinho:2018unz}%
  \BibitemOpen
  \bibfield  {author} {\bibinfo {author} {\bibfnamefont {A.~M.}\ \bibnamefont
  {Pinho}}, \bibinfo {author} {\bibfnamefont {S.}~\bibnamefont {Casas}}, \ and\
  \bibinfo {author} {\bibfnamefont {L.}~\bibnamefont {Amendola}},\ }\href
  {\doibase 10.1088/1475-7516/2018/11/027} {\bibfield  {journal} {\bibinfo
  {journal} {JCAP}\ }\textbf {\bibinfo {volume} {11}},\ \bibinfo {pages} {027}
  (\bibinfo {year} {2018})},\ \Eprint {http://arxiv.org/abs/1805.00027}
  {arXiv:1805.00027 [astro-ph.CO]} \BibitemShut {NoStop}%
\bibitem [{\citenamefont {Beutler}\ \emph {et~al.}(2011)\citenamefont
  {Beutler}, \citenamefont {Blake}, \citenamefont {Colless}, \citenamefont
  {Jones}, \citenamefont {Staveley-Smith}, \citenamefont {Campbell},
  \citenamefont {Parker}, \citenamefont {Saunders},\ and\ \citenamefont
  {Watson}}]{Beutler:2011hx}%
  \BibitemOpen
  \bibfield  {author} {\bibinfo {author} {\bibfnamefont {F.}~\bibnamefont
  {Beutler}}, \bibinfo {author} {\bibfnamefont {C.}~\bibnamefont {Blake}},
  \bibinfo {author} {\bibfnamefont {M.}~\bibnamefont {Colless}}, \bibinfo
  {author} {\bibfnamefont {D.~H.}\ \bibnamefont {Jones}}, \bibinfo {author}
  {\bibfnamefont {L.}~\bibnamefont {Staveley-Smith}}, \bibinfo {author}
  {\bibfnamefont {L.}~\bibnamefont {Campbell}}, \bibinfo {author}
  {\bibfnamefont {Q.}~\bibnamefont {Parker}}, \bibinfo {author} {\bibfnamefont
  {W.}~\bibnamefont {Saunders}}, \ and\ \bibinfo {author} {\bibfnamefont
  {F.}~\bibnamefont {Watson}},\ }\href {\doibase
  10.1111/j.1365-2966.2011.19250.x} {\bibfield  {journal} {\bibinfo  {journal}
  {Mon. Not. Roy. Astron. Soc.}\ }\textbf {\bibinfo {volume} {416}},\ \bibinfo
  {pages} {3017} (\bibinfo {year} {2011})},\ \Eprint
  {http://arxiv.org/abs/1106.3366} {arXiv:1106.3366 [astro-ph.CO]} \BibitemShut
  {NoStop}%
\bibitem [{\citenamefont {Ross}\ \emph {et~al.}(2015)\citenamefont {Ross},
  \citenamefont {Samushia}, \citenamefont {Howlett}, \citenamefont {Percival},
  \citenamefont {Burden},\ and\ \citenamefont {Manera}}]{Ross:2014qpa}%
  \BibitemOpen
  \bibfield  {author} {\bibinfo {author} {\bibfnamefont {A.~J.}\ \bibnamefont
  {Ross}}, \bibinfo {author} {\bibfnamefont {L.}~\bibnamefont {Samushia}},
  \bibinfo {author} {\bibfnamefont {C.}~\bibnamefont {Howlett}}, \bibinfo
  {author} {\bibfnamefont {W.~J.}\ \bibnamefont {Percival}}, \bibinfo {author}
  {\bibfnamefont {A.}~\bibnamefont {Burden}}, \ and\ \bibinfo {author}
  {\bibfnamefont {M.}~\bibnamefont {Manera}},\ }\href {\doibase
  10.1093/mnras/stv154} {\bibfield  {journal} {\bibinfo  {journal} {Mon. Not.
  Roy. Astron. Soc.}\ }\textbf {\bibinfo {volume} {449}},\ \bibinfo {pages}
  {835} (\bibinfo {year} {2015})},\ \Eprint {http://arxiv.org/abs/1409.3242}
  {arXiv:1409.3242 [astro-ph.CO]} \BibitemShut {NoStop}%
\bibitem [{\citenamefont {Alam}\ \emph
  {et~al.}(2017{\natexlab{b}})\citenamefont {Alam} \emph
  {et~al.}}]{Alam:2016hwk}%
  \BibitemOpen
  \bibfield  {author} {\bibinfo {author} {\bibfnamefont {S.}~\bibnamefont
  {Alam}} \emph {et~al.} (\bibinfo {collaboration} {BOSS}),\ }\href {\doibase
  10.1093/mnras/stx721} {\bibfield  {journal} {\bibinfo  {journal} {Mon. Not.
  Roy. Astron. Soc.}\ }\textbf {\bibinfo {volume} {470}},\ \bibinfo {pages}
  {2617} (\bibinfo {year} {2017}{\natexlab{b}})},\ \Eprint
  {http://arxiv.org/abs/1607.03155} {arXiv:1607.03155 [astro-ph.CO]}
  \BibitemShut {NoStop}%
\bibitem [{\citenamefont {de~Sainte~Agathe}\ \emph {et~al.}(2019)\citenamefont
  {de~Sainte~Agathe} \emph {et~al.}}]{Agathe:2019vsu}%
  \BibitemOpen
  \bibfield  {author} {\bibinfo {author} {\bibfnamefont {V.}~\bibnamefont
  {de~Sainte~Agathe}} \emph {et~al.},\ }\href {\doibase
  10.1051/0004-6361/201935638} {\bibfield  {journal} {\bibinfo  {journal}
  {Astron. Astrophys.}\ }\textbf {\bibinfo {volume} {629}},\ \bibinfo {pages}
  {A85} (\bibinfo {year} {2019})},\ \Eprint {http://arxiv.org/abs/1904.03400}
  {arXiv:1904.03400 [astro-ph.CO]} \BibitemShut {NoStop}%
\bibitem [{\citenamefont {Blomqvist}\ \emph {et~al.}(2019)\citenamefont
  {Blomqvist} \emph {et~al.}}]{Blomqvist:2019rah}%
  \BibitemOpen
  \bibfield  {author} {\bibinfo {author} {\bibfnamefont {M.}~\bibnamefont
  {Blomqvist}} \emph {et~al.},\ }\href {\doibase 10.1051/0004-6361/201935641}
  {\bibfield  {journal} {\bibinfo  {journal} {Astron. Astrophys.}\ }\textbf
  {\bibinfo {volume} {629}},\ \bibinfo {pages} {A86} (\bibinfo {year}
  {2019})},\ \Eprint {http://arxiv.org/abs/1904.03430} {arXiv:1904.03430
  [astro-ph.CO]} \BibitemShut {NoStop}%
\bibitem [{\citenamefont {Scolnic}\ \emph {et~al.}(2018)\citenamefont {Scolnic}
  \emph {et~al.}}]{Scolnic:2017caz}%
  \BibitemOpen
  \bibfield  {author} {\bibinfo {author} {\bibfnamefont {D.~M.}\ \bibnamefont
  {Scolnic}} \emph {et~al.},\ }\href {\doibase 10.3847/1538-4357/aab9bb}
  {\bibfield  {journal} {\bibinfo  {journal} {Astrophys. J.}\ }\textbf
  {\bibinfo {volume} {859}},\ \bibinfo {pages} {101} (\bibinfo {year}
  {2018})},\ \Eprint {http://arxiv.org/abs/1710.00845} {arXiv:1710.00845
  [astro-ph.CO]} \BibitemShut {NoStop}%
\bibitem [{\citenamefont {Jimenez}\ and\ \citenamefont
  {Loeb}(2002)}]{Jimenez:2001gg}%
  \BibitemOpen
  \bibfield  {author} {\bibinfo {author} {\bibfnamefont {R.}~\bibnamefont
  {Jimenez}}\ and\ \bibinfo {author} {\bibfnamefont {A.}~\bibnamefont {Loeb}},\
  }\href {\doibase 10.1086/340549} {\bibfield  {journal} {\bibinfo  {journal}
  {Astrophys. J.}\ }\textbf {\bibinfo {volume} {573}},\ \bibinfo {pages} {37}
  (\bibinfo {year} {2002})},\ \Eprint {http://arxiv.org/abs/astro-ph/0106145}
  {arXiv:astro-ph/0106145} \BibitemShut {NoStop}%
\bibitem [{\citenamefont {Jimenez}\ \emph {et~al.}(2003)\citenamefont
  {Jimenez}, \citenamefont {Verde}, \citenamefont {Treu},\ and\ \citenamefont
  {Stern}}]{Jimenez:2003iv}%
  \BibitemOpen
  \bibfield  {author} {\bibinfo {author} {\bibfnamefont {R.}~\bibnamefont
  {Jimenez}}, \bibinfo {author} {\bibfnamefont {L.}~\bibnamefont {Verde}},
  \bibinfo {author} {\bibfnamefont {T.}~\bibnamefont {Treu}}, \ and\ \bibinfo
  {author} {\bibfnamefont {D.}~\bibnamefont {Stern}},\ }\href {\doibase
  10.1086/376595} {\bibfield  {journal} {\bibinfo  {journal} {Astrophys. J.}\
  }\textbf {\bibinfo {volume} {593}},\ \bibinfo {pages} {622} (\bibinfo {year}
  {2003})},\ \Eprint {http://arxiv.org/abs/astro-ph/0302560}
  {arXiv:astro-ph/0302560} \BibitemShut {NoStop}%
\bibitem [{\citenamefont {Simon}\ \emph {et~al.}(2005)\citenamefont {Simon},
  \citenamefont {Verde},\ and\ \citenamefont {Jimenez}}]{Simon:2004tf}%
  \BibitemOpen
  \bibfield  {author} {\bibinfo {author} {\bibfnamefont {J.}~\bibnamefont
  {Simon}}, \bibinfo {author} {\bibfnamefont {L.}~\bibnamefont {Verde}}, \ and\
  \bibinfo {author} {\bibfnamefont {R.}~\bibnamefont {Jimenez}},\ }\href
  {\doibase 10.1103/PhysRevD.71.123001} {\bibfield  {journal} {\bibinfo
  {journal} {Phys. Rev. D}\ }\textbf {\bibinfo {volume} {71}},\ \bibinfo
  {pages} {123001} (\bibinfo {year} {2005})},\ \Eprint
  {http://arxiv.org/abs/astro-ph/0412269} {arXiv:astro-ph/0412269} \BibitemShut
  {NoStop}%
\bibitem [{\citenamefont {Stern}\ \emph {et~al.}(2010)\citenamefont {Stern},
  \citenamefont {Jimenez}, \citenamefont {Verde}, \citenamefont
  {Kamionkowski},\ and\ \citenamefont {Stanford}}]{Stern:2009ep}%
  \BibitemOpen
  \bibfield  {author} {\bibinfo {author} {\bibfnamefont {D.}~\bibnamefont
  {Stern}}, \bibinfo {author} {\bibfnamefont {R.}~\bibnamefont {Jimenez}},
  \bibinfo {author} {\bibfnamefont {L.}~\bibnamefont {Verde}}, \bibinfo
  {author} {\bibfnamefont {M.}~\bibnamefont {Kamionkowski}}, \ and\ \bibinfo
  {author} {\bibfnamefont {S.}~\bibnamefont {Stanford}},\ }\href {\doibase
  10.1088/1475-7516/2010/02/008} {\bibfield  {journal} {\bibinfo  {journal}
  {JCAP}\ }\textbf {\bibinfo {volume} {02}},\ \bibinfo {pages} {008} (\bibinfo
  {year} {2010})},\ \Eprint {http://arxiv.org/abs/0907.3149} {arXiv:0907.3149
  [astro-ph.CO]} \BibitemShut {NoStop}%
\bibitem [{\citenamefont {Moresco}\ \emph {et~al.}(2012)\citenamefont
  {Moresco}, \citenamefont {Verde}, \citenamefont {Pozzetti}, \citenamefont
  {Jimenez},\ and\ \citenamefont {Cimatti}}]{Moresco:2012by}%
  \BibitemOpen
  \bibfield  {author} {\bibinfo {author} {\bibfnamefont {M.}~\bibnamefont
  {Moresco}}, \bibinfo {author} {\bibfnamefont {L.}~\bibnamefont {Verde}},
  \bibinfo {author} {\bibfnamefont {L.}~\bibnamefont {Pozzetti}}, \bibinfo
  {author} {\bibfnamefont {R.}~\bibnamefont {Jimenez}}, \ and\ \bibinfo
  {author} {\bibfnamefont {A.}~\bibnamefont {Cimatti}},\ }\href {\doibase
  10.1088/1475-7516/2012/07/053} {\bibfield  {journal} {\bibinfo  {journal}
  {JCAP}\ }\textbf {\bibinfo {volume} {07}},\ \bibinfo {pages} {053} (\bibinfo
  {year} {2012})},\ \Eprint {http://arxiv.org/abs/1201.6658} {arXiv:1201.6658
  [astro-ph.CO]} \BibitemShut {NoStop}%
\bibitem [{\citenamefont {Zhang}\ \emph {et~al.}(2014)\citenamefont {Zhang},
  \citenamefont {Zhang}, \citenamefont {Yuan}, \citenamefont {Zhang},\ and\
  \citenamefont {Sun}}]{Zhang:2012mp}%
  \BibitemOpen
  \bibfield  {author} {\bibinfo {author} {\bibfnamefont {C.}~\bibnamefont
  {Zhang}}, \bibinfo {author} {\bibfnamefont {H.}~\bibnamefont {Zhang}},
  \bibinfo {author} {\bibfnamefont {S.}~\bibnamefont {Yuan}}, \bibinfo {author}
  {\bibfnamefont {T.-J.}\ \bibnamefont {Zhang}}, \ and\ \bibinfo {author}
  {\bibfnamefont {Y.-C.}\ \bibnamefont {Sun}},\ }\href {\doibase
  10.1088/1674-4527/14/10/002} {\bibfield  {journal} {\bibinfo  {journal} {Res.
  Astron. Astrophys.}\ }\textbf {\bibinfo {volume} {14}},\ \bibinfo {pages}
  {1221} (\bibinfo {year} {2014})},\ \Eprint {http://arxiv.org/abs/1207.4541}
  {arXiv:1207.4541 [astro-ph.CO]} \BibitemShut {NoStop}%
\bibitem [{\citenamefont {Moresco}(2015)}]{Moresco:2015cya}%
  \BibitemOpen
  \bibfield  {author} {\bibinfo {author} {\bibfnamefont {M.}~\bibnamefont
  {Moresco}},\ }\href {\doibase 10.1093/mnrasl/slv037} {\bibfield  {journal}
  {\bibinfo  {journal} {Mon. Not. Roy. Astron. Soc.}\ }\textbf {\bibinfo
  {volume} {450}},\ \bibinfo {pages} {L16} (\bibinfo {year} {2015})},\ \Eprint
  {http://arxiv.org/abs/1503.01116} {arXiv:1503.01116 [astro-ph.CO]}
  \BibitemShut {NoStop}%
\bibitem [{\citenamefont {Moresco}\ \emph {et~al.}(2016)\citenamefont
  {Moresco}, \citenamefont {Pozzetti}, \citenamefont {Cimatti}, \citenamefont
  {Jimenez}, \citenamefont {Maraston}, \citenamefont {Verde}, \citenamefont
  {Thomas}, \citenamefont {Citro}, \citenamefont {Tojeiro},\ and\ \citenamefont
  {Wilkinson}}]{Moresco:2016mzx}%
  \BibitemOpen
  \bibfield  {author} {\bibinfo {author} {\bibfnamefont {M.}~\bibnamefont
  {Moresco}}, \bibinfo {author} {\bibfnamefont {L.}~\bibnamefont {Pozzetti}},
  \bibinfo {author} {\bibfnamefont {A.}~\bibnamefont {Cimatti}}, \bibinfo
  {author} {\bibfnamefont {R.}~\bibnamefont {Jimenez}}, \bibinfo {author}
  {\bibfnamefont {C.}~\bibnamefont {Maraston}}, \bibinfo {author}
  {\bibfnamefont {L.}~\bibnamefont {Verde}}, \bibinfo {author} {\bibfnamefont
  {D.}~\bibnamefont {Thomas}}, \bibinfo {author} {\bibfnamefont
  {A.}~\bibnamefont {Citro}}, \bibinfo {author} {\bibfnamefont
  {R.}~\bibnamefont {Tojeiro}}, \ and\ \bibinfo {author} {\bibfnamefont
  {D.}~\bibnamefont {Wilkinson}},\ }\href {\doibase
  10.1088/1475-7516/2016/05/014} {\bibfield  {journal} {\bibinfo  {journal}
  {JCAP}\ }\textbf {\bibinfo {volume} {05}},\ \bibinfo {pages} {014} (\bibinfo
  {year} {2016})},\ \Eprint {http://arxiv.org/abs/1601.01701} {arXiv:1601.01701
  [astro-ph.CO]} \BibitemShut {NoStop}%
\bibitem [{\citenamefont {Ratsimbazafy}\ \emph {et~al.}(2017)\citenamefont
  {Ratsimbazafy}, \citenamefont {Loubser}, \citenamefont {Crawford},
  \citenamefont {Cress}, \citenamefont {Bassett}, \citenamefont {Nichol},\ and\
  \citenamefont {V\"ais\"anen}}]{Ratsimbazafy:2017vga}%
  \BibitemOpen
  \bibfield  {author} {\bibinfo {author} {\bibfnamefont {A.}~\bibnamefont
  {Ratsimbazafy}}, \bibinfo {author} {\bibfnamefont {S.}~\bibnamefont
  {Loubser}}, \bibinfo {author} {\bibfnamefont {S.}~\bibnamefont {Crawford}},
  \bibinfo {author} {\bibfnamefont {C.}~\bibnamefont {Cress}}, \bibinfo
  {author} {\bibfnamefont {B.}~\bibnamefont {Bassett}}, \bibinfo {author}
  {\bibfnamefont {R.}~\bibnamefont {Nichol}}, \ and\ \bibinfo {author}
  {\bibfnamefont {P.}~\bibnamefont {V\"ais\"anen}},\ }\href {\doibase
  10.1093/mnras/stx301} {\bibfield  {journal} {\bibinfo  {journal} {Mon. Not.
  Roy. Astron. Soc.}\ }\textbf {\bibinfo {volume} {467}},\ \bibinfo {pages}
  {3239} (\bibinfo {year} {2017})},\ \Eprint {http://arxiv.org/abs/1702.00418}
  {arXiv:1702.00418 [astro-ph.CO]} \BibitemShut {NoStop}%
\bibitem [{\citenamefont {Mossa}\ \emph {et~al.}(2020)\citenamefont {Mossa}
  \emph {et~al.}}]{Mossa:2020gjc}%
  \BibitemOpen
  \bibfield  {author} {\bibinfo {author} {\bibfnamefont {V.}~\bibnamefont
  {Mossa}} \emph {et~al.},\ }\href {\doibase 10.1038/s41586-020-2878-4}
  {\bibfield  {journal} {\bibinfo  {journal} {Nature}\ }\textbf {\bibinfo
  {volume} {587}},\ \bibinfo {pages} {210} (\bibinfo {year}
  {2020})}\BibitemShut {NoStop}%
\bibitem [{\citenamefont {Handley}(2021)}]{Handley:2019tkm}%
  \BibitemOpen
  \bibfield  {author} {\bibinfo {author} {\bibfnamefont {W.}~\bibnamefont
  {Handley}},\ }\href {\doibase 10.1103/PhysRevD.103.L041301} {\bibfield
  {journal} {\bibinfo  {journal} {Phys. Rev. D}\ }\textbf {\bibinfo {volume}
  {103}},\ \bibinfo {pages} {L041301} (\bibinfo {year} {2021})},\ \Eprint
  {http://arxiv.org/abs/1908.09139} {arXiv:1908.09139 [astro-ph.CO]}
  \BibitemShut {NoStop}%
\bibitem [{\citenamefont {Di~Valentino}\ \emph {et~al.}(2019)\citenamefont
  {Di~Valentino}, \citenamefont {Melchiorri},\ and\ \citenamefont
  {Silk}}]{DiValentino:2019qzk}%
  \BibitemOpen
  \bibfield  {author} {\bibinfo {author} {\bibfnamefont {E.}~\bibnamefont
  {Di~Valentino}}, \bibinfo {author} {\bibfnamefont {A.}~\bibnamefont
  {Melchiorri}}, \ and\ \bibinfo {author} {\bibfnamefont {J.}~\bibnamefont
  {Silk}},\ }\href {\doibase 10.1038/s41550-019-0906-9} {\bibfield  {journal}
  {\bibinfo  {journal} {Nature Astron.}\ }\textbf {\bibinfo {volume} {4}},\
  \bibinfo {pages} {196} (\bibinfo {year} {2019})},\ \Eprint
  {http://arxiv.org/abs/1911.02087} {arXiv:1911.02087 [astro-ph.CO]}
  \BibitemShut {NoStop}%
\bibitem [{\citenamefont {Ryan}\ \emph {et~al.}(2018)\citenamefont {Ryan},
  \citenamefont {Doshi},\ and\ \citenamefont {Ratra}}]{Ryan:2018aif}%
  \BibitemOpen
  \bibfield  {author} {\bibinfo {author} {\bibfnamefont {J.}~\bibnamefont
  {Ryan}}, \bibinfo {author} {\bibfnamefont {S.}~\bibnamefont {Doshi}}, \ and\
  \bibinfo {author} {\bibfnamefont {B.}~\bibnamefont {Ratra}},\ }\href
  {\doibase 10.1093/mnras/sty1922} {\bibfield  {journal} {\bibinfo  {journal}
  {Mon. Not. Roy. Astron. Soc.}\ }\textbf {\bibinfo {volume} {480}},\ \bibinfo
  {pages} {759} (\bibinfo {year} {2018})},\ \Eprint
  {http://arxiv.org/abs/1805.06408} {arXiv:1805.06408 [astro-ph.CO]}
  \BibitemShut {NoStop}%
\bibitem [{\citenamefont {Park}\ and\ \citenamefont
  {Ratra}(2019)}]{Park:2018tgj}%
  \BibitemOpen
  \bibfield  {author} {\bibinfo {author} {\bibfnamefont {C.-G.}\ \bibnamefont
  {Park}}\ and\ \bibinfo {author} {\bibfnamefont {B.}~\bibnamefont {Ratra}},\
  }\href {\doibase 10.1007/s10509-019-3627-8} {\bibfield  {journal} {\bibinfo
  {journal} {Astrophys. Space Sci.}\ }\textbf {\bibinfo {volume} {364}},\
  \bibinfo {pages} {134} (\bibinfo {year} {2019})},\ \Eprint
  {http://arxiv.org/abs/1809.03598} {arXiv:1809.03598 [astro-ph.CO]}
  \BibitemShut {NoStop}%
\bibitem [{\citenamefont {Efstathiou}\ and\ \citenamefont
  {Gratton}(2020)}]{Efstathiou:2020wem}%
  \BibitemOpen
  \bibfield  {author} {\bibinfo {author} {\bibfnamefont {G.}~\bibnamefont
  {Efstathiou}}\ and\ \bibinfo {author} {\bibfnamefont {S.}~\bibnamefont
  {Gratton}},\ }\href {\doibase 10.1093/mnrasl/slaa093} {\bibfield  {journal}
  {\bibinfo  {journal} {Mon. Not. Roy. Astron. Soc.}\ }\textbf {\bibinfo
  {volume} {496}},\ \bibinfo {pages} {L91} (\bibinfo {year} {2020})},\ \Eprint
  {http://arxiv.org/abs/2002.06892} {arXiv:2002.06892 [astro-ph.CO]}
  \BibitemShut {NoStop}%
\bibitem [{\citenamefont {Chudaykin}\ \emph {et~al.}(2021)\citenamefont
  {Chudaykin}, \citenamefont {Dolgikh},\ and\ \citenamefont
  {Ivanov}}]{Chudaykin:2020ghx}%
  \BibitemOpen
  \bibfield  {author} {\bibinfo {author} {\bibfnamefont {A.}~\bibnamefont
  {Chudaykin}}, \bibinfo {author} {\bibfnamefont {K.}~\bibnamefont {Dolgikh}},
  \ and\ \bibinfo {author} {\bibfnamefont {M.~M.}\ \bibnamefont {Ivanov}},\
  }\href {\doibase 10.1103/PhysRevD.103.023507} {\bibfield  {journal} {\bibinfo
   {journal} {Phys. Rev. D}\ }\textbf {\bibinfo {volume} {103}},\ \bibinfo
  {pages} {023507} (\bibinfo {year} {2021})},\ \Eprint
  {http://arxiv.org/abs/2009.10106} {arXiv:2009.10106 [astro-ph.CO]}
  \BibitemShut {NoStop}%
\bibitem [{\citenamefont {Vagnozzi}\ \emph
  {et~al.}(2020{\natexlab{b}})\citenamefont {Vagnozzi}, \citenamefont
  {Di~Valentino}, \citenamefont {Gariazzo}, \citenamefont {Melchiorri},
  \citenamefont {Mena},\ and\ \citenamefont {Silk}}]{Vagnozzi:2020zrh}%
  \BibitemOpen
  \bibfield  {author} {\bibinfo {author} {\bibfnamefont {S.}~\bibnamefont
  {Vagnozzi}}, \bibinfo {author} {\bibfnamefont {E.}~\bibnamefont
  {Di~Valentino}}, \bibinfo {author} {\bibfnamefont {S.}~\bibnamefont
  {Gariazzo}}, \bibinfo {author} {\bibfnamefont {A.}~\bibnamefont
  {Melchiorri}}, \bibinfo {author} {\bibfnamefont {O.}~\bibnamefont {Mena}}, \
  and\ \bibinfo {author} {\bibfnamefont {J.}~\bibnamefont {Silk}},\ }\href@noop
  {} {\  (\bibinfo {year} {2020}{\natexlab{b}})},\ \Eprint
  {http://arxiv.org/abs/2010.02230} {arXiv:2010.02230 [astro-ph.CO]}
  \BibitemShut {NoStop}%
\bibitem [{\citenamefont {Vagnozzi}\ \emph {et~al.}(2021)\citenamefont
  {Vagnozzi}, \citenamefont {Loeb},\ and\ \citenamefont
  {Moresco}}]{Vagnozzi:2020dfn}%
  \BibitemOpen
  \bibfield  {author} {\bibinfo {author} {\bibfnamefont {S.}~\bibnamefont
  {Vagnozzi}}, \bibinfo {author} {\bibfnamefont {A.}~\bibnamefont {Loeb}}, \
  and\ \bibinfo {author} {\bibfnamefont {M.}~\bibnamefont {Moresco}},\ }\href
  {\doibase 10.3847/1538-4357/abd4df} {\bibfield  {journal} {\bibinfo
  {journal} {Astrophys. J.}\ }\textbf {\bibinfo {volume} {908}},\ \bibinfo
  {pages} {84} (\bibinfo {year} {2021})},\ \Eprint
  {http://arxiv.org/abs/2011.11645} {arXiv:2011.11645 [astro-ph.CO]}
  \BibitemShut {NoStop}%
\bibitem [{\citenamefont {Cao}\ \emph {et~al.}(2021)\citenamefont {Cao},
  \citenamefont {Ryan},\ and\ \citenamefont {Ratra}}]{Cao:2021ldv}%
  \BibitemOpen
  \bibfield  {author} {\bibinfo {author} {\bibfnamefont {S.}~\bibnamefont
  {Cao}}, \bibinfo {author} {\bibfnamefont {J.}~\bibnamefont {Ryan}}, \ and\
  \bibinfo {author} {\bibfnamefont {B.}~\bibnamefont {Ratra}},\ }\href@noop {}
  {\  (\bibinfo {year} {2021})},\ \Eprint {http://arxiv.org/abs/2101.08817}
  {arXiv:2101.08817 [astro-ph.CO]} \BibitemShut {NoStop}%
\bibitem [{\citenamefont {Di~Valentino}\ \emph
  {et~al.}(2021{\natexlab{b}})\citenamefont {Di~Valentino}, \citenamefont
  {Melchiorri},\ and\ \citenamefont {Silk}}]{DiValentino:2020hov}%
  \BibitemOpen
  \bibfield  {author} {\bibinfo {author} {\bibfnamefont {E.}~\bibnamefont
  {Di~Valentino}}, \bibinfo {author} {\bibfnamefont {A.}~\bibnamefont
  {Melchiorri}}, \ and\ \bibinfo {author} {\bibfnamefont {J.}~\bibnamefont
  {Silk}},\ }\href {\doibase 10.3847/2041-8213/abe1c4} {\bibfield  {journal}
  {\bibinfo  {journal} {Astrophys. J. Lett.}\ }\textbf {\bibinfo {volume}
  {908}},\ \bibinfo {pages} {L9} (\bibinfo {year} {2021}{\natexlab{b}})},\
  \Eprint {http://arxiv.org/abs/2003.04935} {arXiv:2003.04935 [astro-ph.CO]}
  \BibitemShut {NoStop}%
\bibitem [{\citenamefont {Blas}\ \emph {et~al.}(2011)\citenamefont {Blas},
  \citenamefont {Lesgourgues},\ and\ \citenamefont {Tram}}]{Blas:2011rf}%
  \BibitemOpen
  \bibfield  {author} {\bibinfo {author} {\bibfnamefont {D.}~\bibnamefont
  {Blas}}, \bibinfo {author} {\bibfnamefont {J.}~\bibnamefont {Lesgourgues}}, \
  and\ \bibinfo {author} {\bibfnamefont {T.}~\bibnamefont {Tram}},\ }\href
  {\doibase 10.1088/1475-7516/2011/07/034} {\bibfield  {journal} {\bibinfo
  {journal} {JCAP}\ }\textbf {\bibinfo {volume} {07}},\ \bibinfo {pages} {034}
  (\bibinfo {year} {2011})},\ \Eprint {http://arxiv.org/abs/1104.2933}
  {arXiv:1104.2933 [astro-ph.CO]} \BibitemShut {NoStop}%
\bibitem [{\citenamefont {Audren}\ \emph {et~al.}(2013)\citenamefont {Audren},
  \citenamefont {Lesgourgues}, \citenamefont {Benabed},\ and\ \citenamefont
  {Prunet}}]{Audren:2012wb}%
  \BibitemOpen
  \bibfield  {author} {\bibinfo {author} {\bibfnamefont {B.}~\bibnamefont
  {Audren}}, \bibinfo {author} {\bibfnamefont {J.}~\bibnamefont {Lesgourgues}},
  \bibinfo {author} {\bibfnamefont {K.}~\bibnamefont {Benabed}}, \ and\
  \bibinfo {author} {\bibfnamefont {S.}~\bibnamefont {Prunet}},\ }\href
  {\doibase 10.1088/1475-7516/2013/02/001} {\bibfield  {journal} {\bibinfo
  {journal} {JCAP}\ }\textbf {\bibinfo {volume} {02}},\ \bibinfo {pages} {001}
  (\bibinfo {year} {2013})},\ \Eprint {http://arxiv.org/abs/1210.7183}
  {arXiv:1210.7183 [astro-ph.CO]} \BibitemShut {NoStop}%
\bibitem [{\citenamefont {Brinckmann}\ and\ \citenamefont
  {Lesgourgues}(2019)}]{Brinckmann:2018cvx}%
  \BibitemOpen
  \bibfield  {author} {\bibinfo {author} {\bibfnamefont {T.}~\bibnamefont
  {Brinckmann}}\ and\ \bibinfo {author} {\bibfnamefont {J.}~\bibnamefont
  {Lesgourgues}},\ }\href {\doibase 10.1016/j.dark.2018.100260} {\bibfield
  {journal} {\bibinfo  {journal} {Phys. Dark Univ.}\ }\textbf {\bibinfo
  {volume} {24}},\ \bibinfo {pages} {100260} (\bibinfo {year} {2019})},\
  \Eprint {http://arxiv.org/abs/1804.07261} {arXiv:1804.07261 [astro-ph.CO]}
  \BibitemShut {NoStop}%
\bibitem [{\citenamefont {Lesgourgues}(2011)}]{Lesgourgues:2011re}%
  \BibitemOpen
  \bibfield  {author} {\bibinfo {author} {\bibfnamefont {J.}~\bibnamefont
  {Lesgourgues}},\ }\href@noop {} {\  (\bibinfo {year} {2011})},\ \Eprint
  {http://arxiv.org/abs/1104.2932} {arXiv:1104.2932 [astro-ph.IM]} \BibitemShut
  {NoStop}%
\bibitem [{\citenamefont {Gelman}\ and\ \citenamefont
  {Rubin}(1992)}]{Gelman:1992zz}%
  \BibitemOpen
  \bibfield  {author} {\bibinfo {author} {\bibfnamefont {A.}~\bibnamefont
  {Gelman}}\ and\ \bibinfo {author} {\bibfnamefont {D.~B.}\ \bibnamefont
  {Rubin}},\ }\href {\doibase 10.1214/ss/1177011136} {\bibfield  {journal}
  {\bibinfo  {journal} {Statist. Sci.}\ }\textbf {\bibinfo {volume} {7}},\
  \bibinfo {pages} {457} (\bibinfo {year} {1992})}\BibitemShut {NoStop}%
\bibitem [{\citenamefont {Hildebrandt}\ \emph {et~al.}(2017)\citenamefont
  {Hildebrandt} \emph {et~al.}}]{Hildebrandt:2016iqg}%
  \BibitemOpen
  \bibfield  {author} {\bibinfo {author} {\bibfnamefont {H.}~\bibnamefont
  {Hildebrandt}} \emph {et~al.},\ }\href {\doibase 10.1093/mnras/stw2805}
  {\bibfield  {journal} {\bibinfo  {journal} {Mon. Not. Roy. Astron. Soc.}\
  }\textbf {\bibinfo {volume} {465}},\ \bibinfo {pages} {1454} (\bibinfo {year}
  {2017})},\ \Eprint {http://arxiv.org/abs/1606.05338} {arXiv:1606.05338
  [astro-ph.CO]} \BibitemShut {NoStop}%
\bibitem [{\citenamefont {Addison}\ \emph {et~al.}(2016)\citenamefont
  {Addison}, \citenamefont {Huang}, \citenamefont {Watts}, \citenamefont
  {Bennett}, \citenamefont {Halpern}, \citenamefont {Hinshaw},\ and\
  \citenamefont {Weiland}}]{Addison:2015wyg}%
  \BibitemOpen
  \bibfield  {author} {\bibinfo {author} {\bibfnamefont {G.~E.}\ \bibnamefont
  {Addison}}, \bibinfo {author} {\bibfnamefont {Y.}~\bibnamefont {Huang}},
  \bibinfo {author} {\bibfnamefont {D.~J.}\ \bibnamefont {Watts}}, \bibinfo
  {author} {\bibfnamefont {C.~L.}\ \bibnamefont {Bennett}}, \bibinfo {author}
  {\bibfnamefont {M.}~\bibnamefont {Halpern}}, \bibinfo {author} {\bibfnamefont
  {G.}~\bibnamefont {Hinshaw}}, \ and\ \bibinfo {author} {\bibfnamefont
  {J.~L.}\ \bibnamefont {Weiland}},\ }\href {\doibase
  10.3847/0004-637X/818/2/132} {\bibfield  {journal} {\bibinfo  {journal}
  {Astrophys. J.}\ }\textbf {\bibinfo {volume} {818}},\ \bibinfo {pages} {132}
  (\bibinfo {year} {2016})},\ \Eprint {http://arxiv.org/abs/1511.00055}
  {arXiv:1511.00055 [astro-ph.CO]} \BibitemShut {NoStop}%
\bibitem [{\citenamefont {Karpenka}\ \emph {et~al.}(2015)\citenamefont
  {Karpenka}, \citenamefont {Feroz},\ and\ \citenamefont
  {Hobson}}]{Karpenka:2014moa}%
  \BibitemOpen
  \bibfield  {author} {\bibinfo {author} {\bibfnamefont {N.~V.}\ \bibnamefont
  {Karpenka}}, \bibinfo {author} {\bibfnamefont {F.}~\bibnamefont {Feroz}}, \
  and\ \bibinfo {author} {\bibfnamefont {M.~P.}\ \bibnamefont {Hobson}},\
  }\href {\doibase 10.1093/mnras/stv415} {\bibfield  {journal} {\bibinfo
  {journal} {Mon. Not. Roy. Astron. Soc.}\ }\textbf {\bibinfo {volume} {449}},\
  \bibinfo {pages} {2405} (\bibinfo {year} {2015})},\ \Eprint
  {http://arxiv.org/abs/1407.5496} {arXiv:1407.5496 [astro-ph.IM]} \BibitemShut
  {NoStop}%
\bibitem [{\citenamefont {Lin}\ and\ \citenamefont
  {Ishak}(2017{\natexlab{a}})}]{Lin:2017ikq}%
  \BibitemOpen
  \bibfield  {author} {\bibinfo {author} {\bibfnamefont {W.}~\bibnamefont
  {Lin}}\ and\ \bibinfo {author} {\bibfnamefont {M.}~\bibnamefont {Ishak}},\
  }\href {\doibase 10.1103/PhysRevD.96.023532} {\bibfield  {journal} {\bibinfo
  {journal} {Phys. Rev.}\ }\textbf {\bibinfo {volume} {D96}},\ \bibinfo {pages}
  {023532} (\bibinfo {year} {2017}{\natexlab{a}})},\ \Eprint
  {http://arxiv.org/abs/1705.05303} {arXiv:1705.05303 [astro-ph.CO]}
  \BibitemShut {NoStop}%
\bibitem [{\citenamefont {Lin}\ and\ \citenamefont
  {Ishak}(2017{\natexlab{b}})}]{Lin:2017bhs}%
  \BibitemOpen
  \bibfield  {author} {\bibinfo {author} {\bibfnamefont {W.}~\bibnamefont
  {Lin}}\ and\ \bibinfo {author} {\bibfnamefont {M.}~\bibnamefont {Ishak}},\
  }\href {\doibase 10.1103/PhysRevD.96.083532} {\bibfield  {journal} {\bibinfo
  {journal} {Phys. Rev.}\ }\textbf {\bibinfo {volume} {D96}},\ \bibinfo {pages}
  {083532} (\bibinfo {year} {2017}{\natexlab{b}})},\ \Eprint
  {http://arxiv.org/abs/1708.09813} {arXiv:1708.09813 [astro-ph.CO]}
  \BibitemShut {NoStop}%
\bibitem [{\citenamefont {Adhikari}\ and\ \citenamefont
  {Huterer}(2019)}]{Adhikari:2018wnk}%
  \BibitemOpen
  \bibfield  {author} {\bibinfo {author} {\bibfnamefont {S.}~\bibnamefont
  {Adhikari}}\ and\ \bibinfo {author} {\bibfnamefont {D.}~\bibnamefont
  {Huterer}},\ }\href {\doibase 10.1088/1475-7516/2019/01/036} {\bibfield
  {journal} {\bibinfo  {journal} {JCAP}\ }\textbf {\bibinfo {volume} {1901}},\
  \bibinfo {pages} {036} (\bibinfo {year} {2019})},\ \Eprint
  {http://arxiv.org/abs/1806.04292} {arXiv:1806.04292 [astro-ph.CO]}
  \BibitemShut {NoStop}%
\bibitem [{\citenamefont {Raveri}\ and\ \citenamefont
  {Hu}(2019)}]{Raveri:2018wln}%
  \BibitemOpen
  \bibfield  {author} {\bibinfo {author} {\bibfnamefont {M.}~\bibnamefont
  {Raveri}}\ and\ \bibinfo {author} {\bibfnamefont {W.}~\bibnamefont {Hu}},\
  }\href {\doibase 10.1103/PhysRevD.99.043506} {\bibfield  {journal} {\bibinfo
  {journal} {Phys. Rev.}\ }\textbf {\bibinfo {volume} {D99}},\ \bibinfo {pages}
  {043506} (\bibinfo {year} {2019})},\ \Eprint
  {http://arxiv.org/abs/1806.04649} {arXiv:1806.04649 [astro-ph.CO]}
  \BibitemShut {NoStop}%
\bibitem [{\citenamefont {Nicola}\ \emph {et~al.}(2019)\citenamefont {Nicola},
  \citenamefont {Amara},\ and\ \citenamefont {Refregier}}]{Nicola:2018rcd}%
  \BibitemOpen
  \bibfield  {author} {\bibinfo {author} {\bibfnamefont {A.}~\bibnamefont
  {Nicola}}, \bibinfo {author} {\bibfnamefont {A.}~\bibnamefont {Amara}}, \
  and\ \bibinfo {author} {\bibfnamefont {A.}~\bibnamefont {Refregier}},\ }\href
  {\doibase 10.1088/1475-7516/2019/01/011} {\bibfield  {journal} {\bibinfo
  {journal} {JCAP}\ }\textbf {\bibinfo {volume} {1901}},\ \bibinfo {pages}
  {011} (\bibinfo {year} {2019})},\ \Eprint {http://arxiv.org/abs/1809.07333}
  {arXiv:1809.07333 [astro-ph.CO]} \BibitemShut {NoStop}%
\bibitem [{\citenamefont {Handley}\ and\ \citenamefont
  {Lemos}(2019{\natexlab{a}})}]{Handley:2019wlz}%
  \BibitemOpen
  \bibfield  {author} {\bibinfo {author} {\bibfnamefont {W.}~\bibnamefont
  {Handley}}\ and\ \bibinfo {author} {\bibfnamefont {P.}~\bibnamefont
  {Lemos}},\ }\href {\doibase 10.1103/PhysRevD.100.043504} {\bibfield
  {journal} {\bibinfo  {journal} {Phys. Rev.}\ }\textbf {\bibinfo {volume}
  {D100}},\ \bibinfo {pages} {043504} (\bibinfo {year} {2019}{\natexlab{a}})},\
  \Eprint {http://arxiv.org/abs/1902.04029} {arXiv:1902.04029 [astro-ph.CO]}
  \BibitemShut {NoStop}%
\bibitem [{\citenamefont {Handley}\ and\ \citenamefont
  {Lemos}(2019{\natexlab{b}})}]{Handley:2019pqx}%
  \BibitemOpen
  \bibfield  {author} {\bibinfo {author} {\bibfnamefont {W.}~\bibnamefont
  {Handley}}\ and\ \bibinfo {author} {\bibfnamefont {P.}~\bibnamefont
  {Lemos}},\ }\href {\doibase 10.1103/PhysRevD.100.023512} {\bibfield
  {journal} {\bibinfo  {journal} {Phys. Rev.}\ }\textbf {\bibinfo {volume}
  {D100}},\ \bibinfo {pages} {023512} (\bibinfo {year} {2019}{\natexlab{b}})},\
  \Eprint {http://arxiv.org/abs/1903.06682} {arXiv:1903.06682 [astro-ph.CO]}
  \BibitemShut {NoStop}%
\bibitem [{\citenamefont {Garcia-Quintero}\ \emph {et~al.}(2019)\citenamefont
  {Garcia-Quintero}, \citenamefont {Ishak}, \citenamefont {Fox},\ and\
  \citenamefont {Lin}}]{Garcia-Quintero:2019cgt}%
  \BibitemOpen
  \bibfield  {author} {\bibinfo {author} {\bibfnamefont {C.}~\bibnamefont
  {Garcia-Quintero}}, \bibinfo {author} {\bibfnamefont {M.}~\bibnamefont
  {Ishak}}, \bibinfo {author} {\bibfnamefont {L.}~\bibnamefont {Fox}}, \ and\
  \bibinfo {author} {\bibfnamefont {W.}~\bibnamefont {Lin}},\ }\href {\doibase
  10.1103/PhysRevD.100.123538} {\bibfield  {journal} {\bibinfo  {journal}
  {Phys. Rev.}\ }\textbf {\bibinfo {volume} {D100}},\ \bibinfo {pages} {123538}
  (\bibinfo {year} {2019})},\ \Eprint {http://arxiv.org/abs/1910.01608}
  {arXiv:1910.01608 [astro-ph.CO]} \BibitemShut {NoStop}%
\bibitem [{\citenamefont {Lemos}\ \emph {et~al.}(2020)\citenamefont {Lemos},
  \citenamefont {K\"ohlinger}, \citenamefont {Handley}, \citenamefont
  {Joachimi}, \citenamefont {Whiteway},\ and\ \citenamefont
  {Lahav}}]{Lemos:2019txn}%
  \BibitemOpen
  \bibfield  {author} {\bibinfo {author} {\bibfnamefont {P.}~\bibnamefont
  {Lemos}}, \bibinfo {author} {\bibfnamefont {F.}~\bibnamefont {K\"ohlinger}},
  \bibinfo {author} {\bibfnamefont {W.}~\bibnamefont {Handley}}, \bibinfo
  {author} {\bibfnamefont {B.}~\bibnamefont {Joachimi}}, \bibinfo {author}
  {\bibfnamefont {L.}~\bibnamefont {Whiteway}}, \ and\ \bibinfo {author}
  {\bibfnamefont {O.}~\bibnamefont {Lahav}},\ }\href {\doibase
  10.1093/mnras/staa1836} {\bibfield  {journal} {\bibinfo  {journal} {Mon. Not.
  Roy. Astron. Soc.}\ }\textbf {\bibinfo {volume} {496}},\ \bibinfo {pages}
  {4647} (\bibinfo {year} {2020})},\ \Eprint {http://arxiv.org/abs/1910.07820}
  {arXiv:1910.07820 [astro-ph.CO]} \BibitemShut {NoStop}%
\bibitem [{\citenamefont {Raveri}\ \emph {et~al.}(2020)\citenamefont {Raveri},
  \citenamefont {Zacharegkas},\ and\ \citenamefont {Hu}}]{Raveri:2019gdp}%
  \BibitemOpen
  \bibfield  {author} {\bibinfo {author} {\bibfnamefont {M.}~\bibnamefont
  {Raveri}}, \bibinfo {author} {\bibfnamefont {G.}~\bibnamefont {Zacharegkas}},
  \ and\ \bibinfo {author} {\bibfnamefont {W.}~\bibnamefont {Hu}},\ }\href
  {\doibase 10.1103/PhysRevD.101.103527} {\bibfield  {journal} {\bibinfo
  {journal} {Phys. Rev. D}\ }\textbf {\bibinfo {volume} {101}},\ \bibinfo
  {pages} {103527} (\bibinfo {year} {2020})},\ \Eprint
  {http://arxiv.org/abs/1912.04880} {arXiv:1912.04880 [astro-ph.CO]}
  \BibitemShut {NoStop}%
\bibitem [{\citenamefont {Sakr}\ \emph {et~al.}(2018)\citenamefont {Sakr},
  \citenamefont {Ili\'c}, \citenamefont {Blanchard}, \citenamefont {Bittar},\
  and\ \citenamefont {Farah}}]{Sakr:2018new}%
  \BibitemOpen
  \bibfield  {author} {\bibinfo {author} {\bibfnamefont {Z.}~\bibnamefont
  {Sakr}}, \bibinfo {author} {\bibfnamefont {S.}~\bibnamefont {Ili\'c}},
  \bibinfo {author} {\bibfnamefont {A.}~\bibnamefont {Blanchard}}, \bibinfo
  {author} {\bibfnamefont {J.}~\bibnamefont {Bittar}}, \ and\ \bibinfo {author}
  {\bibfnamefont {W.}~\bibnamefont {Farah}},\ }\href {\doibase
  10.1051/0004-6361/201833151} {\bibfield  {journal} {\bibinfo  {journal}
  {Astron. Astrophys.}\ }\textbf {\bibinfo {volume} {620}},\ \bibinfo {pages}
  {A78} (\bibinfo {year} {2018})},\ \Eprint {http://arxiv.org/abs/1803.11170}
  {arXiv:1803.11170 [astro-ph.CO]} \BibitemShut {NoStop}%
\bibitem [{\citenamefont {Zubeldia}\ and\ \citenamefont
  {Challinor}(2019)}]{Zubeldia:2019brr}%
  \BibitemOpen
  \bibfield  {author} {\bibinfo {author} {\bibfnamefont {I.~n.}\ \bibnamefont
  {Zubeldia}}\ and\ \bibinfo {author} {\bibfnamefont {A.}~\bibnamefont
  {Challinor}},\ }\href {\doibase 10.1093/mnras/stz2153} {\bibfield  {journal}
  {\bibinfo  {journal} {Mon. Not. Roy. Astron. Soc.}\ }\textbf {\bibinfo
  {volume} {489}},\ \bibinfo {pages} {401} (\bibinfo {year} {2019})},\ \Eprint
  {http://arxiv.org/abs/1904.07887} {arXiv:1904.07887 [astro-ph.CO]}
  \BibitemShut {NoStop}%
\bibitem [{\citenamefont {Perivolaropoulos}\ and\ \citenamefont
  {Kazantzidis}(2019)}]{Perivolaropoulos:2019vkb}%
  \BibitemOpen
  \bibfield  {author} {\bibinfo {author} {\bibfnamefont {L.}~\bibnamefont
  {Perivolaropoulos}}\ and\ \bibinfo {author} {\bibfnamefont {L.}~\bibnamefont
  {Kazantzidis}},\ }\href {\doibase 10.1142/S021827181942001X} {\bibfield
  {journal} {\bibinfo  {journal} {Int. J. Mod. Phys. D}\ }\textbf {\bibinfo
  {volume} {28}},\ \bibinfo {pages} {1942001} (\bibinfo {year} {2019})},\
  \Eprint {http://arxiv.org/abs/1904.09462} {arXiv:1904.09462 [gr-qc]}
  \BibitemShut {NoStop}%
\bibitem [{\citenamefont {Leauthaud}\ \emph {et~al.}(2017)\citenamefont
  {Leauthaud} \emph {et~al.}}]{Leauthaud:2016jdb}%
  \BibitemOpen
  \bibfield  {author} {\bibinfo {author} {\bibfnamefont {A.}~\bibnamefont
  {Leauthaud}} \emph {et~al.},\ }\href {\doibase 10.1093/mnras/stx258}
  {\bibfield  {journal} {\bibinfo  {journal} {Mon. Not. Roy. Astron. Soc.}\
  }\textbf {\bibinfo {volume} {467}},\ \bibinfo {pages} {3024} (\bibinfo {year}
  {2017})},\ \Eprint {http://arxiv.org/abs/1611.08606} {arXiv:1611.08606
  [astro-ph.CO]} \BibitemShut {NoStop}%
\bibitem [{\citenamefont {Lange}\ \emph {et~al.}(2019)\citenamefont {Lange},
  \citenamefont {Yang}, \citenamefont {Guo}, \citenamefont {Luo},\ and\
  \citenamefont {van~den Bosch}}]{Lange:2019nya}%
  \BibitemOpen
  \bibfield  {author} {\bibinfo {author} {\bibfnamefont {J.~U.}\ \bibnamefont
  {Lange}}, \bibinfo {author} {\bibfnamefont {X.}~\bibnamefont {Yang}},
  \bibinfo {author} {\bibfnamefont {H.}~\bibnamefont {Guo}}, \bibinfo {author}
  {\bibfnamefont {W.}~\bibnamefont {Luo}}, \ and\ \bibinfo {author}
  {\bibfnamefont {F.~C.}\ \bibnamefont {van~den Bosch}},\ }\href {\doibase
  10.1093/mnras/stz2124} {\bibfield  {journal} {\bibinfo  {journal} {Mon. Not.
  Roy. Astron. Soc.}\ }\textbf {\bibinfo {volume} {488}},\ \bibinfo {pages}
  {5771} (\bibinfo {year} {2019})},\ \Eprint {http://arxiv.org/abs/1906.08680}
  {arXiv:1906.08680 [astro-ph.CO]} \BibitemShut {NoStop}%
\bibitem [{\citenamefont {Yuan}\ \emph
  {et~al.}(2020{\natexlab{a}})\citenamefont {Yuan}, \citenamefont
  {Eisenstein},\ and\ \citenamefont {Leauthaud}}]{Yuan:2019fbi}%
  \BibitemOpen
  \bibfield  {author} {\bibinfo {author} {\bibfnamefont {S.}~\bibnamefont
  {Yuan}}, \bibinfo {author} {\bibfnamefont {D.~J.}\ \bibnamefont
  {Eisenstein}}, \ and\ \bibinfo {author} {\bibfnamefont {A.}~\bibnamefont
  {Leauthaud}},\ }\href {\doibase 10.1093/mnras/staa634} {\bibfield  {journal}
  {\bibinfo  {journal} {Mon. Not. Roy. Astron. Soc.}\ }\textbf {\bibinfo
  {volume} {493}},\ \bibinfo {pages} {5551} (\bibinfo {year}
  {2020}{\natexlab{a}})},\ \Eprint {http://arxiv.org/abs/1907.05909}
  {arXiv:1907.05909 [astro-ph.CO]} \BibitemShut {NoStop}%
\bibitem [{\citenamefont {Zu}(2020)}]{Zu:2020uqo}%
  \BibitemOpen
  \bibfield  {author} {\bibinfo {author} {\bibfnamefont {Y.}~\bibnamefont
  {Zu}},\ }\href@noop {} {\  (\bibinfo {year} {2020})},\ \Eprint
  {http://arxiv.org/abs/2010.01143} {arXiv:2010.01143 [astro-ph.CO]}
  \BibitemShut {NoStop}%
\bibitem [{\citenamefont {Yuan}\ \emph
  {et~al.}(2020{\natexlab{b}})\citenamefont {Yuan}, \citenamefont {Hadzhiyska},
  \citenamefont {Bose}, \citenamefont {Eisenstein},\ and\ \citenamefont
  {Guo}}]{Yuan:2020xlk}%
  \BibitemOpen
  \bibfield  {author} {\bibinfo {author} {\bibfnamefont {S.}~\bibnamefont
  {Yuan}}, \bibinfo {author} {\bibfnamefont {B.}~\bibnamefont {Hadzhiyska}},
  \bibinfo {author} {\bibfnamefont {S.}~\bibnamefont {Bose}}, \bibinfo {author}
  {\bibfnamefont {D.~J.}\ \bibnamefont {Eisenstein}}, \ and\ \bibinfo {author}
  {\bibfnamefont {H.}~\bibnamefont {Guo}},\ }\href {\doibase
  10.1093/mnras/stab235} {\  (\bibinfo {year} {2020}{\natexlab{b}}),\
  10.1093/mnras/stab235},\ \Eprint {http://arxiv.org/abs/2010.04182}
  {arXiv:2010.04182 [astro-ph.CO]} \BibitemShut {NoStop}%
\bibitem [{\citenamefont {Lange}\ \emph {et~al.}(2020)\citenamefont {Lange},
  \citenamefont {Leauthaud}, \citenamefont {Singh}, \citenamefont {Guo},
  \citenamefont {Zhou}, \citenamefont {Smith},\ and\ \citenamefont
  {Cyr-Racine}}]{Lange:2020mnl}%
  \BibitemOpen
  \bibfield  {author} {\bibinfo {author} {\bibfnamefont {J.~U.}\ \bibnamefont
  {Lange}}, \bibinfo {author} {\bibfnamefont {A.}~\bibnamefont {Leauthaud}},
  \bibinfo {author} {\bibfnamefont {S.}~\bibnamefont {Singh}}, \bibinfo
  {author} {\bibfnamefont {H.}~\bibnamefont {Guo}}, \bibinfo {author}
  {\bibfnamefont {R.}~\bibnamefont {Zhou}}, \bibinfo {author} {\bibfnamefont
  {T.~L.}\ \bibnamefont {Smith}}, \ and\ \bibinfo {author} {\bibfnamefont
  {F.-Y.}\ \bibnamefont {Cyr-Racine}},\ }\href {\doibase 10.1093/mnras/stab189}
  {\  (\bibinfo {year} {2020}),\ 10.1093/mnras/stab189},\ \Eprint
  {http://arxiv.org/abs/2011.02377} {arXiv:2011.02377 [astro-ph.CO]}
  \BibitemShut {NoStop}%
\bibitem [{\citenamefont {Liu}\ and\ \citenamefont {Hill}(2015)}]{Liu:2015xfa}%
  \BibitemOpen
  \bibfield  {author} {\bibinfo {author} {\bibfnamefont {J.}~\bibnamefont
  {Liu}}\ and\ \bibinfo {author} {\bibfnamefont {J.~C.}\ \bibnamefont {Hill}},\
  }\href {\doibase 10.1103/PhysRevD.92.063517} {\bibfield  {journal} {\bibinfo
  {journal} {Phys. Rev. D}\ }\textbf {\bibinfo {volume} {92}},\ \bibinfo
  {pages} {063517} (\bibinfo {year} {2015})},\ \Eprint
  {http://arxiv.org/abs/1504.05598} {arXiv:1504.05598 [astro-ph.CO]}
  \BibitemShut {NoStop}%
\bibitem [{\citenamefont {Giannantonio}\ \emph {et~al.}(2016)\citenamefont
  {Giannantonio} \emph {et~al.}}]{Giannantonio:2015ahz}%
  \BibitemOpen
  \bibfield  {author} {\bibinfo {author} {\bibfnamefont {T.}~\bibnamefont
  {Giannantonio}} \emph {et~al.} (\bibinfo {collaboration} {DES}),\ }\href
  {\doibase 10.1093/mnras/stv2678} {\bibfield  {journal} {\bibinfo  {journal}
  {Mon. Not. Roy. Astron. Soc.}\ }\textbf {\bibinfo {volume} {456}},\ \bibinfo
  {pages} {3213} (\bibinfo {year} {2016})},\ \Eprint
  {http://arxiv.org/abs/1507.05551} {arXiv:1507.05551 [astro-ph.CO]}
  \BibitemShut {NoStop}%
\bibitem [{\citenamefont {Kuntz}(2015)}]{Kuntz:2015wza}%
  \BibitemOpen
  \bibfield  {author} {\bibinfo {author} {\bibfnamefont {A.}~\bibnamefont
  {Kuntz}},\ }\href {\doibase 10.1051/0004-6361/201526940} {\bibfield
  {journal} {\bibinfo  {journal} {Astron. Astrophys.}\ }\textbf {\bibinfo
  {volume} {584}},\ \bibinfo {pages} {A53} (\bibinfo {year} {2015})},\ \Eprint
  {http://arxiv.org/abs/1510.00398} {arXiv:1510.00398 [astro-ph.CO]}
  \BibitemShut {NoStop}%
\bibitem [{\citenamefont {Giusarma}\ \emph {et~al.}(2018)\citenamefont
  {Giusarma}, \citenamefont {Vagnozzi}, \citenamefont {Ho}, \citenamefont
  {Ferraro}, \citenamefont {Freese}, \citenamefont {Kamen-Rubio},\ and\
  \citenamefont {Luk}}]{Giusarma:2018jei}%
  \BibitemOpen
  \bibfield  {author} {\bibinfo {author} {\bibfnamefont {E.}~\bibnamefont
  {Giusarma}}, \bibinfo {author} {\bibfnamefont {S.}~\bibnamefont {Vagnozzi}},
  \bibinfo {author} {\bibfnamefont {S.}~\bibnamefont {Ho}}, \bibinfo {author}
  {\bibfnamefont {S.}~\bibnamefont {Ferraro}}, \bibinfo {author} {\bibfnamefont
  {K.}~\bibnamefont {Freese}}, \bibinfo {author} {\bibfnamefont
  {R.}~\bibnamefont {Kamen-Rubio}}, \ and\ \bibinfo {author} {\bibfnamefont
  {K.-B.}\ \bibnamefont {Luk}},\ }\href {\doibase 10.1103/PhysRevD.98.123526}
  {\bibfield  {journal} {\bibinfo  {journal} {Phys. Rev. D}\ }\textbf {\bibinfo
  {volume} {98}},\ \bibinfo {pages} {123526} (\bibinfo {year} {2018})},\
  \Eprint {http://arxiv.org/abs/1802.08694} {arXiv:1802.08694 [astro-ph.CO]}
  \BibitemShut {NoStop}%
\bibitem [{\citenamefont {Darwish}\ \emph {et~al.}(2020)\citenamefont {Darwish}
  \emph {et~al.}}]{Darwish:2020fwf}%
  \BibitemOpen
  \bibfield  {author} {\bibinfo {author} {\bibfnamefont {O.}~\bibnamefont
  {Darwish}} \emph {et~al.},\ }\href {\doibase 10.1093/mnras/staa3438}
  {\bibfield  {journal} {\bibinfo  {journal} {Mon. Not. Roy. Astron. Soc.}\
  }\textbf {\bibinfo {volume} {500}},\ \bibinfo {pages} {2250} (\bibinfo {year}
  {2020})},\ \Eprint {http://arxiv.org/abs/2004.01139} {arXiv:2004.01139
  [astro-ph.CO]} \BibitemShut {NoStop}%
\bibitem [{\citenamefont {Abbott}\ \emph {et~al.}(2020)\citenamefont {Abbott}
  \emph {et~al.}}]{Abbott:2020knk}%
  \BibitemOpen
  \bibfield  {author} {\bibinfo {author} {\bibfnamefont {T.~M.~C.}\
  \bibnamefont {Abbott}} \emph {et~al.} (\bibinfo {collaboration} {DES}),\
  }\href {\doibase 10.1103/PhysRevD.102.023509} {\bibfield  {journal} {\bibinfo
   {journal} {Phys. Rev. D}\ }\textbf {\bibinfo {volume} {102}},\ \bibinfo
  {pages} {023509} (\bibinfo {year} {2020})},\ \Eprint
  {http://arxiv.org/abs/2002.11124} {arXiv:2002.11124 [astro-ph.CO]}
  \BibitemShut {NoStop}%
\bibitem [{\citenamefont {Abazajian}\ \emph {et~al.}(2016)\citenamefont
  {Abazajian} \emph {et~al.}}]{Abazajian:2016yjj}%
  \BibitemOpen
  \bibfield  {author} {\bibinfo {author} {\bibfnamefont {K.~N.}\ \bibnamefont
  {Abazajian}} \emph {et~al.} (\bibinfo {collaboration} {CMB-S4}),\ }\href@noop
  {} {\  (\bibinfo {year} {2016})},\ \Eprint {http://arxiv.org/abs/1610.02743}
  {arXiv:1610.02743 [astro-ph.CO]} \BibitemShut {NoStop}%
\bibitem [{\citenamefont {Ade}\ \emph {et~al.}(2019)\citenamefont {Ade} \emph
  {et~al.}}]{Ade:2018sbj}%
  \BibitemOpen
  \bibfield  {author} {\bibinfo {author} {\bibfnamefont {P.}~\bibnamefont
  {Ade}} \emph {et~al.} (\bibinfo {collaboration} {Simons Observatory}),\
  }\href {\doibase 10.1088/1475-7516/2019/02/056} {\bibfield  {journal}
  {\bibinfo  {journal} {JCAP}\ }\textbf {\bibinfo {volume} {02}},\ \bibinfo
  {pages} {056} (\bibinfo {year} {2019})},\ \Eprint
  {http://arxiv.org/abs/1808.07445} {arXiv:1808.07445 [astro-ph.CO]}
  \BibitemShut {NoStop}%
\bibitem [{\citenamefont {Abitbol}\ \emph {et~al.}(2019)\citenamefont {Abitbol}
  \emph {et~al.}}]{Abitbol:2019nhf}%
  \BibitemOpen
  \bibfield  {author} {\bibinfo {author} {\bibfnamefont {M.~H.}\ \bibnamefont
  {Abitbol}} \emph {et~al.} (\bibinfo {collaboration} {Simons Observatory}),\
  }\href@noop {} {\bibfield  {journal} {\bibinfo  {journal} {Bull. Am. Astron.
  Soc.}\ }\textbf {\bibinfo {volume} {51}},\ \bibinfo {pages} {147} (\bibinfo
  {year} {2019})},\ \Eprint {http://arxiv.org/abs/1907.08284} {arXiv:1907.08284
  [astro-ph.IM]} \BibitemShut {NoStop}%
\bibitem [{\citenamefont {Ivezi\'c}\ \emph {et~al.}(2019)\citenamefont
  {Ivezi\'c} \emph {et~al.}}]{Ivezic:2008fe}%
  \BibitemOpen
  \bibfield  {author} {\bibinfo {author} {\bibfnamefont {v.}~\bibnamefont
  {Ivezi\'c}} \emph {et~al.} (\bibinfo {collaboration} {LSST}),\ }\href
  {\doibase 10.3847/1538-4357/ab042c} {\bibfield  {journal} {\bibinfo
  {journal} {Astrophys. J.}\ }\textbf {\bibinfo {volume} {873}},\ \bibinfo
  {pages} {111} (\bibinfo {year} {2019})},\ \Eprint
  {http://arxiv.org/abs/0805.2366} {arXiv:0805.2366 [astro-ph]} \BibitemShut
  {NoStop}%
\bibitem [{\citenamefont {Aghamousa}\ \emph {et~al.}(2016)\citenamefont
  {Aghamousa} \emph {et~al.}}]{Aghamousa:2016zmz}%
  \BibitemOpen
  \bibfield  {author} {\bibinfo {author} {\bibfnamefont {A.}~\bibnamefont
  {Aghamousa}} \emph {et~al.} (\bibinfo {collaboration} {DESI}),\ }\href@noop
  {} {\  (\bibinfo {year} {2016})},\ \Eprint {http://arxiv.org/abs/1611.00036}
  {arXiv:1611.00036 [astro-ph.IM]} \BibitemShut {NoStop}%
\bibitem [{\citenamefont {Weltman}\ \emph {et~al.}(2020)\citenamefont {Weltman}
  \emph {et~al.}}]{Bull:2018lat}%
  \BibitemOpen
  \bibfield  {author} {\bibinfo {author} {\bibfnamefont {A.}~\bibnamefont
  {Weltman}} \emph {et~al.},\ }\href {\doibase 10.1017/pasa.2019.42} {\bibfield
   {journal} {\bibinfo  {journal} {Publ. Astron. Soc. Austral.}\ }\textbf
  {\bibinfo {volume} {37}},\ \bibinfo {pages} {e002} (\bibinfo {year}
  {2020})},\ \Eprint {http://arxiv.org/abs/1810.02680} {arXiv:1810.02680
  [astro-ph.CO]} \BibitemShut {NoStop}%
\bibitem [{\citenamefont {Amendola}\ \emph
  {et~al.}(2013{\natexlab{b}})\citenamefont {Amendola} \emph
  {et~al.}}]{Amendola:2012ys}%
  \BibitemOpen
  \bibfield  {author} {\bibinfo {author} {\bibfnamefont {L.}~\bibnamefont
  {Amendola}} \emph {et~al.} (\bibinfo {collaboration} {Euclid Theory Working
  Group}),\ }\href {\doibase 10.12942/lrr-2013-6} {\bibfield  {journal}
  {\bibinfo  {journal} {Living Rev. Rel.}\ }\textbf {\bibinfo {volume} {16}},\
  \bibinfo {pages} {6} (\bibinfo {year} {2013}{\natexlab{b}})},\ \Eprint
  {http://arxiv.org/abs/1206.1225} {arXiv:1206.1225 [astro-ph.CO]} \BibitemShut
  {NoStop}%
\end{thebibliography}%

\end{document}